\documentclass[reprint,superscriptaddress,amsmath,amssymb,aps,pre]{revtex4-2}
\usepackage{graphicx}
\usepackage{dcolumn}
\usepackage{bm}
\usepackage{color}
\usepackage{units}
\usepackage{wasysym}
\usepackage{MnSymbol}
\usepackage{comment}
\usepackage{times}
\usepackage{upgreek}
\usepackage [english]{babel}
\usepackage[mathlines]{lineno}
\usepackage[unicode=true,pdfusetitle,bookmarks=false,colorlinks=true,citecolor=blue,urlcolor=blue,linkcolor=blue]{hyperref}
\begin{document}

\title{The Mechanics of Nucleation and Growth and the Surface Tensions of Active Matter}

\author{Luke Langford}
\affiliation{Department of Materials Science and Engineering, University of California, Berkeley, California 94720, USA}

\author{Ahmad K. Omar}
\email{aomar@berkeley.edu}
\affiliation{Department of Materials Science and Engineering, University of California, Berkeley, California 94720, USA}
\affiliation{Materials Sciences Division, Lawrence Berkeley National Laboratory, Berkeley, California 94720, USA}

\begin{abstract}
Homogeneous nucleation, a textbook transition path for phase transitions, is typically understood on thermodynamic grounds through the prism of classical nucleation theory.
However, recent studies have suggested the applicability of classical nucleation theory to systems far from equilibrium.
In this Article, we formulate a purely mechanical perspective of homogeneous nucleation and growth, elucidating the criteria for the properties of a critical nucleus without appealing to equilibrium notions.
Applying this theory to active fluids undergoing motility-induced phase separation, we find that nucleation proceeds in a qualitatively similar fashion to equilibrium systems, with concepts such as the Gibbs-Thomson effect and nucleation barriers remaining valid.
We further demonstrate that the recovery of such concepts allows us to extend classical theories of nucleation rates and coarsening dynamics to active systems upon using the mechanically-derived definitions of the nucleation barrier and surface tensions.
Three distinct surface tensions -- the mechanical, capillary, and Ostwald tensions -- play a central role in our theory.
While these three surface tensions are identical in equilibrium, our work highlights the distinctive role of each tension in the stability of active interfaces and the nucleation and growth of motility-induced phases. 
\end{abstract}

\maketitle

\section{Introduction} 
In recent years it has become clear that many of the intriguing patterns and phases formed by active systems do not occur spontaneously but require a rate-determining rare event~\cite{Richard2016NucleationParticles,Redner2016,Levis2017ActiveCoexistence,Cates2023ClassicalSeparation}. 
The rare events needed to initiate a variety of active phase transitions often appear remarkably similar to those of the corresponding process in passive systems.
Phase separation and crystallization of homogeneous active fluids, under certain conditions, require the formation of a \textit{spherical critical nucleus}~\cite{Redner2016, Omar2021}. 
Upon forming, the nuclei can continue to grow and coarsen through the familiar coarsening mechanisms of coalescence and Ostwald ripening~\cite{Liu2019Self-DrivenFormation,Zhang2021ActiveDensity,Redner2013StructureFluid}.
These observations suggest that \textit{classical nucleation theory }(CNT) and established coarsening theories may be applicable to active systems~\cite{Becker1935KinetischeDampfen,Cahn1959FreeFluid,Lifshitz1961TheSolutions,Oxtoby1992HomogeneousExperiment,Baldan2002ReviewTheories,Desai2009DynamicsStructures,Debenedetti2020MetastableLiquids}. 
CNT has shaped and framed our understanding of the transition rate of a variety of activated processes in passive systems including the rupture of biological membranes~\cite{Ting2011MinimumRupture}, the formation of dust clouds on exoplanets~\cite{Kiefer2023TheNucleation}, and the motion of ferroelectric domain walls~\cite{Shin2007NucleationMotion}. 
In its canonical form, CNT implicitly leverages microscopic time-reversal symmetry (TRS) to posit a \textit{free energy barrier} for the rare event. 
In the case of a metastable fluid that will ultimately crystallize or phase separate, the rare event required to initiate the phase transition is the formation of a spherical finite-sized nucleus. 
Straightforward thermodynamic arguments weighing the competing effects of the volumetric driving force for forming a nucleus and the associated interfacial penalty allow for the determination of the barrier height and the critical nucleus' density and radius using bulk equations of state (e.g.,~the chemical potential) and the surface energy.

The naive application of the above ideas to active systems results in immediate issues: concepts such as chemical potential~\cite{Takatori2015TowardsMatter,Solon2018GeneralizedMatter} and surface energy~\cite{Bialke2015,Lee2017InterfaceSeparations,Tjhung2018,Hermann2019b,Fausti2021CapillarySeparation,Langford2024TheoryPhases} are not uniquely defined in active systems.
Nevertheless, a number of simulations and numerical studies in the last decade have made it clear that CNT and post-nucleation coarsening theories may be applicable far from equilibrium.
Zakine~\textit{et al.} developed~\cite{Zakine2023Minimum-ActionTransitions} and applied~\cite{Zakine2024UnveilingMethod} action minimizing methods to stochastic field dynamics described by Active Model B~\cite{Wittkowski2014ScalarSeparation} and found that the transition from homogeneous to inhomogeneous states is well described by the rate-limiting formation of a radially symmetric nucleus.
Particle-based simulations of metastable active fluids undergoing motility-induced phase separation (MIPS) conducted by Redner~\textit{et al.}~\cite{Redner2016} were consistent with the cluster rate formation statistics posited by Becker-D\"{o}ring theory~\cite{Becker1935KinetischeDampfen,Oxtoby1992HomogeneousExperiment,Debenedetti2020MetastableLiquids}. 
These observations suggest that one may be able to construct a theory for the nucleation of active phases with similar physical assumptions as CNT -- but without invoking a thermodynamic variational principle.

Significant progress towards generalizing CNT to nonequilibrium fluids has been recently made by Cates, Nardini, and co-workers~\cite{Tjhung2018,Cates2023ClassicalSeparation}. 
Introducing the minimal stochastic field theory, Active Model B+ (AMB+), they demonstrated that the size and density of the critical nucleus can be solely determined from dynamics~\cite{Tjhung2018} and elucidated the conditions at which the critical nucleus represents an unstable fixed point.
Furthermore, they derived the radial growth dynamics of active nuclei, finding theoretical justification for an effective free energy landscape that controls the size distribution of pre-critical nuclei~\cite{Cates2023ClassicalSeparation}.
Although these efforts have yielded crucial insights towards the nucleation of active phases, explicitly mapping microscopic models of active systems to the phenomenological coefficients of AMB+ may not always be possible.
Moreover, the absence of a unique mechanical interpretation of active field theories makes it difficult to understand the physical significance of the \textit{negative mechanical surface tension}~\cite{Bialke2015} (which \textit{does} control the Laplace pressure difference of active bubbles and droplets~\cite{Solon2018GeneralizedEnsembles}) of active particles in nucleation.

In this Article, we seek to answer the following questions beginning from a microscopic description and without appealing to equilibrium notions: Provided the saturation conditions, how might one determine the size ($R_C$) and density ($\rho^{\rm in}$) of the critical nucleus?; what is the precise role of the various active surface tensions in the nucleation process?; and how might one describe the post-nucleation growth dynamics of coarsening active fluids? 
To address these questions, we begin by deriving the conditions for determining the density and radius of a critical nucleus from \textit{purely mechanical considerations}.
We apply this framework to fluid nucleation of active Brownian particles (ABPs) and determine the critical nucleus properties as a function of activity and the saturation density.
We then combine the recently developed perspective of Cates and Nardini~\cite{Cates2023ClassicalSeparation} with our fluctuating hydrodynamics of ABPs~\cite{Langford2024TheoryPhases} to describe the time evolution of the critical nucleus size, finding that droplets of ABPs are associated with a far higher nucleation barrier than bubbles.
This nucleation barrier follows the familiar equilibrium intuition of an energetic \textit{penalty} scaling with the surface area of the nucleus competing with an energetic \textit{driving force} scaling with the volume of the nucleus.
Building off our active CNT, we are able to extend classic perspectives on coarsening, including those of Lifhitz, Sylozov, and Wagner~\cite{Lifshitz1961TheSolutions} as well as Voorhees and Glicksman~\cite{Voorhees1984SolutionTheory,Voorhees1984SolutionSimulations}, to systems far from equilibrium.
We are thus able to numerically investigate large-scale, late-stage coarsening dynamics that would otherwise be prohibitive for particle-based simulations.

Surprisingly, we find that the properties and dynamics of the critical nucleus of ABPs are dependent on \textit{three} different surface tensions.
These tensions include the negative mechanical surface tension~\cite{Bialke2015}, which controls the pressure difference between the nucleus and parent phase, the capillary-wave tension~\cite{Fausti2021CapillarySeparation,Langford2024TheoryPhases}, which drives interfaces to minimize surface area and results in a spherical nucleus geometry, and the Ostwald tension~\cite{Tjhung2018}, which plays a central role in both the nucleation barrier and coarsening dynamics. 
At equilibrium, all three of these tensions must be identical and are defined by the energetic cost of increasing interfacial area, but far from equilibrium they are free to take on disparate values.
Despite this nonequilibrium peculiarity, we demonstrate the recovery of phenomena typically associated with equilibrium, such as the Gibbs-Thomson effect and nucleation barriers, from dynamics that microscopically violate time reversal symmetry.

\section{Results and Discussion}
\subsection{Equilibrium Nucleation Theory}
\label{sec:EQCNT}
We briefly recapitulate the thermodynamic arguments of CNT applied to bubble/droplet nucleation, the focus of this work. 
The complete derivation can be found in Section 3 of the Supplementary Material (SM). 
We consider a macroscopic homogeneous single-component system prepared with a uniform number density $\rho^{\rm sat}$ within a fixed system volume $V$ (the total number of particles is thus also fixed at $N = V\rho^{\rm sat}$).
We focus on saturation densities $\rho^{\rm sat}$ within the liquid-gas binodal of the phase diagram but far from the spinodal boundaries, limiting our analysis to be far from the critical point.
This limitation is necessary in order to ensure that the pathway of the phase transition is unambiguously nucleation and growth, rather than spinodal decomposition.
In this region of the phase diagram, a system prepared with a spatially uniform density of $\rho^{\rm sat}$ will remain linearly stable against density fluctuations until a critical nucleus with density $\rho^{\rm in}$ and volume $V^{\rm nuc}$ is formed. 
We assume that the formation of the critical nucleus is a sufficiently rare event that we may consider a single nucleation event in isolation and a nucleus volume, $V^{\rm nuc}$, that is much smaller than the system volume.
These assumptions provide additional limits on the range of $\rho^{\rm sat}$ in which CNT is applicable. 

The extensive Helmholtz free energy for the system is simply {$\mathcal{F} = V^{\rm nuc} f(\rho^{\rm in}) + \left(V - V^{\rm nuc}\right)f(\rho^{\rm out}) + \partial V^{\rm nuc}\upgamma$} where $f(\rho)$ is the free energy density of a uniform fluid with density $\rho$, the density of the parent phase surrounding the nucleus is $\rho^{\rm out}$, and $\upgamma(>0)$ is the surface energy.
The interfacial penalty ensures that the mean-field nucleus shape will be spherical in order to reduce surface area, which we have labeled as $\partial V^{\rm nuc}$.
Conservation of the total particle number results in a constraint of {$\rho^{\rm in}V^{\rm nuc} + \rho^{\rm out}\left(V - V^{\rm nuc}\right) = \rho^{\rm sat}V$}.
Note that while the spatial density profile must smoothly vary from the nucleus density $\rho^{\rm in}$ to the parent phase density $\rho^{\rm out}$ this interfacial width is much smaller than the nucleus size and its contribution to the particle conservation constraint is negligible.
This allows us to focus on the coexisting interior and exterior densities rather than the complete spatial profile. 
We then extremize the free energy (subject to the particle conservation constraint) with respect to the coexisting densities and nucleus size and find the following conditions:
\begin{subequations}
\label{eq:thermocoexist}
    \begin{align}
        \mu(\rho^{\rm in}) &= \mu(\rho^{\rm out}),\\ p(\rho^{\rm in}) &= p(\rho^{\rm out}) +  \frac{2\upgamma}{R_C},\label{eq:thermocritradius}
    \end{align}
\end{subequations}
where we have defined the chemical potential $\mu(\rho)\equiv \partial f/\partial \rho$ and pressure $p(\rho) \equiv -f(\rho) + \rho\mu(\rho)$.
Here, $R_C$ is the critical radius of the spherical nucleus. 
These criteria are the expected conditions of chemical potential equality and the Young-Laplace pressure difference which, due to the positive surface energy, results in a higher nucleus pressure relative to the parent phase. 
Together with the particle conservation constraint, Eqs.~\eqref{eq:thermocoexist} can be used to solve for $\rho^{\rm in}$, $\rho^{\rm out}$ and $R_C$.

We can make use of the condition that the droplet volume is much smaller than the system volume such that $\rho^{\rm out} \approx \rho^{\rm sat}$. 
In this limit, we have effectively changed our thermodynamic ensemble as the nucleus is now in contact with a particle reservoir with fixed chemical potential $\mu(\rho^{\rm sat}) \equiv \mu^{\rm sat}$ and thus $\rho^{\rm in}$ can be found directly from  $\mu(\rho^{\rm in}) = \mu^{\rm sat}$.
The critical radius is then simply ${R_C = 2\upgamma/(\mu^{\rm sat}\Delta\rho - \Delta f})$ where the difference between an arbitrary property, $a$, in the interior and exterior of the nucleus is defined as $\Delta a \equiv a^{\rm in} - a^{\rm sat}$.
The free energy barrier associated with the critical nucleus can be expressed as the sum of the volumetric and interfacial contributions ${F^{\rm \dag} =4\pi R_C^2\upgamma - \frac{4}{3}\pi R_C^3 (\mu^{\rm sat}\Delta \rho - \Delta f) }$ or equivalently, ${F^{\rm \dag} = \frac{4}{3}\pi \upgamma R_C^2}$.
The probability of observing such a critical nucleus is thus expected to be $P(R_C)\sim \text{exp}\left[F^{\dag}/k_BT\right]$.

It is clear that in passive systems, \textit{thermodynamics alone} provides a guide for determining the density, size, and statistics of the critical nucleus.
Given a saturation density, Eqs.~\eqref{eq:thermocoexist} are the two equations that allow for the determination of the two unknowns: $\rho^{\rm in}$ and $R_C$. 
In Sec.~\ref{sec:NUCMECH}, we derive analogous equations capable of determining $\rho^{\rm in}$ and $R_C$ for systems \textit{arbitrarily far from equilibrium} by adopting a mechanical perspective. 

\subsection{Mechanics of a Critical Nucleus}
\label{sec:NUCMECH}
The pathway for transitioning from a homogeneous supersaturated active fluid to a state of two-fluid coexistence appears eerily reminiscent to those of passive systems~\cite{Redner2016, Zakine2024UnveilingMethod}.
In both scenarios, the initial activated process is the formation of a \textit{spherical nucleus} with a specific critical radius $R_C$ and internal density $\rho^{\rm in}$.
In the absence of a free energy, the traditional arguments previously outlined used to establish passive classical nucleation theory are no longer applicable. 
We therefore take a generalized \textit{mechanical approach} with the aim of extending CNT to systems arbitrarily far from equilibrium.
We again limit our analysis to saturation densities near the liquid and gas binodal densities such that the density of the fluid surrounding a critical nucleus is approximately $\rho^{\rm sat}$ and we may focus on a \textit{single} nucleus coexisting with this saturated fluid. 

Let us consider the dynamical definition of a critical nucleus.
When the critical nucleus is an unstable fixed point~\cite{Note1}, nuclei smaller (larger) than $R_C$ will be associated with a particle flux normal to the interface such that the nuclei shrink (grow).
$R_C$ thus represents a critical size at which the direction of this flux switches, so the system must have zero flux normal to the interface precisely when a nucleus has a radius of $R_C$.
For a single realization of an active fluid, the stochastic noise plays a crucial role in the density flux, which we will discuss further in Sec.~\ref{sec:NUCLANG}. 
For the purposes of determining the properties of the critical nucleus, we can consider an \textit{ensemble} of supersaturated states. 
We choose a spherical coordinate system with an origin coinciding with that of the nucleus.
The ensemble-averaged nucleus shape will be spherical and the ensemble-averaged density flux will no longer be stochastic.
We will determine the critical nucleus size by searching for states of zero radial flux. 

Two flux-free states are well-established: the macroscopically phase separated state that the system will eventually reach and the initial metastable homogeneous state.
The critical nucleus will correspond to an additional state in which a \textit{finite-size} minority phase with zero radial flux coexists with a parent phase.
The condition for determining a stationary state of a density profile is given by the linear momentum balance (see Section 1 of the SM and Ref.~\cite{Omar2023b}) which governs the dynamics of the density flux.  
On timescales where inertia can be neglected (or for all times for overdamped particle dynamics), the radial-component of the linear momentum balance is simply $\left(\boldsymbol{\nabla}\cdot\bm{\sigma} + \mathbf{b}\right)\cdot\mathbf{e}_r = 0$, where $\bm{\sigma}$ is the stress tensor, $\mathbf{b}$ are the body forces, and $\mathbf{e}_r$ is the unit vector in the radial dimension.
We note that exact microscopic expressions for both $\bm{\sigma}$ and $\mathbf{b}$ can be determined through an Irving-Kirkwood-Noll procedure~\cite{Irving1950,Noll1955,Lehoucq2010}, an approach that we will later adopt in applying this theory to active Brownian particles. 
In these formulations, terms often appear to be non-local, requiring integrals over all space to, for instance, compute the stress  at a particular point in space arising from particle interactions. 
Nevertheless, these terms are often well-approximated (e.g.,~when interactions are short-ranged and partial spatial correlations decay quickly) by a series of local terms, the coefficients of which can be systematically determined~\cite{Lindell1993DeltaMethod}.
We thus focus on systems in which $\bm{\sigma}$ and $\mathbf{b}$ can be fully expressed as local density gradient expansions.
Such a focus may become limiting for application to systems with truly long-ranged forces that prevent a local expansion of stresses/body forces.

Up to a divergence-free term, we may absorb the body force into a stress-like contribution such that ${\boldsymbol{\nabla}\cdot\bm{\sigma} + \mathbf{b} = \boldsymbol{\nabla}\cdot\left[\bm{\sigma} + \mathbf{F}\right] = \boldsymbol{\nabla}\cdot\bm{\Sigma}}$, where $\mathbf{F}$ is the Coulomb integral of $\mathbf{b}$ and $\bm{\Sigma} \equiv \bm{\sigma} + \mathbf{F}$ is the dynamic stress tensor.
While $\bm{\sigma}$ and $\mathbf{b}$ are fully local, the Coulomb integral may result in non-local contributions to $\mathbf{F}$.
The momentum balance (${\boldsymbol{\nabla}\cdot\bm{\Sigma} = \bm{0}}$) can be used to solve for the density profile upon identifying the precise form of the dynamic stress tensor. 
Under spherical symmetry, $\bm{\Sigma}$ has only two independent components and can be expressed as $\bm{\Sigma} = \Sigma_{rr}\mathbf{e}_r\mathbf{e}_{r} + \Sigma_{tt}\left(\mathbf{I} - \mathbf{e}_r\mathbf{e}_r\right)$ where $\mathbf{I}$ is the identity tensor.
We can make analytical progress by expanding the dynamic stress tensor to second order in density gradients and assuming spatial isotropy in the expansion coefficients:
\begin{subequations}
    \label{eq:dynstress}
    \begin{align}
    \end{align}
    \begin{align}
        \Sigma_{tt} = -\mathcal{P}(\rho) + \kappa_1(\rho)\left(\frac{\partial}{\partial r} + \frac{2}{r}\right)\frac{\partial\rho}{\partial r} + \kappa_2(\rho)\left(\frac{\partial\rho}{\partial r}\right)^2 + \mathcal{K},
    \end{align}
\end{subequations}
where $-\mathcal{P}(\rho)$ is the bulk isotropic contribution to the dynamic stress tensor, $\mathcal{K}$ is the non-local term originating from the Coulomb integral of $\mathbf{b}$, and $\{\kappa_i\}$ are interfacial coefficients. 
These interfacial coefficients can be found by formally carrying out the expansion of $\bm{\sigma}$ and $\mathbf{b}$ beginning from the microscopic definition of stress \textit{and} expanding the local terms that arise from the Coulomb integral of $\mathbf{b}$.
The procedure for determining the precise forms of $\{\kappa_i\}$ and $\mathcal{K}$ from the expansion coefficients of $\bm{\sigma}$ and $\mathbf{b}$ is outlined in Section 1.7 of the SM.
In particular, a form for the nonlocal $\mathcal{K}$ term is found. We note that this form is free of any terms with explicit dependence on $1/r$. 
In a passive system $\mathcal{P}(\rho)$ is simply the bulk pressure and $\{\kappa_i\}$ are given by the Korteweg expansion~\cite{Korteweg1904} while non-local terms can be well-approximated by local terms for single-component systems with short-ranged interactions. 
More generally, the bulk pressure and interfacial coefficients can contain dynamical contributions otherwise absent in passive systems~\cite{Langford2024TheoryPhases}.
We also note that under the spherically symmetric configuration and stresses assumed here,  setting the radial flux to zero also implies zero flux in the angular directions.
In two dimensions, it is possible for a critical nucleus with radial symmetry to develop a finite angular flux if the underlying microscopic dynamics violate spatial parity. 
Describing the critical nucleus in  chiral active matter~\cite{Soni2019TheFluid,Ma2022DynamicalParticles}, would require a different expansion of the stress tensor such as those presented in Refs.~\cite{Klymko2017StatisticalHydrodynamics,Langford2025PhaseMatter}. 

Given the above physical picture, we now outline our process to find the zero radial flux states associated with a finite-sized minority phase.
Our spherically symmetric geometry results in the linear momentum balance simplifying to:
\begin{equation}
    0 = \frac{\partial \Sigma_{rr}}{\partial r} - \frac{2}{r}\mathcal{G}(\rho),\label{eq:Jstresssymmetry}
\end{equation}
where we have defined $\mathcal{G}(\rho) = \Sigma_{tt} - \Sigma_{rr}$ as the ``excess stress'' in the non-radial directions.
Deep within the both the nucleus and parent phase, where density gradients vanish, the dynamic stress tensor must be isotropic, implying $\mathcal{G}(\rho)$ is \textit{zero everywhere except within the interfacial region}. 
If we assume that the width of the interfacial region, $w$, is small compared to the radius of the critical nucleus, integration of Eq.~\eqref{eq:Jstresssymmetry} from $r=0$ to $r=\infty$ results in:
\begin{subequations}
\label{eq:coexistcriteria1}
\begin{equation}
\label{eq:activelaplacepress}
    \Delta\mathcal{P} = \mathcal{K}\bigr|_{r=0} + \frac{2\upgamma_{\rm mech}}{R_C} + \mathcal{O}\left(\frac{1}{R_C^2}\right).
\end{equation}
Here, $\upgamma_{\rm mech}$ is the \textit{mechanical surface tension} as defined by Kirkwood and Buff~\cite{Kirkwood1949}:
\begin{align}
    \label{eq:mechtensiondef}
    \upgamma_{\rm mech} &= \int_0^{\infty} \left[\Sigma_{tt} - \Sigma_{rr}\right] \text{d}r, \nonumber \\ &=  -\int_0^{\infty}\left[\kappa_3(\rho)\left(\frac{\partial \rho}{\partial r}\right)^2 + \kappa_4(\rho)\frac{\partial^2\rho}{\partial r^2} \right]\text{d}r,
\end{align}
\end{subequations}
where the second equality utilizes our specific form of the dynamic stress [see Eqs.~\eqref{eq:dynstress}].
Equation~\eqref{eq:activelaplacepress} is a generalized Young-Laplace equation -- with the equilibrium surface energy and pressure more generally replaced by the mechanical tension and dynamic pressure, respectively, and an additional contribution arising from the non-local terms.
Equation~\eqref{eq:mechtensiondef} carries the familiar mechanical intuition of surface tension: it measures the excess stress parallel to the interface.
This mechanical tension was determined to be \textit{strongly negative} for phase-separated ABPs~\cite{Bialke2015} and has been the subject of some controversy regarding its physical relevance.
Solon~\textit{et al.}~\cite{Solon2018GeneralizedEnsembles} correctly identified that the mechanical surface tension of Kirkwood and Buff controls the Laplace pressure difference of droplets consistent with a fully local (dynamic) stress.
More generally, the nonlocal term $\mathcal{K}$ may also contribute to the Laplace pressure difference.
Note that Eq.~\eqref{eq:activelaplacepress} ignores terms higher order in $1/R_C$. 
This approximation becomes increasingly accurate as the saturation is reduced [see Fig.~\ref{fig:figure2}].

While the generalized Young-Laplace equation provides a mechanical condition remarkably similar to its equilibrium counterpart, we still require a second condition to allow us to determine our two unknowns, $\rho^{\rm in}$ and $R_C$.
For passive fluids, this condition is equality of chemical potential between the nucleus and the surrounding phase. 
The search for chemical potential-like quantities for nonequilibrium systems has been the subject of much interest in the last decade in the context of determining macroscopic binodals~\cite{Takatori2015TowardsMatter, Solon2015, Solon2018GeneralizedMatter, Omar2023b}.
These works have resulted in generalized Maxwell constructions~\cite{Solon2018GeneralizedMatter, Omar2023b} that are derived from performing the appropriate integration of the mechanical criteria of coexistence~\cite{Aifantis1983TheRule, Omar2023b}.

We follow similar arguments used in constructing the nonequilibrium coexistence criteria of macroscopic binodals~\cite{Aifantis1983TheRule, Omar2023b}. 
Beginning from the mechanical balance [Eq.~\eqref{eq:Jstresssymmetry}] we perform an indefinite integral to find an integro-differential equation for $\Sigma_{rr}$ with the Laplace pressure used as a boundary condition. 
We then introduce the generalized Maxwell construction pseudovariable~\cite{Aifantis1983TheRule,Solon2018GeneralizedMatter,Omar2023b} $\mathcal{E}$ defined by the differential equation ${(\kappa_1+\kappa_4)\partial^2 \mathcal{E}/\partial \rho^2 = (2(\kappa_2+\kappa_3) - \partial (\kappa_1+\kappa_4)/\partial \rho) \partial \mathcal{E}/\partial \rho }$, which is found to be $\mathcal{E} \sim 1/\rho$ for a passive system. 
This pseudovariable allows us to eliminate nonlinear terms (just as it does in the case of macroscopic coexistence~\cite{Aifantis1983TheRule,Solon2018GeneralizedMatter,Omar2023b}) by multiplying the antiderivative of Eq.~\eqref{eq:Jstresssymmetry} by $d\mathcal{E}/dr$ and integrating from $r=0$ to $r=\infty$ to arrive at:
\begin{subequations}
\label{eq:coexistcriteria2}
\begin{align}
    \int_{\mathcal{E}^{\rm in}}^{\mathcal{E}^{\rm sat}}\text{d}\mathcal{E}\left[\mathcal{P}(\rho) - \mathcal{P}(\rho^{\rm sat})\right] + &\int_0^{\infty}\left(\mathcal{E}^{\rm in}-\mathcal{E}\right)\frac{\partial\mathcal{K}}{\partial r}\text{d}r\nonumber  \\ = &\frac{2\upgamma_{\rm ost}}{R_C} \frac{\rho^{\rm sat}\Delta\mathcal{E}}{\Delta\rho} + \mathcal{O}\left(\frac{1}{R_C^2}\right),\label{eq:activecoexist2}
\end{align}
where, analogously to Tjhung \textit{et al.}~\cite{Tjhung2018}, we have defined the ``Ostwald'' tension as
\begin{align}
    \upgamma_{\rm ost} = \frac{\Delta\rho}{\rho^{\rm sat}\Delta\mathcal{E}}\Biggl[\int_0^{\infty}\Bigl(\kappa_1(\rho)\frac{\partial\mathcal{E}}{\partial \rho}\left(\frac{\partial\rho}{\partial r}\right)^2    + \mathcal{G}\left(\mathcal{E}-\mathcal{E}^{\rm in}\right) \Bigr)\text{d}r\Biggr].\label{eq:osttensiondef}
\end{align}
\end{subequations}
Note that $\upgamma_{\rm ost}$ has been defined such that it has the same units as mechanical surface tension and becomes identical to $\upgamma_{\rm mech}$ in a passive system.
$\upgamma_{\rm ost}$ is named for its importance in understanding the evaporation-condensation mechanism of active coarsening, which we will discuss in Sec.~\ref{sec:ACD}.
As $\rho^{\rm sat}$ approaches a binodal of the phase diagram, the right hand side of Eq.~\eqref{eq:activecoexist2} vanishes and we are left with one of the macroscopic coexistence criteria.
Interestingly, a term depending on gradients (and therefore the interface) remains.
At equilibrium, the details of the interface do not modify the coexistence criteria as interfacial effects are generally sub-extensive in the free energy. 
However, outside of equilibrium no such guarantee exists. 
This interfacial term vanishes as the derivative of $\mathcal{K}$ approaches zero, and one may refer to the expansion of $\mathcal{K}$ in the SM for a precise expression of this limit.

The connection between Eqs.~\eqref{eq:activelaplacepress}~and~\eqref{eq:activecoexist2} to the thermodynamic criteria shown in Eqs.~\eqref{eq:thermocoexist} can be made explicit.
As described by Solon~\textit{et al.}, the generalized Maxwell constructions on the pressure can also be expressed as a difference in chemical potential-like quantities between coexisting phases~\cite{Solon2018GeneralizedEnsembles}. 
It was subsequently shown that this potential satisfies a generalized Gibbs-Duhem relation~\cite{Evans2023}, which, for the single-component system under consideration, is expressed as $\text{d}g = \mathcal{E}\text{d}\mathcal{P}$ where $g$ is the generalized chemical potential.  
The generalized Gibbs-Duhem relation allows Eq.~\eqref{eq:activecoexist2} to be expressed as a difference of chemical potentials with the nonequilibrium critical nucleus criteria now taking the form:
\begin{subequations}
\label{eq:coexistrestate}
    \begin{align}
         \Delta g =  \mathcal{E}^{\rm in}\mathcal{K}\bigr|_{r=0} &+ \int_0^{\infty}\left(2\mathcal{E}^{\rm in}-\mathcal{E}\right)\frac{\partial\mathcal{K}}{\partial r}\text{d}r \nonumber \\  & +  \frac{2}{R_C}\left[\mathcal{E}^{\rm in}\upgamma_{\rm mech} + \frac{\rho^{\rm sat}\Delta\mathcal{E}}{\Delta\rho}\upgamma_{\rm ost}\right],\label{eq:chempot}
    \end{align}
    \begin{equation}
        \Delta\mathcal{P} = \mathcal{K}\bigr|_{r=0} +  \frac{2\upgamma_{\rm mech}}{R_C}.
    \end{equation}
\end{subequations}
Additionally, by defining a pseudopotential ${\Phi \equiv \int \mathcal{P} \text{d}\mathcal{E}}$, Eq.~\eqref{eq:activecoexist2} can be rearranged to express the critical radius as ${R_C = 2\upgamma_{\rm ost}\rho^{\rm sat}\Delta\mathcal{E}/\Delta\rho\left(\Delta\mathcal{E}\mathcal{P}^{\rm sat} - \Delta\Phi\right)}$.
For passive systems without non-local terms, $\upgamma_{\rm mech} = \upgamma_{\rm ost}$ and $\mathcal{E} \sim 1/\rho$.
One can use these relations to readily verify that $g^{\rm in} = g^{\rm sat}$ and that Eq.~\eqref{eq:activecoexist2} recovers the thermodynamically derived expression for $R_C$.
Our mechanical criteria for the critical nucleus [Eqs.~\eqref{eq:coexistrestate}] thus recover and generalize the thermodynamic criteria [Eq.~\eqref{eq:thermocoexist}].

Equations~\eqref{eq:coexistrestate} [or, equivalently, Eqs.~\eqref{eq:coexistcriteria1} and~\eqref{eq:coexistcriteria2}] are the \textit{finite-size} coexistence criteria for the critical nucleus.
In Ref.~\cite{Tjhung2018}, finite-sized coexistence criteria were derived which appear to be distinct from those of the present work.
Fundamentally, this is because pressure (referred to as ``pseudopressure''), surface tension (referred to as ``pseudotension''), and chemical potential have different physical interpretations in Ref.~\cite{Tjhung2018} than in this work.
Indeed, Ref.~\cite{Tjhung2018} explicitly cautioned against interpreting pseudopressure and pseudotension in the same manner as mechanical pressure and mechanical surface tension.
In Section 1.6 of the SM we formally map AMB+ to our mechanical treatment of finite-size coexistence.
Despite their ambiguous physical interpretation, we find that the finite-sized coexistence criteria presented in Ref.~\cite{Tjhung2018} are fully consistent with those found here.
The unambiguous physical interpretation of the mechanical quantities used in our work represents an advantage of a mechanical perspective which also has a clear connection to microscopics through the Irving-Kirkwood-Noll procedure~\cite{Irving1950,Noll1955,Lehoucq2010}.

Given a saturation density, one may use Eqs.~\eqref{eq:coexistrestate} to solve for the unknowns $\rho^{\rm in}$ and $R_C$.
We note that as $\rho^{\rm sat}$ approaches either binodal density, $R_C$ will diverge; Eq.~\eqref{eq:activelaplacepress} and Eq.~\eqref{eq:activecoexist2} then simplify to the macroscopic phase coexistence criteria presented in Ref.~\cite{Omar2023b}.
As one approaches this limit and $R_C$ grows, $\mathcal{O}\left(R_C^{-2}\right)$ corrections to the coexistence criteria become negligible and we expect our theory to be most accurate.
Solution of the coexistence criteria is complicated by the presence of $\upgamma_{\rm mech}$ and $\upgamma_{\rm  ost}$ in Eqs.~\eqref{eq:coexistrestate}, which depend on the density profile, which in turn depends on $R_C$ and $\rho^{\rm in}$.
Therefore, solution of Eqs.~\eqref{eq:coexistrestate} requires \textit{self-consistently} solving for the density profile using a procedure detailed in Section 1 of the SM. 
In Sec.~\ref{sec:NUCABP}, we apply our mechanical theory to a canonical model system for active matter, active Brownian particles.

\subsection{Nucleation of Motility-Induced Phases} 
\label{sec:NUCABP}
We consider a system of $N$ interacting ABPs in which the time variation of the position $\mathbf{r}_i$ and orientation $\mathbf{q}_i$ of the $i$th particle follow overdamped Langevin equations:
\begin{subequations}
  \label{eq:eom}
    \begin{align}
        \Dot{\mathbf{r}}_i &= U_o \mathbf{q}_i + \frac{1}{\zeta}\sum_{j\neq i}^N\mathbf{F}_{ij}\label{eq:rdot}, \\ \Dot{\mathbf{q}}_i &= \mathbf{q}_i\times \bm{\Omega}_i\label{eq:qdot},
    \end{align} 
\end{subequations}
where $\mathbf{F}_{ij}$ is the (pairwise) interparticle force, $U_o$ is the intrinsic active speed, $\zeta$ is the translational drag coefficient, and $\bm{\Omega}_i$ is a stochastic angular velocity with zero mean and variance $\langle \bm{\Omega}_i (t) \bm{\Omega}_j(t')\rangle = 2D_R\delta_{ij}\delta(t-t')\mathbf{I}$. 
Here, $D_R$ is the rotational diffusivity, which may be athermal in origin, and is inversely related to the orientational relaxation time $\tau_R \equiv D_R^{-1}$.
For the remainder of this Article, we consider purely repulsive $\mathbf{F}_{ij}$, corresponding to a sufficiently stiff interaction potential such that hard-sphere statistics are closely approximated with an effective hard-sphere diameter of $d_{\rm hs}$~\cite{Weeks1971,Omar2021}.
One can then define an intrinsic run length, $\ell_o/d_{\rm hs} = U_o\tau_R$, the typical distance an ideal particle travels before reorienting.

\begin{figure}
	\centering
	\includegraphics[width=0.48\textwidth]{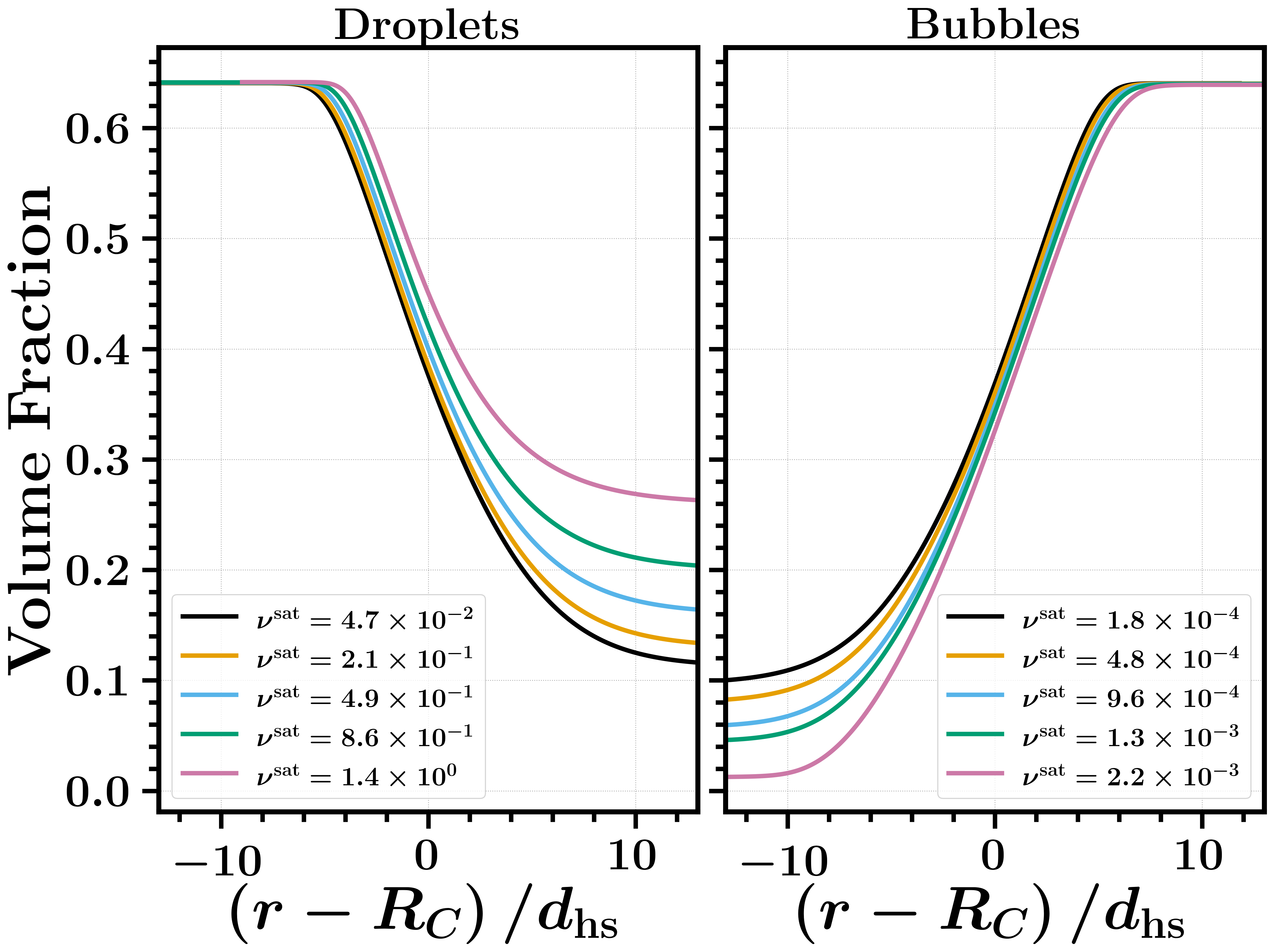}
	\caption{\protect\small{{Density profiles of critical bubbles and droplets at $\ell_o/d_{\rm hs} = 150$. All profiles have been shifted by their corresponding critical radius, given in Fig.~\ref{fig:figure2}. Here, the volume fraction is defined as the number density multiplied by the effective volume of a particle $\rho \pi d_{\rm hs}^3 / 6$.}}}
	\label{fig:figure1}
\end{figure}

In three dimensions, the form of $\mathcal{P}(\rho)$ and $\{\kappa_i(\rho)\}$ for ABPs are given by~\cite{Omar2023b,Langford2024TheoryPhases}:
\begin{subequations}
    \label{eq:abpconstitutive}
    \begin{align}
         \mathcal{P}(\rho) &= p_C(\rho) + \frac{\rho\zeta U_o\ell_o \overline{U}(\rho)}{6}\label{eq:dynpress}, \\ \kappa_1(\rho) &= \frac{\ell_o^2}{20}\left(\overline{U}(\rho)\right)^2\frac{\partial p_C}{\partial \rho}\label{eq:adef},   \\ \kappa_3(\rho) &= \frac{\ell_o^2}{20}\overline{U}(\rho)\frac{\partial}{\partial \rho}\left[\overline{U}(\rho)\frac{\partial p_C}{\partial \rho}\right]\label{eq:bdef}, \\ \kappa_2(\rho) &= \kappa_4(\rho) =   0, \\ \mathcal{K} &= 0,
    \end{align}
\end{subequations}
where $p_C(\rho)$ is the \textit{conservative} pressure arising from particle interactions and $\overline{U}(\rho)$ is the dimensionless effective active speed.
In Eqs.~\eqref{eq:abpconstitutive} we have ignored the gradient contributions of the interaction potential to $\bm{\Sigma}$, which is justified when $\ell_o/d_{\rm hs} \gg 1$ for purely repulsive interactions~\cite{Solon2018GeneralizedEnsembles,Omar2023b,Evans2023}.
Given our expressions for $\{\kappa_i(\rho)\}$, one can solve for the generalized Maxwell construction pseudovariable $\mathcal{E}\sim p_C(\rho)$~\cite{Omar2023b}.
We also note that because $\mathcal{K} = 0$, Eq.~\eqref{eq:activelaplacepress} reduces to the familiar Laplace pressure difference (i.e. no non-local corrections) and the negative sign of $\upgamma_{\rm mech}$~\cite{Bialke2015} implies that the pressure inside of a droplet is \textit{lower} than the pressure outside of the droplet~\cite{Solon2018GeneralizedEnsembles}.
The phase diagram, dependent on $\ell_o/d_{\rm hs}$ and volume fraction $\phi \equiv \rho \pi d_{\rm hs}^3/6$, has been well established computationally~\cite{Omar2021} and theoretically~\cite{Omar2023b} with the latter predicting a critical point at $\phi \approx 0.49$ and $\ell_o \approx 16.2 d_{\rm hs}$

\begin{figure}
	\centering
	\includegraphics[width=0.48\textwidth]{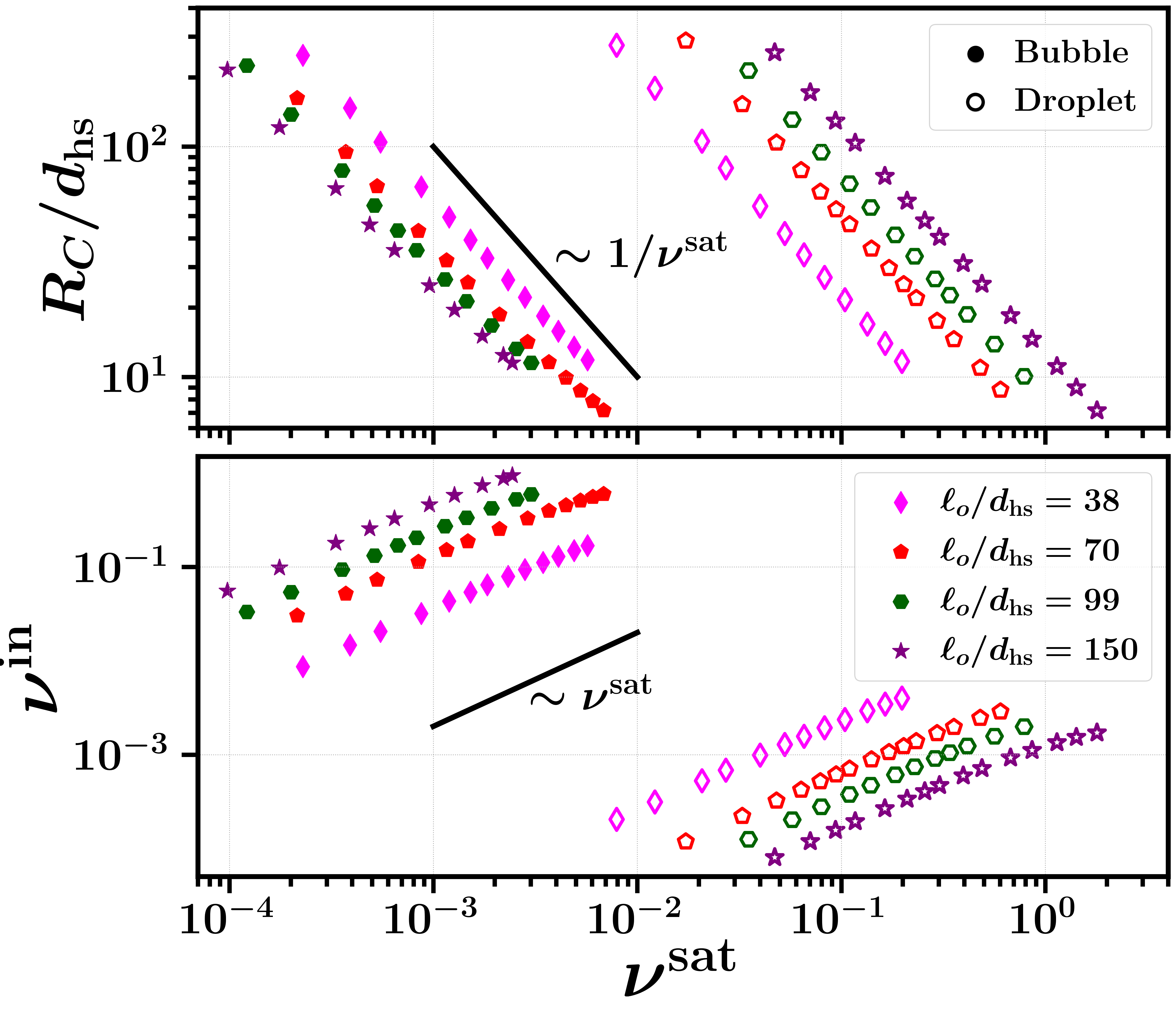}
	\caption{\protect\small{{Critical radius and relative change in nucleus density (${\nu^{\rm in} \equiv |\rho^{\rm in} - \rho^{\rm gas, \rm liq}| / \rho^{\rm gas, \rm liq}}$) as a function of supersaturation (${\nu^{\rm sat} \equiv |\rho^{\rm sat} - \rho^{\rm gas, \rm liq}| / \rho^{\rm gas, \rm liq}}$){\color{black}, as calculated from Eq.~\eqref{eq:coexistrestate} using the numerical procedure outlined in Section 1.3 of the SM. Each data point requires self-consistent solution of the corresponding full spatial profile, e.g. those shown Fig.~\ref{fig:figure1}. Linear dependence of $\nu^{\rm in}$ on $\nu^{\rm sat}$ and inverse dependence of $R_C/d_{\rm hs}$ on $\nu^{\rm sat}$ implies a Gibbs-Thomson relation with $\nu^{\rm {in}}\propto 1/R$ }.}}}
	\label{fig:figure2}
\end{figure}

With the dynamic stress of ABPs in hand, we are able to self-consistently solve our generalized critical nucleus conditions [Eqs.~\eqref{eq:coexistrestate}], with the resulting density profiles for both bubbles and droplets shown in Fig.~\ref{fig:figure1} for a select activity and various supersaturations.
Values of $R_C$ and $\rho^{\rm in}$ obtained from self-consistent solution of the criteria at various activities and supersaturations are provided in Fig.~\ref{fig:figure2}.
Here, we define a dimensionless overall supersaturation density as as the relative departure of the majority phase density from the binodal density, ${\nu^{\rm sat} = |\rho^{\rm sat} - \rho^{\rm gas,liq}|/\rho^{\rm gas,liq}}$ and a dimensionless nucleus density as the relative departure of the minority phase density from the opposite binodal ${\nu^{\rm in} = |\rho^{\rm in} - \rho^{\rm liq,gas}|/\rho^{\rm liq,gas}}$.
Our theory is expected to hold as $\nu^{\rm sat}\to 0$.
We observe a roughly linear relationship between $\nu^{\rm sat}$ and $\nu^{\rm in}$ as well as an inverse relationship between $\nu^{\rm sat}$ and $R_C$. 
These results are therefore in excellent agreement with existing simulation data reporting that ABPs obey a Gibbs-Thomson relation with $\nu^{\rm in} \propto 1/R$~\cite{Lee2017InterfaceSeparations}. 

Figure~\ref{fig:figure2} displays a striking difference between critical bubbles and droplets. 
At all examined values of activity, critical droplets of active Brownian particles are over an order of magnitude larger than critical bubbles under similar saturation conditions. 
This difference between bubbles and droplets widens as activity is increased and the phases become increasingly distinct. 
To contextualize this order of magnitude difference, we calculate the critical nucleus size and density for an \textit{equilibrium} van der Waals fluid. 
While the size of the critical droplets are indeed larger than the size of critical bubbles in passive fluids, they remain within an order of magnitude of each other at similar saturation conditions even far from the critical point. 
In equilibrium, this difference is rooted in the asymmetry of the equations of state outside of the binodal densities: the vapor bubbles are more compressible than the liquid droplets.
This asymmetry results in larger Laplace pressure differences for bubbles in comparison to droplets under similar saturation conditions, leading to the critical radius discrepancy. 
While this asymmetry in the compressibility is also present for active systems, the coexistence criteria contain uniquely nonequilibrium contributions, including a significant discrepancy between the vapor and bubble Ostwald tensions. 
At equilibrium these two quantities are identical, but can differ by an order of magnitude in active systems as shown in Fig.~\ref{fig:figure3}.

{\color{black} For a more physical understanding of the discrepancy in critical nucleus size between bubbles and droplets, we draw on the kinetic arguments independently proposed by Redner \textit{et al.}~\cite{Redner2016} and Lee~\cite{Lee2017InterfaceSeparations}.
Establishing a flux-free critical nucleus with a density (and therefore interaction pressure) distinct from its surroundings requires the generation of an interfacial region of particles with a polarization pointing into the dense phase.
Crucially, the rate at which particles with this orientation arrive to the interface must equal the rate at which they reorient and escape in order to satisfy the flux-free condition.
The arrival rate is controlled by gas phase active speed and density while the escape rate is controlled by the minimum angle a particle must reorient in order to escape, the so-called ``horizon angle~\cite{Redner2016}.'' 
For the same $R_C$, the rate of escape will depend greatly on whether the nucleus is the liquid phase (small horizon angle) or the gas phase (large horizon angle). 
As a result, for a similar supersaturation, the droplet must have a larger critical radius in order to satisfy the flux-free condition.
}

\subsection{Langevin Dynamics of a Nucleus}
\label{sec:NUCLANG}
Further insight into the active nucleation process can be obtained by considering the growth dynamics of nuclei near their critical size.
We assume the dynamics of the nucleus to proceed quasistatically: any given nucleus with radius $R$ (not necessarily equal to $R_C$) will be in a local mechanical balance with its surroundings.
We now distinguish the density just outside the nucleus (now denoted as $\rho^{\rm out}$), and the density far from the nucleus which remains fixed at the saturation density, $\rho^{\rm sat}$.
The mechanical balance between the nucleus and its immediate surroundings will determine the intermediate density which will generally not equal the far away saturation density,  $\rho^{\rm out}\neq \rho^{\rm sat}$.
The inequality between the intermediate and far away densities will drive the transport of particles and will cause $\rho^{\rm out}$ to evolve in time.
This, in turn, will cause $R$ to also change in time as determined by the local mechanical balance.
Cates and Nardini~\cite{Cates2023ClassicalSeparation} recently demonstrated that such a description can allow for the determination of effective nucleation barriers of active fluids with dynamics described by AMB+, despite the absence of well-defined free energy for these driven systems.
Deriving the stochastic growth dynamics of the nuclei requires a fluctuating hydrodynamic description of the density field which we recently obtained for interacting ABPs~\cite{Langford2024TheoryPhases}:
\begin{subequations}
    \label{eq:flucthydro}
    \begin{align}
        \frac{\partial \rho}{\partial t} &= -\boldsymbol{\nabla}\cdot\mathbf{J}\label{eq:continuity},\\ \mathbf{J} &= \frac{1}{\zeta}\boldsymbol{\nabla}\cdot\bm{\Sigma} + \bm{\eta}^{\rm act},\label{eq:hydroflux}
    \end{align}
where $\rho$ now is a \textit{stochastic} density field (while in the previous sections it was an ensemble average), $\bm{\Sigma}$ is defined identically as the dynamic stress tensor of Sec.~\ref{sec:NUCABP} and $\bm{\eta^{\rm act}}$ is an athermal stochastic contribution to the flux with zero mean and a variance of:
\begin{align}
     \langle \bm{\eta}^{\rm act}(\mathbf{r},t)\bm{\eta}^{\rm act}(\mathbf{r}',t') \rangle = &2\frac{k_BT^{\rm act}}{\zeta}\left(\rho\mathbf{I} - \frac{d}{d-1}\mathbf{Q}'\right)\nonumber \\ &\times\delta(t-t')\delta(\mathbf{r}-\mathbf{r}') \ \label{eq:etavar}.
\end{align}
\end{subequations}
Here $\mathbf{Q}'$ is the traceless nematic order and we have defined $k_BT^{\rm act} = \ell_o \zeta U_o /d(d-1)$ as the active energy scale.

We recently established a surface-area minimizing principle for liquid-gas interfaces of ABPs resulting from a positive capillary-wave tension, $\upgamma_{\rm cw}$, which resists fluctuations to the area-minimizing interfacial shape.
The strength of the fluctuations were found to be set by the athermal active energy scale $k_BT^{\rm act}$~\cite{Langford2024TheoryPhases}.
The ratio of the capillary tension to this energy scale -- the active interfacial stiffness -- controls the amplitude of long-wavelength interfacial fluctuations.
A fluctuating nucleus can be well-approximated as a sphere when the average radius $R$ satisfies  $(R^2\upgamma_{\rm cw}/k_BT^{\rm act})^{1/2} \gg 1$.
Under this condition, the stochastic density field dynamics can be connected to the radial growth dynamics of the nucleus through the following radially symmetric ansatz~\cite{Bray1994TheoryKinetics,Cates2023ClassicalSeparation}:
\begin{equation}
    \rho(\mathbf{r},t) = \upvarphi_R\left[r - R(t)\right],\label{eq:ansatz}
\end{equation}
where $\upvarphi_R$ is the noise-averaged density profile associated with a nucleus of radius $R$. 
The density profiles calculated in Sec.~\ref{sec:NUCABP} correspond to $\upvarphi_R$ when $R=R_C$. 
The ansatz may be substituted into our fluctuating hydrodynamics equations [Eqs.~\eqref{eq:flucthydro}], allowing us to obtain the following stochastic radial growth dynamics:
\begin{subequations}
    \label{eq:langevinnucleus}
    \begin{align}
        \zeta_{\rm eff}\frac{\partial R}{\partial t} &= F^{D}(R) + \tilde{\chi}(t),\label{eq:Revo}\\ \langle \tilde{\chi}(t)\rangle &= 0,\label{eq:chiav} \\ \langle \tilde{\chi}(t)\tilde{\chi}(t') \rangle & = 2k_BT^{\rm act}\zeta_{\rm eff}\delta(t-t'),\label{eq:chivar} \\ \zeta_{\rm eff} &= \frac{4\pi\Delta\rho^2R^3\zeta}{\rho^{\rm sat}},\label{eq:mobility} \\ F^D(R) &= -8\pi \upgamma_{\rm ost}R\left[1-\frac{R}{R_C} \right].\label{eq:conforce}
    \end{align}
\end{subequations}
The radial growth dynamics take the form of a one-dimensional overdamped Langevin equation containing three distinct forces: a drag force with drag coefficient $\zeta_{\rm eff}$, a stochastic force with an athermal temperature scale of $k_BT^{\rm act}$~\cite{vanKampen1981ItoStratonovich}, and a deterministic force $F^{D}(R)$ proportional to the Ostwald tension.  
This deterministic force vanishes when $R=R_C$ where the density outside nucleus is identical to the far away saturation density $\rho^{\rm out} = \rho^{\rm sat}$.
Crucially, the noise statistics satisfy the fluctuation-dissipation theorem (FDT) despite the underlying microscopic dynamics [Eqs.~\eqref{eq:eom}] violating FDT. 
We can plainly see that $\upgamma_{\rm ost}$ is the quantity controlling the dynamics of nucleation and ripening. 
As first argued by Thjung \textit{et al.}~\cite{Tjhung2018} a positive $\upgamma_{\rm ost}$ implies $R_C$ is an \text{unstable} fixed point in which nuclei with $R>R_C$ tend to grow while nuclei with $R<R_C$ tend to shrink. 
\textit{A priori}, there is no reason to assume $\upgamma_{\rm ost}$ is generally positive.
In the case of a negative $\upgamma_{\rm ost}$, $R_C$ becomes a \textit{stable} fixed point. 
{\color{black}In such a scenario, one may expect phenomonology ranging from microphase separation to a ``bubbly liquid'' phase coexisting with a gas phase~\cite{Tjhung2018}.
However, for ABPs we do not find a negative $\upgamma_{\rm ost}$ [see Fig.~\ref{fig:figure4}] -- consistent with existing simulation work on \textit{three-dimensional} ABPs~\cite{Stenhammar2014PhaseDimensionality,Omar2021,Turci2021PhaseParticles,Langford2024TheoryPhases}.
In two dimensions, some simulations of ABPs have reported phenomenology consistent with bubbly liquids~\cite{Stenhammar2014PhaseDimensionality,Shi2020Self-OrganizedParticles}. 
Whether these bubbly phases are a result of a negative $\upgamma_{\rm ost}$ or other effects~\cite{Caporusso2020Motility-InducedSystem} remains an open question.}

\begin{figure}
	\centering
	\includegraphics[width=0.48\textwidth]{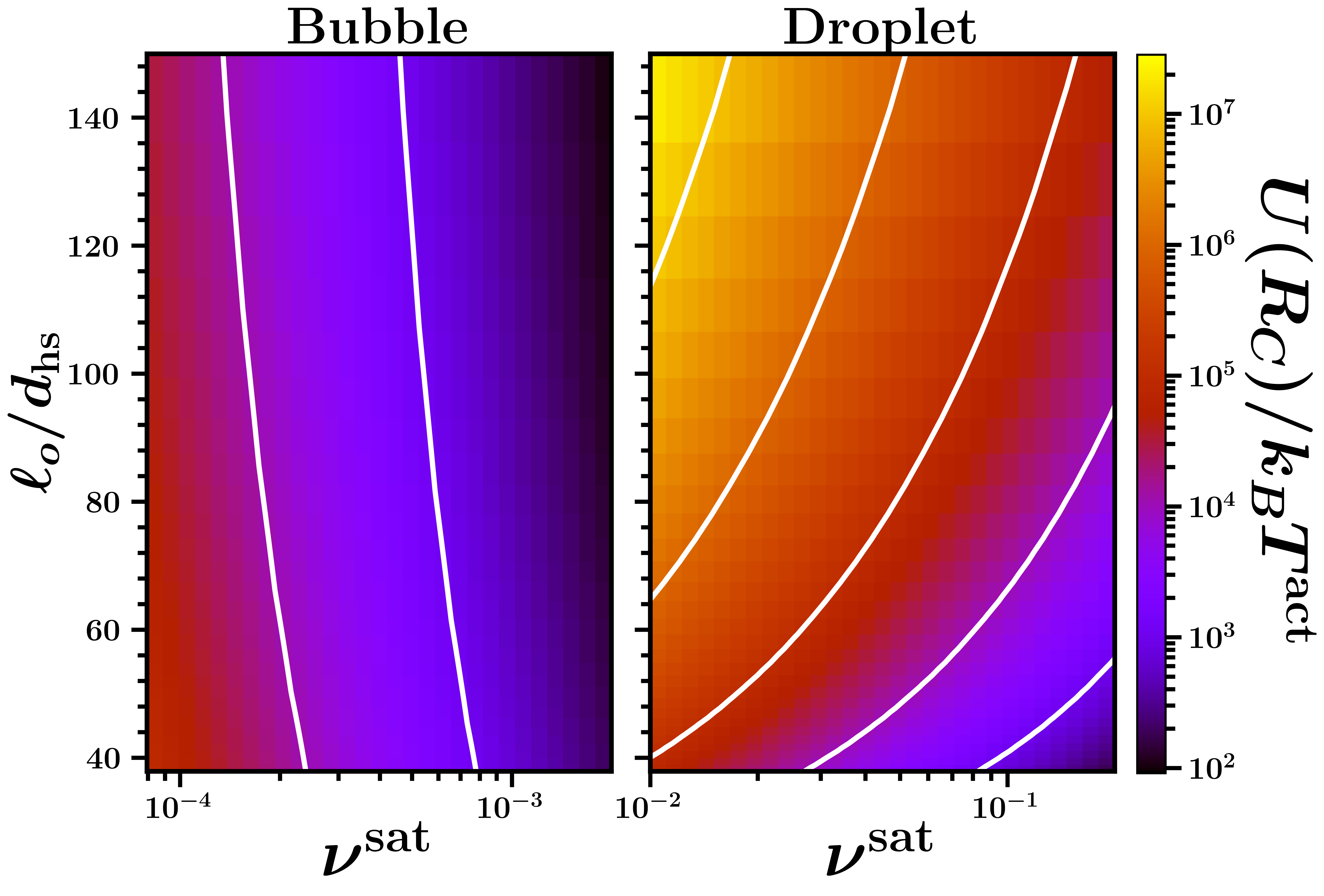}
	\caption{\protect\small{{Barrier $U(R_C)$ relative to active energy scale as a function of activity and supersaturation. The distinct white contour lines distinguish values of $U(R_C)/k_BT^{\rm act}$ separated by orders of magnitude.}}}
	\label{fig:figure3}
\end{figure}

As our deterministic force is a function of a single variable, $R$, we can define (within an integration constant) an effective conservative potential $U(R)$:
\begin{equation}
    U(R) = 8\pi \upgamma_{\rm ost}\left[\frac{R^2}{2}-\frac{R^3}{3R_C}\right]
\end{equation}
with $F^D(R) = - \partial U/\partial R$. 
In defining this potential, we have treated $\upgamma_{\rm ost}$ as independent of $R$ (just as in Ref.~\cite{Cates2023ClassicalSeparation}), an approximation that appears to be well justified while $R\sim R_C$.
This potential may be rearranged into a form familiar to equilibrium intuition:
\begin{subequations}
\label{eq:potentialintuitive}
    \begin{equation}
        U(R) = \underbrace{4\pi R^2\upgamma_{\rm ost}}_{\rm{Interface}} - \underbrace{\frac{4}{3}\pi R^3 \mathcal{V}}_{\rm{Bulk}},
    \end{equation}
where:
    \begin{equation}
        \mathcal{V} = \frac{\mathcal{P}(\rho^{\rm sat})\Delta\rho}{\rho^{\rm sat}} - \frac{\Delta\Phi\Delta\rho}{\rho^{\rm sat}\Delta\mathcal{E}}.
    \end{equation}   
\end{subequations}
At equilibrium, $\mathcal{V}$ simplifies to the expected $\mu(\rho^{\rm sat})\Delta\rho - \Delta f$.
From Eq.~\eqref{eq:potentialintuitive}, one can clearly appreciate that the barrier contains two contributions: a \textit{positive} term which depends solely on \textit{interfacial} properties scaling as $R^2$ and a \textit{negative} term which depends solely on \textit{bulk} properties scaling as $R^3$.
Thus the familiar picture of homogeneous nucleation where interfacial effects compete with bulk effects in order to define a critical radius and nucleation barrier is recovered but now contains generalized volumetric driving and interfacial penalty terms.

Evaluating the potential at the critical radius allows us to compare the relative strengths of the nucleation barrier across activities and saturations, as shown in Fig.~\ref{fig:figure3}.
Here we see a decrease in the nucleation barrier as $\nu^{\rm sat}$ increases, consistent with equilibrium intuition.
We further observe that the nucleation barrier height is much higher for droplets in comparison to bubbles. 
The disparity in nucleation barrier between droplets and bubbles has an attractive physical interpretation: much like the capillary tension of fluids comprised of ABPs~\cite{Langford2024TheoryPhases}, $\upgamma_{\rm ost}$ is rooted in \textit{nematic} flows.
As we discuss in recent work~\cite{Langford2024TheoryPhases}, these nematic flows lead to particles accumulating at \textit{convex} liquid surfaces and escaping from \textit{concave} liquid surfaces. 
Then for \textit{small} nuclei $R<R_C$, where curvature is high, droplets (concave liquid surfaces) are less favorable to grow than bubbles (convex liquid surfaces). 
This demonstrates that while Eqs.~\eqref{eq:langevinnucleus} satisfies the fluctuation dissipation theorem and allows for the definition of a conservative potential, the underlying physics of this conservative potential remain rooted in nonequilibrium effects.

\begin{figure}
	\centering
	\includegraphics[width=0.48\textwidth]{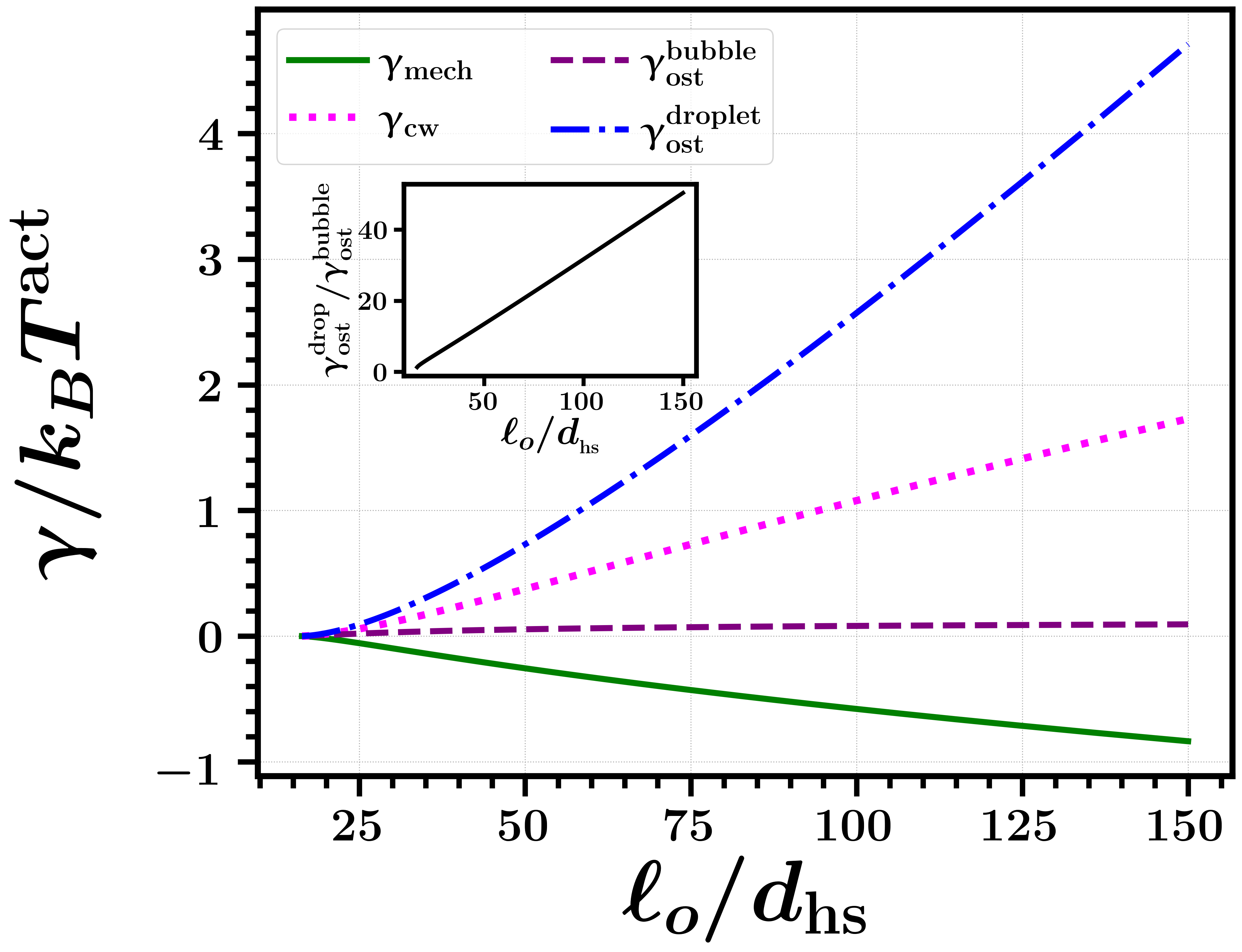}
	\caption{\protect\small{{Surface tensions as defined by $\upgamma_{\rm mech}$, $k\to 0$ limit of $\upgamma_{\rm cw}$, and $R\to\infty$ limit of $\upgamma_{\rm ost}$ for both bubbles and droplets as a function of run length. }}}
	\label{fig:figure4}
\end{figure}

The recovery of an effective detailed balance with respect to the size of a sub-critical nucleus directly implies that the probability of observing a nucleus of size $R<R_C$ is proportional to ${\text{exp}\left[-U(R)/k_BT^{\rm act}\right]}$~\cite{Cates2023ClassicalSeparation}.
We note that the potential defined here relied on the assumption that the density profiles and quantities such as $\upgamma_{\rm ost}$ are constant with $R$ an approximation that clearly breaks down when $R << R_C$.
Even in equilibrium, the nucleation potential formulated using bulk thermodynamic quantities (e.g~ the so-called "capillary approximation") is often used for all $R<R_C$ despite likely being beyond the range of applicability of bulk thermodynamic quantities~\cite{Oxtoby1992HomogeneousExperiment,Merikanto2007OriginClusters,Royall2024ColloidalMysteries}.
Nevertheless, use of this potential for all $R$ allows one to straightforwardly extend the arguments of Becker-D\"{o}ring theory~\cite{Becker1935KinetischeDampfen,Oxtoby1992HomogeneousExperiment,Debenedetti2020MetastableLiquids} to extract the following kinetic expression for the rate at which an active droplet nucleates per unit volume, as outlined in Appendix A:
\begin{equation}
    \mathcal{R} = \frac{\left(\rho^{\rm sat}\right)^2}{\rho^{\rm in}} U_o \sqrt{\frac{\pi\upgamma_{\rm ost}}{2k_BT^{\rm act}}} \text{exp}\left[-\frac{U(R_C)}{k_BT^{\rm act}}\right].\label{eq:nucleationrate}
\end{equation}
The characteristic timescale to observe a nucleation event is given by $\tau_{\rm nuc} = (V\mathcal{R})^{-1}$ where $V$ is the total volume of the system. 
Thus the work of Redner~\textit{et al.}~\cite{Redner2016}, which inferred the existence of a nucleation barrier by observing clustering statistics consistent with Becker-D\"{o}ring theory from particle simulation, is well-justified on theoretical grounds, {\color{black} although our calculated barrier is for three dimensions while the simulations of Ref.~\cite{Redner2016} are in two}.
We note that the rate [Eq.~\eqref{eq:nucleationrate}] is applicable to the nucleation of active droplets but not to active bubbles -- the physical interpretation of the ``clusters'' posited in the derivation of Eq.~\eqref{eq:nucleationrate} becomes unclear~\cite{Oxtoby1992HomogeneousExperiment} in this context.

\subsection{Active Coarsening Dynamics}
\label{sec:ACD}
While Eqs.~\eqref{eq:coexistrestate} and Eqs.~\eqref{eq:langevinnucleus} were motivated to describe the earliest stages of nucleation, i.e.~the structure and dynamics of a single nucleus forming in a supersaturated solution, they may also describe late-stage coarsening dynamics long after all nucleation events have occurred.
At this stage, the system will contain several nuclei of different sizes. 
The coarsening of these domains can proceed through a combination of several different mechanisms, including evaporation-condensation (Ostwald ripening) and coalescence.
At sufficiently low $\nu^{\rm sat}$, the nuclei will be separated by large enough distances that collisions between them are exceedingly rare and coarsening will be dominated by Ostwald ripening.
In such a scenario, Eqs.~\eqref{eq:langevinnucleus} can describe the dynamics of each nucleus such that they are only coupled via the density of the majority phase, $\rho^{\rm sat}$, which now evolves with time. 
The finite-size coexistence criteria presented in this work imply minor differences in density of the majority phase just outside nuclei of different sizes, inversely proportional to their radii.
The majority phase therefore has an inhomogeneous density, driving transport of mass away from nuclei with enriched outside densities towards nuclei with depleted outside densities.
For passive systems as well as ABPs [see Fig.~\ref{fig:figure1}], smaller droplets have enriched outside densities while smaller bubbles have depleted outside densities.
Noting that transport of mass away from a nucleus would cause a droplet to shrink and a bubble to grow, this implies that large nuclei of minority phase will grow at the expense of smaller nuclei until the system reaches a final steady state with a single macroscopic domain.
Nuclei compete to absorb (or inject) particles from the majority phase and therefore $R_C$, the radius below which nuclei shrink and above which nuclei grow, is now a function of time.
While we do not have profiles for a particular system in which $\upgamma_{\rm ost} < 0$, we suspect they would reveal smaller droplets having depleted outside densities and smaller bubbles having enriched outside densities.
This would result in smaller nuclei growing at the expense of larger ones, consistent with stable radial dynamics of Eq.~\eqref{eq:langevinnucleus}.

With the above physical picture in mind, we can determine the time-dependence of the post-nucleation critical radius in a similar manner as the Lifshitz-Slyozov-Wagner (LSW) theory~\cite{Lifshitz1961TheSolutions}. 
In the limit of a vanishing phase fraction of the minority phase, the nuclei can be approximated as independent with each satisfying the Langevin equation shown in Eq.~\eqref{eq:langevinnucleus}.
The Fokker-Planck equation consistent with Eq.~\eqref{eq:langevinnucleus} can be straightforwardly found and its long-time asymptotic solution (while discarding terms of higher order than $R^{-3}$) can be determined analogously to LSW theory, as detailed in Appendix B. 
In doing so, we find that the critical radius evolves in time according to:
\begin{equation}
    R_C(t) = \left[R^3_C(t_o) + \frac{6\rho^{\rm sat}}{\zeta\Delta\rho^2}\upgamma_{\rm ost}\mathcal{C}t\right]^{1/3},\label{eq:LSWgrowth}
\end{equation}
where $\mathcal{C}$ is a dimensionless constant and $t_o$ is an arbitrary time origin. 
We note that this long-time asymptotic solution also implies that the average radius is always within a dimensionless multiplicative constant of $R_C$~\cite{Lifshitz1961TheSolutions,Desai2009DynamicsStructures}.
Equation~\eqref{eq:LSWgrowth} demonstrates that the $t^{1/3}$ growth law predicted by LSW theory should generically hold out of equilibrium while $\upgamma_{\rm ost} > 0$.
This $1/3$ power law has been measured experimentally in active systems~\cite{Liu2019Self-DrivenFormation,Zhang2021ActiveDensity} with disparate microscopic details, suggesting that this growth law may in fact be robust.
We note that the simulations of Redner \textit{et al.}~\cite{Redner2013StructureFluid} measured the average number of particles per droplet (in two-dimensions) to grow slightly faster than $t^{1/2}$ {\color{black} at late times}. 
This is consistent with a radial growth rate faster than $t^{1/4}$ and is perhaps consistent with our theoretical prediction.

{\color{black}Curiously, Ref.~\cite{Redner2013StructureFluid} also measured a much faster coarsening rate at early times where nuclei are spaced closely. 
While in this work we do not consider coarsening driven by nucleus collision, it is well-established for passive systems to also result in a $t^{1/3}$ growth law~\cite{Binder1974TheoryMixtures}.
A transition from fast coarsening when nuclei are closely spaced to slow coarsening when nuclei are distant is eerily reminiscent of coarsening in systems with hydrodynamic interactions, in which collisions of nuclei (and subsequent relaxation to spherical geometry) induce flows that cause further collisions~\cite{Tanaka1996CoarseningMixtures,Nikolayev1996NewCoarsening}.
While standard ABP models do not consider hydrodynamic interactions, a detailed accounting of the polarization field at the early stages of coarsening may reveal non-diffusive contributions. 
Furthermore, if $\ell_o$ greatly exceeds the spacing between nuclei, particles can be ballistically exchanged between nuclei, an effect that is also not accounted for in our perspective.
A systematic treatment of early-stage active coarsening would help bridge the gap in our understanding between nucleation and late-stage coarsening and is the subject of future work.
}

We also find that Eqs.~\eqref{eq:coexistrestate} may be used to treat coarsening problems with a finite phase fraction of minority phase, which is beyond the scope of LSW theory.
To demonstrate this, we adopt the perspective and numerical scheme of Voorhees and Glicksman~\cite{Voorhees1984SolutionTheory,Voorhees1984SolutionSimulations}, originally developed to study precipitation of solid phases in alloys. 
We now consider timescales at which the minority phase has nucleated and coarsened to the extent that the final phase fraction has been reached.
All that remains to complete the phase separation process is for the disparate nuclei to coarsen to a single domain such that a single interface exists.

In the numerical scheme, the initial sizes and positions of nuclei in a fully periodic domain are prescribed. 
While the positions remain fixed, the sizes of the nuclei evolve with time via diffusion of particles through the majority phase.
{\color{black} Additionally, the majority phase will relax much faster than the individual nuclei grow or shrink.
We therefore consider the diffusive dynamics of \textit{only the majority phase} (i.e.,~excluding the regions occupied by the nuclei) with the nuclei simply acting as particle generation terms (sources or sinks of particles). 
Considering only the majority phase dynamics implies that large density-gradient interfaces are ignored; we therefore discard gradient terms from $\bm{\Sigma}$ and arrive at the quasi-stationary expression for the majority phase's density field (see Appendix C for further details):
\begin{equation}
    0 = \boldsymbol{\nabla}^2\mathcal{P}(\rho) + \zeta \sum_{i=1}^{N_{\rm nuc}}S_i(\mathbf{r}),\label{eq:vgsink}
\end{equation}
where $N_{\rm nuc}$ is the number of nuclei and $S_i(\mathbf{r})$ is a function quantifying its injection/absorption of particles into the majority phase.
Using Eq.~\eqref{eq:vgsink} to solve for the pressure field is analogous to solving for the electrostatic potential resulting from $N_{\rm nuc}$ charge densities. 
Approximating each nucleus as spherically symmetric} allows one to treat the ``charge densities'' as ``point charges'', i.e. $S_i(\mathbf{r}) \sim \delta(\mathbf{r}-\mathbf{r}_i)$. 
Then the effective diffusive interactions between nuclei, responsible for Ostwald ripening, can be calculated analogously to the electrostatic energy of a collection of point charges~\cite{Voorhees1984SolutionTheory}.

This perspective, while initially developed for passive systems, only relies on thermodynamic assumptions by employing an equilibrium Gibbs-Thomson relation as a boundary condition between each nucleus and the majority phase{\color{black}, allowing for the determination of the magnitude of each $S_i$}.
Our derived Eqs.~\eqref{eq:coexistrestate} are the analogue of the Gibbs-Thomson relation for an active system, allowing us to extend the perspective of Ref.~\cite{Voorhees1984SolutionTheory} to active matter.

\begin{figure}
	\centering
	\includegraphics[width=0.48\textwidth]{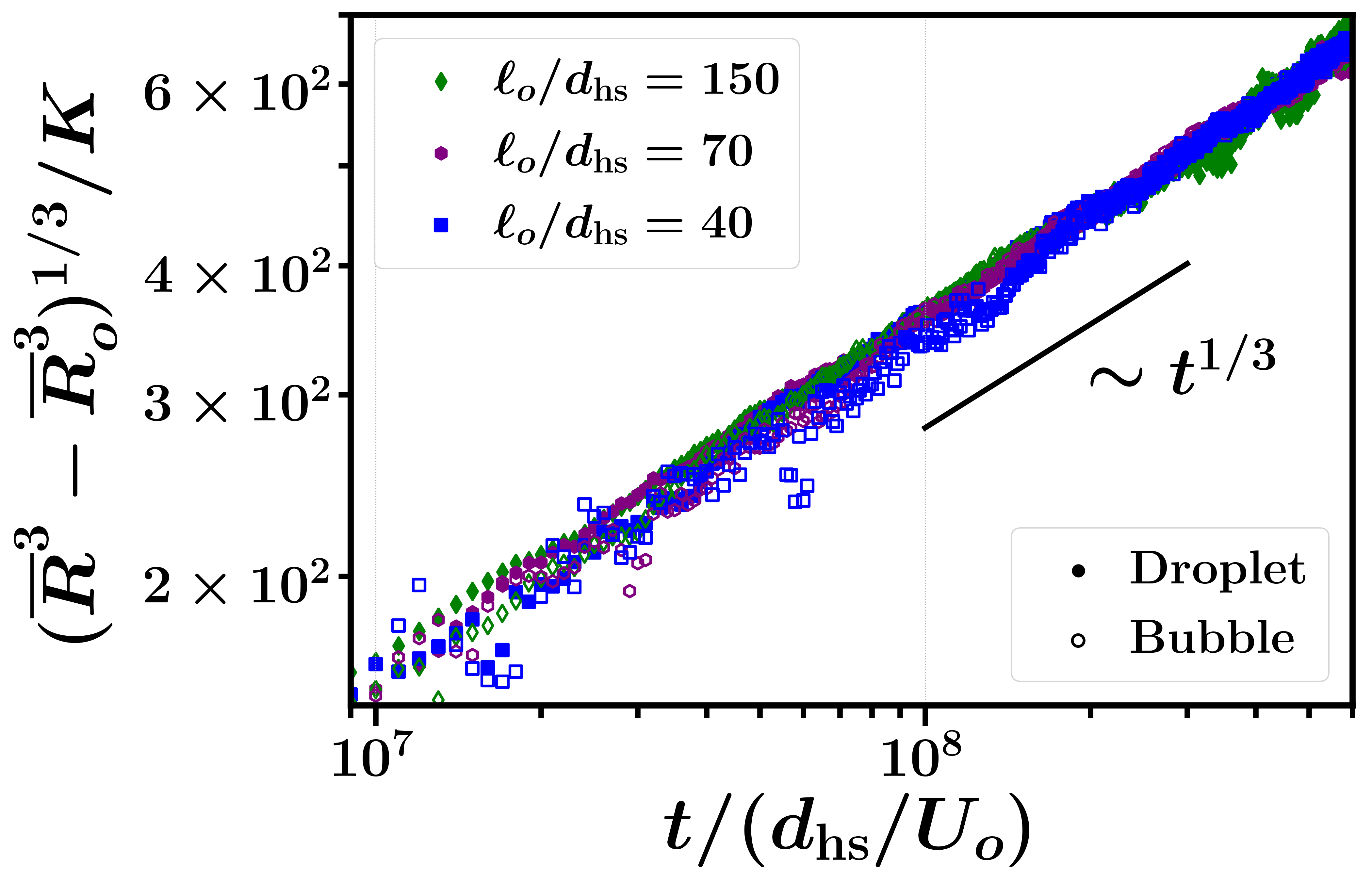}
	\caption{\protect\small{{Average radius growth as a function of time for nuclei initialized with an LSW distribution and average radius $750 d_{\rm hs}$. Radius growth curves across different activities, as well as between bubbles and droplets, collapse onto one another when scaled by prefactor ${K\equiv \left(\rho^{\rm sat}\upgamma_{\rm ost}d_{\rm hs}/\zeta U_o\Delta\rho^2\right)^{1/3}}$.}}}
	\label{fig:figure5}
\end{figure}

For demonstration, we implement the active Voorhees-Glicksman theory to simulate the ripening of a periodic cell of active nuclei. 
Simulations of $250$ nuclei with random positions~\cite{Martinez2009PACKMOL:Simulations} and sizes were executed with the nuclei separated by exceedingly large distances. 
It is therefore expected for the coarsening dynamics to be in agreement with LSW theory~\cite{Lifshitz1961TheSolutions}.
Noting again that the average radius is proportional to the critical radius, we expect the average radius across multiple activities to collapse to the same curve when scaled by the prefactor of the growth rate found in Eq.~\eqref{eq:LSWgrowth}, ${K\equiv \left(\rho^{\rm sat}\upgamma_{\rm ost}d_{\rm hs}/\zeta U_o\Delta\rho^2\right)^{1/3}}$.
The average nucleus size from these simulations scaled by $K$ are shown in Fig.~\ref{fig:figure5}.
Our numerical simulations reveal that the average nucleus size is an excellent agreement with the predicted $\sim t^{1/3}$ scaling with curves for both bubbles and droplets collapsing for all activities.
Plots of the distribution of nucleus sizes, which display evidence of evolving towards an LSW distribution~\cite{Lifshitz1961TheSolutions}, are provided in Section 4 of the SM.

Although the Voorhees-Glicksman theory can be conveniently applied to late-stage active coarsening dynamics via Eqs.~\eqref{eq:coexistrestate}, it has significant limitations.
In particular, the positions of the nuclei are fixed in time while in reality nuclei may diffuse, which may have significant impact on the inter-nuclei diffusion interactions and in extreme cases may lead to collision of nuclei (coalescence).
More sophisticated coarsening theories~\cite{Baldan2002ReviewTheories} attempting to account for these effects, such as the work of Enomoto~\textit{et al.}~\cite{Enomoto1986FiniteRipening} also only draw on thermodynamic arguments in the form of a Gibbs-Thomson relation.
Our work establishes a Gibbs-Thomson relation for single-component phase separation purely from mechanics, without invoking equilibrium assumptions.
Then established perspectives on classical coarsening may be extended to nonequilibrium systems as long as a nonequilibrium Gibbs-Thomson relation, such as that found in this work, is used. 

\section{Discussion and conclusions} 
We derive a mechanical scheme for computing the properties of a critical nucleus arbitrarily far from equilibrium and apply this perspective to motility-induced phases comprised of active Brownian particles. 
From this theory we are able to derive generalized Gibbs-Thompson equations for active systems, a result which has been measured numerically but lacked theoretical justification~\cite{Lee2017InterfaceSeparations}.
Expanding on this theory, we use the framework proposed by Cates and Nardini~\cite{Cates2023ClassicalSeparation} to find the Langevin dynamics of a nucleus. 
These dynamics allow us to define an effective nucleation barrier, revealing that the kinetic barrier to nucleating a droplet is substantially higher than that of a bubble.
In addition, we find that the critical radius is an \textit{unstable} fixed point for all activities where motility-induced phase separations occurs, demonstrating the nucleation dynamics of ABPs occur similarly to a passive system. 
By following the arguments first presented by Becker and D\"{o}ring~\cite{Becker1935KinetischeDampfen}, we also determine the expected rate of nucleation for active droplets~\cite{Redner2016}.
Our generalization of a Gibbs-Thompson relation to mechanical nonequilibrium systems allows a wide body of existing work studying the coarsening dynamics~\cite{Lifshitz1961TheSolutions,Baldan2002ReviewTheories} of passive systems to be applied to active matter.
As an example, we demonstrated the applicability of Voorhees-Glicksman theory~\cite{Voorhees1984SolutionTheory,Voorhees1984SolutionSimulations} to active Brownian particles.

Our theory identifies physical implications of three independently defined surface tensions, shown in Fig.~\ref{fig:figure4}.
The mechanical tension, $\upgamma_{\rm mech}$, which can be negative~\cite{Bialke2015}, controls the pressure difference between a nucleus and its surroundings \textit{only} in the absence of non-local contributions to the dynamic stress.
The capillary tension, $\upgamma_{\rm cw}$, establishes a surface area minimizing principle for an interface, which allows us to approximate fluctuating nuclei as spherically symmetric~\cite{Fausti2021CapillarySeparation,Cates2023ClassicalSeparation,Langford2024TheoryPhases}.
The Ostwald tension, $\upgamma_{\rm ost}$, plays a role in the finite-sized coexistence criteria in addition to controlling the dynamics of a nucleus~\cite{Tjhung2018,Cates2023ClassicalSeparation}. 
All three of these surface tensions are \textit{identical} and \textit{positive} for an equilibrium system but may take disparate, and even negative, values in active systems.
Our results point to several new avenues for future work. 
One interesting direction would be to find rigorous connections between $\upgamma_{\rm ost}$ and more palpable kinetic arguments such as those proposed by Redner~\textit{et al.}~\cite{Redner2013StructureFluid,Redner2016} and 
Lee~\cite{Lee2017InterfaceSeparations}, potentially aided by measurements from particle based simulations.
Another would be to thoroughly investigate the physical implications of the nonlocal stress term $\mathcal{K}$ identified here, which may lead to macroscopic coexistence criteria explicitly dependent on interfacial properties.

The controversy surrounding a negative $\upgamma_{\rm mech}$ has sparked a number of new definitions of surface tension in nonequilibrium systems.
These methods have been rooted in a variety of perspectives, ranging from mechanical balances~\cite{Omar2020,Li2023} to dynamics~\cite{Lee2017InterfaceSeparations,Fausti2021CapillarySeparation} to power functional theory~\cite{Hermann2019b}.
Here, we argue that outside of equilibrium the surface tension must be defined contextually.
While $\upgamma_{\rm mech}$ and $\upgamma_{\rm ost}$ were not found to enter the dynamics of capillary-waves~\cite{Langford2024TheoryPhases}, they are crucial for describing nucleation.
Therefore one cannot haphazardly apply the definition of surface tension from one context to another; \textit{the physically relevant surface tension must be derived from first principles}.
While this work provides insight to the ongoing debates surrounding nonequilibrium surface tension, it may also serve as a starting point for other puzzles in active matter.
This work could be extended to include multiple order parameters~\cite{Evans2023} in order to understand the nucleation of active crystals.
Furthermore, applying the perspective of this work, in conjunction with the fluctuating hydrodynamics~\cite{Langford2024TheoryPhases}, may lead to identification of microscopic systems where $\upgamma_{\rm ost}$ changes sign and novel nucleation dynamics are uncovered~\cite{Tjhung2018}.

Nucleation events are intrinsically challenging to observe and characterize in simulation and experiment. 
Indeed, measuring the nucleation rates of equilibrium systems as simple as hard spheres has been a longstanding challenge~\cite{Gasser2001Real-SpaceCrystallization,Auer2001PredictionColloids,Filion2010CrystalTechniques,Filion2011SimulationPersists,Gispen2023Brute-forceTheory}.
Theories for nucleation are therefore essential for both guiding computational and experimental investigations and making predictions for conditions in which measurements can be prohibitive. 
To this end, CNT has provided crucial intuition for nucleation studies.
It is our hope that this work, which makes use of a number of important recent developments~\cite{Tjhung2018, Solon2018GeneralizedEnsembles, Cates2023ClassicalSeparation, Fausti2021CapillarySeparation, Langford2024TheoryPhases}, can provide a similar road map for studies of active nucleation by extending CNT and canonical coarsening theories to systems far from equilibrium.
\section{Supplemental Material}
See Supplementary Material for detailed derivation of coexistence criteria, nucleation dynamics, and coarsening behavior.
\section{Acknowledgments}
This material is based upon work supported by the U.S. Department of Energy, Office of Science, under Award Number DE-SC0024900. 
L.L. was supported in part by the Department of Defense (DoD) through the National Defense Science \& Engineering Graduate (NDSEG) Fellowship Program. 
\section{Author Declarations}
The authors have no financial conflicts of interests to report.

\appendix
\section{Becker-D\"{o}ring Nucleation Rate}
The existence of an effective potential and the establishment of nuclei obeying an effective detailed balance allows for one to extend equilibrium Becker-D\"{p}ring theory~\cite{Becker1935KinetischeDampfen,Oxtoby1992HomogeneousExperiment,Debenedetti2020MetastableLiquids} to active particles in order to predict the rate of droplet nucleation.
The nucleation event is characterized by the formation of a cluster with a critical number of particles, so we would like to determine the rate at which such critical clusters form.
We will denote the number of clusters with $\ell$ particles per unit volume as $c_{\ell}$.
Additionally, we will assume that clusters are formed and destroyed by \textit{monomer} attachment and detachment. 
In other words, a size $(\ell + 1)$ cluster is formed by the merging of a size $\ell$ cluster and a size $1$ cluster.
The opposite process would be a size $\ell$ cluster being destroyed by splitting into a size $\ell-1$ cluster and a size $1$ cluster.
The \textit{flux}, $J_{\ell}$, in cluster-size space of $\ell$ clusters is then given by the difference between the rate at which droplets containing $\ell$ particles are formed and the rate at which they are destroyed:
\begin{equation}
    J_{\ell} = c_{\ell-1}S_{\ell-1}\beta_{\ell-1} - c_{\ell}S_{\ell}\alpha_{\ell}; \ell \geq 2.\tag{A1}
\end{equation}
Here $S_{\ell}$ is the surface area of an $\ell$ cluster, $\beta_{\ell}$ is the flux of single particles arriving on the surface of the cluster, and $\alpha_{\ell}$ is the flux of single particles leaving the surface of the cluster.
The evolution of $c_{\ell}$ is then given by:
\begin{equation}
    \frac{\partial c_{\ell}}{\partial t} = J_{\ell} - J_{\ell + 1}.\tag{A2}
\end{equation}
We now need to distinguish between two useful limits. 
One limit, which we call \textit{steady state} requires that all $\partial_tc_{\ell} = 0$ but the individual $J_{\ell}$ may remain finite.
The other limit, which we call \textit{quasi-equilibrium} requires that all $\partial_tc_{\ell} = 0$ but additionally requires that all individual $J_{\ell} = 0$. 
The effective detailed balance established for nucleation dynamics implies that the dynamics which govern quasi-equilibrium must also lead to quasi-equilibrium, as Eq.~\eqref{eq:Revo} obeys an effective time-reversal symmetry.
This implies that the values of $\beta_{\ell}$ and $\alpha_{\ell}$ which are valid at quasi-equilibrium must also be valid in the steady-state.
Cates and Nardini~\cite{Cates2023ClassicalSeparation} established that, to linear approximation, the principle of detailed balance is recovered by nucleating clusters in active system even if no such principle exists microscopically.
Then $\alpha_{\ell}$ is:
\begin{equation}
    \alpha_{\ell} = \frac{c_{\ell -1}^{\rm eq} S_{\ell - 1}\beta_{\ell - 1} } { c_{\ell}^{\rm eq}S_{\ell}},
\end{equation}
and  the flux $J_{\ell}$ at \textit{steady-state} may be expressed as:
\begin{equation}
    J_{\ell} = c_{\ell-1}S_{\ell-1}\beta_{\ell-1} - c_{\ell}S_{\ell}\frac{c_{\ell -1}^{\rm eq} S_{\ell - 1}\beta_{\ell -1} } { c_{\ell}^{\rm eq}S_{\ell}}.
\end{equation}
This flux may be rewritten as:
\begin{equation}
    J_{\ell} = \beta_{\ell-1}S_{\ell-1}c_{\ell-1}^{\rm eq}\left[\frac{c_{\ell - 1}}{c_{\ell - 1}^{\rm eq}} - \frac{c_{\ell}}{c_{\ell}^{\rm eq}}\right].\tag{A4}
\end{equation}
Note that in the steady state all $J_{\ell}$ with $\ell \geq 2$ are equal to $J_{\ell + 1}$. Then we can label $J_{\ell} = \mathcal{R}$ and write:
\begin{equation}
    \frac{\mathcal{R}}{\beta_{\ell - 1}S_{\ell - 1} c^{\rm eq}_{\ell - 1}} = \left[\frac{c_{\ell - 1}}{c_{\ell - 1}^{\rm eq}} - \frac{c_{\ell}}{c_{\ell}^{\rm eq}}\right]\label{seq:ratepresum}.
\end{equation}
We now sum both sides of the equation from $\ell = 2$ to $\ell = \Lambda + 1$, where $\Lambda$ is some very large number (larger than the critical nucleus size).
After summing, the right hand side of Eq.~\eqref{seq:ratepresum} will vanish except for the first term where $\ell = 2$ and the last term where $\ell = \Lambda +  1$:
\begin{equation}
    \mathcal{R} = \left[\frac{c_{1}}{c_{1}^{\rm eq}} - \frac{c_{\Lambda + 1}}{c_{\Lambda + 1}^{\rm eq}}\right]\left[\sum_{\ell = 1}^{\Lambda}\frac{1}{\beta_{\ell}S_{\ell}c_{\ell}^{\rm eq}}\right]^{-1}\label{seq:ratepostsum}.
\end{equation}
We now argue that since the bulk metastable phase (a dilute gas) exists overwhelmingly as single particles, so the quasi-equilibrium and steady-state concentration of single particles are indistinuishable and $c_{1}/c_{1}^{\rm eq} \approx 1$. Additionally, $\Lambda$ is larger than the size of the critical nucleus, so $c_{\Lambda + 1}$ should be zero in the metastable steady state but \textit{finite} at quasi-equilibrium. 
Then Eq.~\eqref{seq:ratepostsum} reduces to:
\begin{equation}
    \mathcal{R} = \left[\sum_{\ell = 1}^{\Lambda}\frac{1}{\beta_{\ell}S_{\ell}c_{\ell}^{\rm eq}}\right]^{-1}.
\end{equation}
Using the effective nucleation potential~\cite{Cates2023ClassicalSeparation}, we know the relative probability of observing a nucleus with radius $R$, and (given a spherical geometry), we therefore we have an expression for $c_{\ell}^{\rm eq}$ as:
\begin{equation}
    c_{\ell}^{\rm eq} = \rho^{\rm sat}\text{exp}\left[\left(k_BT^{\rm act}\right)^{-1}U(R_{\ell})\right],
\end{equation}
where $U(R_{\ell})$ is the effective potential evaluated at a radius corresponding to a sphere of $\ell$ particles. 
Switching from a summation to an integral, we can then approximate the nucleation rate as:
\begin{equation}
    \mathcal{R} = \rho^{\rm sat}\left[\int_{1}^{\Lambda} \text{d}\ell\frac{1}{\beta_{\rm \ell}S_{\ell}} \text{exp}\left[\frac{U(R_{\ell})}{k_BT^{\rm act}}\right]\right]^{-1}.\label{seq:integralpreapprox}
\end{equation}
We now would like to make the connection between radius and cluster size more explicit:
\begin{equation}
    \ell = \frac{4}{3}\pi R^3 \rho^{\rm in}.\label{seq:sizetorad}
\end{equation}
Alternatively, one may rearrange Eq.~\eqref{seq:sizetorad} and obtain an expression for the radius in terms of size:
\begin{equation}
    R = \left[\frac{3\ell}{4\pi\rho^{\rm in}}\right]^{1/3}.\label{seq:radtosize}
\end{equation}
We now consider the derived form of the nucleation barrier, Eq.~\eqref{eq:potentialintuitive}, and express it in terms of cluster size:
\begin{equation}
    U(\ell) = 4\pi \upgamma_{\rm ost} \left[\frac{3\ell}{4\pi \rho^{\rm in}}\right]^{2/3}- \frac{\ell \mathcal{V}}{\rho^{\rm in}}.
\end{equation}
From this form the critical cluster size $\ell^*$ can be solved for by extremizing $U(\ell)$, resulting in:
\begin{equation}
    \ell^* = \frac{32}{3}\pi \rho^{\rm in}\left(\frac{\upgamma_{\rm ost}}{\mathcal{V}}\right)^3.
\end{equation}
We now Taylor expand the the barrier about this maximum to second order,
\begin{equation}
    U(\ell) = U(\ell^*) - \frac{1}{2}\eta \left(\ell - \ell^*\right)^2,\label{seq:taylorexpandedpotential}
\end{equation}
where we have defined $\eta$ as the second derivative of the barrier with respect to $\ell$ and evaluated at $\ell^*$.
Explicitly, this is solved for as:
\begin{equation}
    \eta = \frac{1}{32\pi \left(\rho^{\rm in}\right)^2}\frac{\mathcal{V}^4}{\upgamma_{\rm ost}^3}.
\end{equation}
We now consider the integral from Eq.~\eqref{seq:integralpreapprox}.
Substitution of Eq.~\eqref{seq:taylorexpandedpotential} into Eq.~\eqref{seq:integralpreapprox}, changing integration variables from $\ell$ to $\delta\ell \equiv \ell - \ell^*$, and performing a saddle-point approximation at $\delta\ell = 0$ results in:
\begin{equation}
    \mathcal{R} = \rho^{\rm sat}\beta_{\ell^*}4\pi\left(\frac{3\ell^*}{4\pi \rho^{\rm in}}\right)^{2/3}\sqrt{\frac{\eta}{2\pi k_BT^{\rm act}}}\text{exp}\left[-\frac{U(\ell^*)}{k_BT^{\rm act}}\right].\label{seq:rateprekinetic}
\end{equation}
Everything in Eq.~\eqref{seq:rateprekinetic} can now be evaluated \textit{except} for the rate of single particle attachment on the critical nucleus $\beta_{\ell^*}$, for which we use the simple kinetic model proposed Redner~\cite{Redner2013StructureFluid,Redner2016}.
$\beta_{\ell}$ is the rate of absorption of particles onto the surface of a cluster.
This rate should be given by the density of the gas phase times the speed at which the particles swim times the probability that a given particle will absorb onto the surface:
\begin{equation}
    \beta_{\ell} = \rho^{\rm sat}U_o P^{\rm abs},
\end{equation}
where $P^{\rm abs}$ is the probability a particle absorbs onto the surface.
Assuming that the surface of the nucleus has an inward pointing normal vector $\mathbf{n}$, we model $P^{\rm abs}$ as \textit{zero} if a particle points away from the interface and proportional to $\mathbf{n}\cdot\mathbf{q}$ if pointing in the hemisphere towards interface. 
In other words $P^{\rm abs}$, is given by:
\begin{equation}
P^{\rm abs} = \begin{cases} 
      \mathbf{n}\cdot\mathbf{q} & \mathbf{n}\cdot\mathbf{q} > 0 \\
      0 & \mathbf{n}\cdot\mathbf{q}<0  
   \end{cases}
\end{equation}
We pick a point on the sphere with unit normal vector $-\mathbf{e}_x$, and then represent the orientation of a given particle as:
\begin{equation}
    \mathbf{q} = \sin{\theta}\cos{\phi}\mathbf{e}_{x} + \sin{\theta}\sin{\phi}\mathbf{e}_{y} + \cos{\theta}\mathbf{e}_z.
\end{equation}
Then, assuming that the orientations of the relevant particles are uniformly distributed, the \textit{average} of $\beta_{\ell}$ can be expressed as:
\begin{equation}
    \langle \beta_{\ell} \rangle = \rho^{\rm sat}U_o \frac{1}{4\pi}\int_{0}^{\rm 2\pi} \int_0^{\pi} \sin{\theta}P^{\rm abs} \text{d}\theta \text{d}\phi\label{seq:betaaveragestart}.
\end{equation}
$P^{\rm abs}$ is only nonzero when $\phi$ is between $0$ and $\pi/2$ as well as when $\phi$ is between $3\pi/2$ and $2\pi$, for all values of $\theta$. 
Then Eq.~\eqref{seq:betaaveragestart} reduces to:
\begin{align}
    \langle \beta_{\ell} \rangle =  &\rho^{\rm sat}U_o \frac{1}{2\pi^2}\Biggl[\int_{0}^{\pi/2} \int_0^{\pi} \sin^2{\theta}\cos{\phi}\text{d}\theta \text{d}\phi \nonumber \\   &+ \int_{3\pi/2}^{2\pi} \int_0^{\pi} \sin^2{\theta}\cos{\phi}\text{d}\theta \text{d}\phi \Biggr]\nonumber \\   = &\frac{\rho^{\rm sat}U_o}{4}.\label{seq:betaaveragefinish}
\end{align}
Substitution of Eq.~\eqref{seq:betaaveragefinish} into Eq.~\eqref{seq:rateprekinetic} results in:
\begin{equation}
    \mathcal{R} = \left(\rho^{\rm sat}\right)^2\pi U_o\left(\frac{3\ell^*}{4\pi \rho^{\rm in}}\right)^{2/3}\sqrt{\frac{\eta}{2\pi k_BT^{\rm act}}}\text{exp}\left[-\frac{U(\ell^*)}{k_BT^{\rm act}}\right]\label{seq:nucleationrate}.
\end{equation}
The expected time to observe a nucleation event is given by $\tau_{\rm nuc} = (\mathcal{R}V)^{-1}$, where $V$ is the total volume of the system. 
Following straightforward manipulations, it can be shown that $\eta$ is alternatively expressed as:
\begin{equation}
    \eta = \frac{\upgamma_{\rm ost}}{2\pi (\rho^{\rm in})^2 R_C^4}.\label{seq:etaalternate}
\end{equation}
Substitution of Eq.~\eqref{seq:etaalternate} into Eq.~\eqref{seq:nucleationrate} results in:
\begin{equation}
    \mathcal{R} = \left(\rho^{\rm sat}\right)^2 \pi U_o \sqrt{\frac{\upgamma_{\rm ost}}{2\pi(\rho^{\rm in})^2 k_BT^{\rm act}}} \text{exp}\left[-\frac{U(\ell^*)}{k_BT^{\rm act}}\right].
\end{equation}
\section{Active Lifshitz-Slyozov-Wagner Theory}
We consider the growth of many nuclei in a supersaturated solution.
However, we will also assume that the saturation is sufficiently small that the overall volume fraction of the final phase is effectively zero. 
In this limit the nuclei will evolve independently of one another, and we may use the single nucleus Langevin equation [Eq.~\eqref{eq:Revo}] as the dynamics for all nuclei.
For convenience, we will rewrite Eq.~\eqref{eq:Revo} as:
\begin{equation}
    \frac{\partial R}{\partial t} = \frac{1}{\zeta_{\rm eff}}F^D(R) + g(R)\eta(t),\label{seq:drdtrewrite}
\end{equation}
where $\eta(t)$ is a unit Gaussian white noise with zero mean and variance $\delta(t-t')$ and $g(R)$ is defined as:
\begin{equation}
    g(R) = \sqrt{\frac{2k_BT^{\rm act}\rho^{\rm sat}}{4\pi\zeta\Delta\rho^2R^3}}.\label{seq:glangdef}
\end{equation}
and $F^D(R)$ remains defined as:
\begin{equation}
    F^D(R) = 8\pi R^2\upgamma_{\rm ost}\left[\frac{1}{R_C} - \frac{1}{R}\right].\label{seq:Fdetermindefrestate}
\end{equation}
Although Eq.~\eqref{eq:Revo} contains multiplicative noise, the choice of time discretization scheme (e.g. It\^{o}, Stratonovich, etc.) is inconsequential as the drift term~\cite{vanKampen1981ItoStratonovich} $\frac{1}{2}g(R)g'(R)$ is $\mathcal{O}\left(R^{-4}\right)$.
We now wish to solve for how the distribution of particle sizes evolves with time. 
This can be accomplished by first determining the Fokker-Planck equation associated with Eq.~\eqref{seq:drdtrewrite}:
\begin{equation}
    \frac{\partial P}{\partial t} = -\frac{\partial}{\partial R}\left[\frac{1}{\zeta_{\rm eff}}F^D(R)P(R) + \frac{1}{2}\frac{\partial}{\partial R}\left(g^2(R)P(R)\right)\right].\label{seq:fokkerinit}
\end{equation}
From Eq.~\eqref{seq:glangdef}, we can see that the diffusive term will be negligible in comparison to the drift term for large $R$, resulting in:
\begin{equation}
    \frac{\partial P}{\partial t} = -\frac{\partial}{\partial R}\left[\frac{1}{\zeta_{\rm eff}}F^D(R)P(R;t)\right].\label{seq:fokkerapprox}
\end{equation}
We can now proceed in the same fashion as Lifshitz and Slyozov~\cite{Lifshitz1961TheSolutions}, where it was assumed that, at long times, the asymptotic form of the distribution takes the following form:
\begin{equation}
    P(R;t) = \frac{1}{R_C(t)}\mathcal{N}\left(x\right),\label{seq:lswansatz}
\end{equation}
where we have defined the dimensionless radius $x\equiv R/R_C$ and $\mathcal{N}(x)$ is the dimensionless distribution of radii scaled by $R_C$.
The time-dependence of $P(R;t)$ comes from that of the critical radius. 
Substitution of Eq.~\eqref{seq:lswansatz} into Eq.~\eqref{seq:fokkerapprox} results in:
\begin{align}
    &\frac{1}{R_C^2}\frac{\partial R_C}{\partial t}\left[\mathcal{N}(x) + x\frac{d\mathcal{N}}{dx}\right]\nonumber \\ & = \frac{2\rho^{\rm sat}\upgamma_{\rm ost}}{\zeta R_C^4 \Delta\rho^2}\left[\frac{d\mathcal{N}}{dx}\left(x^{-1} - x^{-2}\right) + \mathcal{N}(x)\left(2x^{-3} - x^{-2}\right)\right]\label{seq:lswbrackets}
\end{align}
In order for the ansatz Eq.~\eqref{seq:lswansatz} to be consistent with Eq.~\eqref{seq:lswbrackets} (i.e., $\mathcal{N}(x)$ \textit{only depends on x}), we require:
\begin{equation}
    R_C^2\frac{\partial R_C}{\partial t} = \frac{2\rho^{\rm sat}}{\zeta\Delta\rho^2} \upgamma_{\rm ost}\mathcal{C},
\end{equation}
where $\mathcal{C}$ is some constant.
This implies that the critical radius evolves in time according to:
\begin{equation}
    R_C(t) = \left[R^3_C(t_o)+ \frac{6\rho^{\rm sat}}{\zeta\Delta\rho^2}\upgamma_{\rm ost}\mathcal{C}t\right]^{1/3}.\label{seq:Rc_dynamics}
\end{equation}
By equating the bracketed terms in Eq.~\eqref{seq:lswbrackets} (within a dimensionless constant) one obtains a differential equation for $\mathcal{N}(x)$ which one can solve to obtain the asymptotic LSW distribution for nucleus size~\cite{Lifshitz1961TheSolutions}.
Here, we are simply interested in how the average nucleus size evolves in time. 
Straightforward insertion of Eq.~\eqref{seq:lswansatz} into ${\langle R \rangle = \int_0^{\infty}P(R;t)RdR}$ reveals that the average radial dynamics are identical to those of $R_C$ [Eq.~\eqref{seq:Rc_dynamics}] within a multiplicative dimensionless constant.
\section{Coarsening Beyond the Zero Saturation Limit}
We consider a system of phase separated active particles where the nucleation of minority phase has completed and the the final phase composition has been reached.
In general, the system may contain many domains of minority phase. 
The overall volume fraction of this system will be finite and thus nuclei may interact with one another.
When $\upgamma_{\rm ost} > 0$, we expect the forward Ostwald ripening process to occur and large domains of minority phase to grow at the expense of small domains until the system reaches a final steady state with a single domain.
Rather than fixing the system to be finite size and considering the influence of boundaries, we consider an infinite tiling of periodic cells.
Due to the large system sizes and long timescales necessary to probe the dynamics of this process which evolves towards larger and larger domains, direct integration of Eq.~\eqref{eq:flucthydro} may be impractical. 
The problem may be greatly simplified by noting that majority phase will relax on timescales much faster than the domains of minority phase grow or shrink.
Ignoring stochastic contributions, the density field dynamics~\eqref{eq:flucthydro} take the form:
\begin{equation}
    \frac{\partial \rho}{\partial t} = -\frac{1}{\zeta}\boldsymbol{\nabla}\cdot\boldsymbol{\nabla}\cdot\bm{\Sigma}.\label{seq:flucthydronoiseless}
\end{equation}
By restricting our perspective to \textit{only describe the majority phase}, we may drop the time dependence from  Eq.~\eqref{seq:flucthydronoiseless} by taking the quasi-static limit. 
From the perspective of the majority phase, the particles injected/adsorbed from the nuclei can be represented with finite-flux (and moving) boundary conditions at the interface between nuclei and the parent phase. 
Here, we take another approach and approximate majority phase particle addition/removal through spatially varying volumetric \textit{sources and sinks}. 
This approach results in the quasi-static continuity equation now taking the following form: 
\begin{equation}
    0 = -\boldsymbol{\nabla}\cdot\boldsymbol{\nabla}\cdot\bm{\Sigma} + \zeta\sum_{u,v,w}\sum_{i=1}^NS_i(\mathbf{r}),
\end{equation}
where we gain the source or sink term $S_i$ in lieu of explicit consideration of the $i$th minority domain. 
Here the sum over $(u,v,w)$ is a summation over all periodic translations of the cell. 
Additionally, coarse-graining the effect of the minority phase on the density field into sources/sinks renders the interface between the nuclei and parent phase beyond our level of description. 
We may therefore disregard the higher order gradient terms from the dynamic stress tensor resulting in:
\begin{equation}
    0 = \boldsymbol{\nabla}^2\mathcal{P}(\mathbf{r}) + \zeta\sum_{u,v,w}\sum_{i=1}^NS_i(\mathbf{r}).\label{seq:vgtheorystart}
\end{equation}

Globally, the number of particles must remain conserved, imposing the following constraint on $S_i$:
\begin{equation}
    \sum_{i=1}^N \int_V S_i(\mathbf{r})\text{d}\mathbf{r} = 0.\label{seq:constraint}
\end{equation}
From Eq.~\eqref{seq:vgtheorystart}, a mathematical analogue between the pressure field and the electrostatic potential due to $N$ charge densities may be drawn. 
By approximating the minority domains as spherically symmetric we may safely collapse the ``charge densities'' $S_i$ into point charges at their centers by defining $S_i(\mathbf{r})\equiv 2\pi B_i\delta(\mathbf{r}-\mathbf{r}_i)/\zeta$.
Equation~\eqref{seq:vgtheorystart} can now be expressed as:
\begin{equation}
    0 = \boldsymbol{\nabla}^2\mathcal{P}(\mathbf{r}) + 4\pi\sum_{u,v,w}\sum_{i=1}^NB_i\delta(\mathbf{r}-\mathbf{r}_i-u\mathbf{a} - v\mathbf{b} - w\mathbf{c}),\label{seq:vgtheorybi}
\end{equation}
where $\mathbf{a}$, $\mathbf{b}$, $\mathbf{c}$ are the lattice vectors and the particle conservation constraint [Eq.~\eqref{seq:constraint}] becomes $\sum B_i = 0$.
The solution to Eq.~\eqref{seq:vgtheorybi} is well-known and given by:
\begin{equation}
    \mathcal{P}(\mathbf{r}) = \mathcal{P}_o + \sum_{u,v,w}\sum_{i=1}^{N}\frac{B_i}{|\mathbf{r} - \mathbf{r}_i - u\mathbf{a} - v\mathbf{b} - w\mathbf{c}|},
    \label{seq:pressurefieldsolution}
\end{equation}
where $\mathcal{P}_o$ is the unknown pressure of the majority phase far from the nuclei.
We now assume that every individual nucleus is in mechanical equilibrium with its immediate surroundings.
Eqs.~\eqref{eq:coexistrestate} allow us to solve for the coexistence densities of a finite sized nucleus with radius $R$. 
We define $\rho^R(R)$ as a function mapping the size of a nucleus with size $R$ to the \textit{outside} coexistence density. 
Then, if $\mathbf{r}^s_j$ is a point on the surface of a nucleus of size $R_j$, we have:
\begin{equation}
    \mathcal{P}(\mathbf{r}^s_j) = \mathcal{P}(\rho^R(R_j))\label{seq:gtboundcon}
\end{equation}
Substitution of the boundary condition [Eq.~\eqref{seq:gtboundcon}] into Eq.~\eqref{seq:pressurefieldsolution} gives:
\begin{align}
    \mathcal{P}(\mathbf{r}^s_j) &= \mathcal{P}(\rho^R(R_j)) \nonumber \\ &= \mathcal{P}_o + \frac{B_j}{R_j} + \sum_{u,v,w}\sum_{i=1}^{N}\frac{B_i}{|\mathbf{r}^s_j - \mathbf{r}_i - u\mathbf{a} - v\mathbf{b} - w\mathbf{c}|},\label{seq:pressureunaveraged}
\end{align}
where the sum ignores the $i=j$ term when $u,v,w = 0$. A key subtlety of Eq.~\eqref{seq:pressureunaveraged} is that $\mathbf{r}^s_j$ is a point on the \textit{surface} of nucleus $j$ while $\mathbf{r}_i$ is the \textit{center} of nucleus $i$. By expanding the summation in terms of Legendre polynomials and averaging over all possible choices of $\mathbf{r}^s_j$ on the surface of the $j$th nucleus, one may show that
\begin{equation}
    \mathcal{P}(\rho^R(R_j)) = \mathcal{P}_o + \frac{B_j}{R_j} + \sum_{u,v,w}\sum_{i=1}^{N}\frac{B_i}{|\mathbf{r}_j - \mathbf{r}_i - u\mathbf{a} - v\mathbf{b} - w\mathbf{c}|},\label{seq:pressureaveraged}
\end{equation}
where $\mathbf{r}_j$ is the center of nucleus $j$. 
Evaluation of the summation over an infinite periodic lattice can be accomplished via standard techniques of Ewald summation, although with the added complication that the sum ignores the $i=j$ case when $u,v,w = 0$, resulting in:
\begin{widetext}
\begin{align}
    \mathcal{P}(\rho^R(R_j)) = &\mathcal{P}(\rho^{\rm mat}) + B_j\left(\frac{1}{R_j} - \sqrt{\frac{1}{\eta \pi}}\right) + \frac{4\pi}{\mathbf{a}\cdot\left(\mathbf{b}\times\mathbf{c}\right)}\sum_{\mathbf{k}\neq \bm{0}}\sum_{i=1}^{N}\frac{B_i}{|\mathbf{k}|^2}\text{exp}\left[-\eta |\mathbf{k}|^2 - i\mathbf{k}\cdot\left(\mathbf{r}_j - \mathbf{r}_i\right)\right] \nonumber \\ &+ \sum_{u,v,w}\sum_{i=1}^{N}\frac{B_i}{|\mathbf{r}_j - \mathbf{r}_i - u\mathbf{a} - v\mathbf{b} - w\mathbf{c}|} \text{erfc}\left(\frac{|\mathbf{r}_j - \mathbf{r}_i - u\mathbf{a} - v\mathbf{b} - w\mathbf{c}|}{2\sqrt{\eta}}\right),
    \label{seq:pressureewaldsum}
\end{align}
\end{widetext}
where $\mathbf{k}$ are the reciprocal lattice vectors corresponding to our periodic cell and $\eta$ is free parameter that controls how quickly the summation converges in real space as opposed to reciprocal space. 
Equation~\eqref{seq:pressureewaldsum} can be expressed as a matrix equation $\mathbf{b} = \mathbf{A}\cdot\mathbf{x}$, where
\begin{subequations}
\label{seq:pressurematrixeq}
\begin{equation}
    \mathbf{b} = \begin{bmatrix}\mathcal{P}(\rho^{R}(R_1))\\ \mathcal{P}(\rho^{R}(R_2))\\ \vdots \\ \mathcal{P}(\rho^{R}(R_N)) \\ 0\end{bmatrix},
\end{equation}
\begin{equation}
    \mathbf{x} = \begin{bmatrix} B_1 \\ B_2 \\ \vdots \\ B_N \\ \mathcal{P}(\rho^{\rm mat})\end{bmatrix},
\end{equation}
and
\begin{equation}
    \mathbf{A} = \begin{bmatrix} \frac{1}{R_1} + D & \mathcal{O}_{12} & \mathcal{O}_{13} & \ldots & 1 \\ \mathcal{O}_{21} & \frac{1}{R_2} + D & \mathcal{O}_{23} & \ldots & 1 \\ \mathcal{O}_{31} & \mathcal{O}_{32} & \frac{1}{R_3} + D & \ldots & 1\\ \vdots & \vdots & \vdots & \ddots & \vdots \\ 1 & 1 & 1 & \ldots & 0\end{bmatrix}.
\end{equation}
\end{subequations}
Here $D$ is given by:
\begin{align}
    D = &-\sqrt{\frac{1}{\eta \pi}} + \frac{4\pi}{L_xL_yL_z}\sum_{\mathbf{k}\neq 0}\frac{1}{|\mathbf{k}|^2}\text{exp}\left(-\eta |\mathbf{k}|^2\right) \nonumber \\ &+ \sum_{u,v,w \neq 0}\frac{1}{|\mathbf{r}_{uvw}|}\text{erfc}\left(\frac{|\mathbf{r}_{uvw}|}{2\sqrt{\eta}}\right),
\end{align}
where we have defined ${|\mathbf{r}_{uvw}| \equiv \left(|u\mathbf{a}|^2 + |v\mathbf{b}|^2 + |w\mathbf{c}|^2\right)^{1/2}}$
and $\mathcal{O}_{ij}$ is given by:
\begin{align}
    \mathcal{O}_{ij} = &\frac{4\pi}{\mathbf{a}\cdot\left(\mathbf{b}\times\mathbf{c}\right)}\sum_{\mathbf{k}\neq\bm{0}}\frac{1}{|\mathbf{k}|^2}\text{exp}\left(-\eta|\mathbf{k}|^2 - i\mathbf{k}\cdot\left(\mathbf{r}_j-\mathbf{r}_i\right)\right) \nonumber \\ &+\sum_{u,v,w} \frac{1}{|\mathbf{r}_{uvw}^{ij}|} \text{erfc}\left(\frac{|\mathbf{r}_{uvw}^{ij}|}{2\sqrt{\eta}}\right),
\end{align}
where we have defined ${|\mathbf{r}^{ij}_{uvw}|\equiv |\mathbf{r}_j - \mathbf{r}_i  - u\mathbf{a} - v\mathbf{b} - w\mathbf{c}|}$.
Given the size and locations of each nucleus in the cell, one may then solve Eqs.~\eqref{seq:pressurematrixeq} in order to find all $B_i$ and the background pressure $\mathcal{P}_o$. Let's now consider the rate of change of the radius of a particular nucleus and how it translates to the nucleus source/sink strength $B_i$. 
If nucleus $i$ changes its radius $R_i$ by an amount $dR_i$ over a time period $dt$, then the rate of mass injected into the majority phase by particle $i$ is given by:
\begin{equation}
    4\pi R_i^2 \frac{dR_i}{dt}\left(\rho^{\rm out} - \rho^{\rm in}\right)
\end{equation}
By the definition of $S_i$, the rate of mass injected into the majority phase by nucleus $i$ is also equal to the integral of $S_i$ across the entire system of volume. Then we have:
\begin{equation}
    4\pi R_i^2 \frac{dR_i}{dt}\left(\rho^{\rm out} - \rho^{\rm in}\right) = \int_V d\mathbf{r} S_i
\end{equation}
From the definition of $B_i$ we then have:
\begin{equation}
    4\pi R_i^2 \frac{dR_i}{dt}\left(\rho^{\rm out} - \rho^{\rm in}\right) =\frac{4\pi}{\zeta} B_i
\end{equation}
Then the rate of change of the radius of nucleus $i$ is given by:
\begin{equation}
    \frac{dR_i}{dt} = \frac{B_i}{\zeta R_i^2 (\rho^{\rm out} - \rho^{\rm in})}
\end{equation}
Therefore solution of the matrix equation defined by Eqs.~\eqref{seq:pressurematrixeq} allows for the extraction of the coarsening rates. 

\end{document}


\maketitle

\pagenumbering{roman}

\setcounter{secnumdepth}{2}
\setcounter{tocdepth}{2}
{
	\hypersetup{
		linkcolor=Black,
		citecolor=Black
	}
	\vspace{-30pt}
	\tableofcontents
}

\noindent\rule{5.0cm}{0.4pt}
	{\footnotesize
	$\\$
	{
	    {$^\dag \,$}aomar@berkeley.edu}
	}

\newpage 
\pagenumbering{arabic}
\section{\label{sec:static}Mechanics of the Critical Nucleus}
In this Section, we detail the theory and numerical procedures used to solve for the stationary density profile of the critical nucles of a system arbitrarily far from equilibrium. 
We first derive the finite-sized phase coexistence criteria for active particles by extending the coexistence theory used to obtain nonequilibrium binodals~\cite{Omar2023b}. 
We then apply this theory to active Brownian particles (ABPs). 
In addition, we provide a step by step walk through of the numerical scheme used to translate phase coexistence criteria into density profiles and critical radii. 
Further, we demonstrate how the generalized Gibbs-Duhem relation found in Ref.~\cite{Evans2023} can be used to rewrite the coexistence criteria in terms of the generalized chemical potential discussed in Ref.~\cite{Solon2018GeneralizedEnsembles}. 
Finally, we conclude this Section with a mapping between the mechanical perspective presented in this work and previous work done in the context of AMB+~\cite{Tjhung2018}.
Throughout this section, as well as the rest of the Supporting Information, we assume three dimensions ($d=3$) unless specified otherwise.

\subsection{\label{sec:coexistcriteria}Nonequilibrium Critical Nucleus Coexistence Criteria}

We aim to describe the structure of a critical nucleus with density $\rho^{\rm in}$ in a system that has been prepared at a supersaturation density $\rho^{\rm sat}$. 
We assume that the critical nucleus is sufficiently small compared to the system size such that the density outside the nucleus is well approximated by the supersaturation density $\rho^{\rm sat}$.
In addition, we approximate the geometry of the critical nucleus as a perfect sphere with radius $R_{C}$.

When the critical nucleus is an unstable fixed point, a nucleus smaller than the critical nucleus will experience a flux normal to its interface such that the nucleus shrinks.
A nucleus larger than the critical nucleus will experience \textit{the opposite} flux normal to the interface such that the nucleus grows.
Therefore, in order to solve for the critical nucleus, we seek states which have zero flux normal to the interface. 
Even if the critical nucleus is a stable fixed point, as recently found in some active systems~\cite{Tjhung2018}, similar arguments may be used to still conclude the critical nucleus will be flux-free.
Two flux-free states, the macroscopically phase separated and metastable homogeneous states, are already known.
The critical nucleus will correspond to a state with a \textit{finite-size} minority phase and zero flux normal to the interface. 
The evolution of the density flux $\mathbf{J} = \rho(\mathbf{r},t)\mathbf{u}(\mathbf{r},t)$, where $\mathbf{u}$ is the number-averaged velocity of particles, can be expressed as a balance of linear momentum:
\begin{equation}
    \frac{\partial \left(m\mathbf{J}\right)}{\partial t} = -\boldsymbol{\nabla}\cdot\left(\frac{m\mathbf{J}\mathbf{J}}{\rho}\right) + \boldsymbol{\nabla}\cdot\bm{\sigma} + \mathbf{b},\label{seq:linmomentum}
\end{equation}
where $m$ is the mass of each particle, $\bm{\sigma}$ is the stress tensor and $\mathbf{b}$ are the body forces acting on the particles.
Additionally, we assume that $\bm{\sigma}$ and $\mathbf{b}$ can be written as local gradient expansions, which is typically valid when interactions are short-ranged.
For an \textit{overdamped} system with spherical symmetry and zero flux in the radial direction, Eq.~\eqref{seq:linmomentum} reduces to $\boldsymbol{\nabla}\cdot\bm{\sigma} + \mathbf{b} = \bm{0}$.
Exact microscopic expressions for both $\bm{\sigma}$ and $\mathbf{b}$ can be determined through an Irving-Kirkwood-Noll procedure~\cite{Irving1950,Noll1955,Lehoucq2010}.
These microscopic expressions can include terms that appear to be non-local, requiring integrals over all space to, for instance, compute the stress  at a particular point in space arising from particle interactions. 
However, these terms are often well-approximated by a local gradient expansion, the coefficients of which can be systematically determined~\cite{Lindell1993DeltaMethod}.
We thus focus on systems in which $\bm{\sigma}$ and $\mathbf{b}$ can be fully expressed as locally.
For spatially isotropic systems, $\bm{\sigma}$ can be expanded to second order in density gradients as:
\begin{equation}
    \bm{\sigma} = \left[\alpha_0(\rho)\mathbf{I} +\alpha_1(\rho)\boldsymbol{\nabla}^2\rho + \alpha_2(\rho)|\boldsymbol{\nabla}\rho|^2\right]\mathbf{I} + \alpha_3(\rho)\boldsymbol{\nabla}\rho\boldsymbol{\nabla}\rho + \alpha_4(\rho)\boldsymbol{\nabla}\boldsymbol{\nabla}\rho,
\end{equation}
and $\mathbf{b}$ can be expanded to third order in density gradients as:
\begin{equation}
    \mathbf{b} = \beta_0(\rho)\boldsymbol{\nabla}\rho + \beta_1(\rho)\boldsymbol{\nabla}\rho\boldsymbol{\nabla}^2\rho + \beta_2(\rho)\boldsymbol{\nabla}|\boldsymbol{\nabla}\rho|^2 + \beta_3(\rho)\boldsymbol{\nabla}\boldsymbol{\nabla}^2\rho + \beta_4(\rho)|\boldsymbol{\nabla}\rho|^2\boldsymbol{\nabla}\rho.
\end{equation}

Up to a divergence-free term, we may absorb the body force into a stress-like contribution such that:
\begin{subequations}
    \begin{equation}
        \boldsymbol{\nabla}\cdot\bm{\sigma} + \mathbf{b} = \boldsymbol{\nabla}\cdot\left[\bm{\sigma} + \mathbf{F} \right] \equiv \boldsymbol{\nabla}\cdot\bm{\Sigma} = \bm{0}, 
    \end{equation}
    \begin{equation}
        \mathbf{F} = \frac{1}{4\pi}\int \text{d}\mathbf{r}' \mathbf{b}(\mathbf{r}')\frac{\mathbf{r}-\mathbf{r}'}{|\mathbf{r}-\mathbf{r}'|^3},\label{seq:Fintdef}
    \end{equation}
\end{subequations}
where we have used the Green's function identity $\boldsymbol{\nabla}\cdot \left((\mathbf{r}-\mathbf{r}')/|\mathbf{r}-\mathbf{r}'|^3\right) = 4\pi\delta(\mathbf{r}-\mathbf{r}')$ in three dimensions.
Details on evaluating the integral Eq.~\eqref{seq:Fintdef} are provided in Section~\ref{sec:nonlocal}.
In general, $\mathbf{F}$ will contain terms which may be expressed as \textit{local} density gradient expansions as well as \textit{non-local} terms which depend on spatial integrals over the entire density field.
Due to spherical symmetry, $\bm{\Sigma}$ has only two independent components and is given by:
\begin{equation}
    \bm{\Sigma} = \begin{bmatrix} \Sigma_{rr} & 0 & 0 \\ 0 & \Sigma_{tt} & 0 \\ 0 & 0 & \Sigma_{tt} \end{bmatrix} = \Sigma_{rr}\mathbf{e}_r\otimes\mathbf{e}_r + \Sigma_{tt}\left(\mathbf{I} - \mathbf{e}_r\otimes\mathbf{e}_r\right).\label{seq:dynstressform}
\end{equation}
We now find the flux-free steady states by demanding the dynamic stress tensor has zero divergence, and we expand the stress tensor to second order in density gradients:

\begin{subequations}
    \label{seq:stationary}
    \begin{align}
         &\boldsymbol{\nabla}\cdot\bm{\Sigma}=\mathbf{0} \label{seq:Jstress}, \\ \Sigma_{rr} = -\mathcal{P}(\rho) + \kappa_1(\rho)\left(\frac{\partial}{\partial r} + \frac{2}{r}\right)\frac{\partial\rho}{\partial r} &+ \left(\kappa_2(\rho) + \kappa_3(\rho)\right)\left(\frac{\partial\rho}{\partial r}\right)^2 + \kappa_4(\rho)\frac{\partial^2\rho}{\partial r^2} + \mathcal{K}\label{seq:dynstressrr},\\ \Sigma_{tt} = -\mathcal{P}(\rho)+ \kappa_1(\rho)&\left(\frac{\partial}{\partial r} + \frac{2}{r}\right)\frac{\partial\rho}{\partial r} + \kappa_2(\rho)\left(\frac{\partial\rho}{\partial r}\right)^2 + \mathcal{K}, \\ \mathcal{P} = -&\alpha_0(\rho) - \int\beta_0(\rho)\text{d}\rho, \\ \kappa_1(\rho) &= \alpha_1(\rho) + \beta_3(\rho), \\ \kappa_2(\rho) = \alpha_2(\rho) + &\frac{1}{2}\frac{\partial\beta_3(\rho)}{\partial\rho} + \beta_2(\rho) - \frac{1}{2}\beta_1(\rho), \\ \kappa_3(\rho) = &\alpha_3(\rho) + \beta_1(\rho) - \frac{\partial\beta_3}{\partial\rho}, \\ &\kappa_4(\rho) = \alpha_4(\rho), \\ \mathcal{K} = \int_r^{\infty}\Biggl[-\beta_4(\rho) + \frac{\partial\beta_2}{\partial\rho} &+ \frac{1}{2}\frac{\partial\beta_1}{\partial\rho} - \frac{1}{2}\frac{\partial^2\beta_3}{\partial\rho^2} \Biggr] \left(\frac{\partial\rho}{\partial r'}\right)^3\text{d}r'.
    \end{align}
\end{subequations}
Here, $\mathcal{P}(\rho)$ is the bulk isotropic contribution to $\bm{\Sigma}$, which is the dynamic pressure, and $\mathcal{K}(r)$ is the non-local terms obtained from evaluating Eq.~\eqref{seq:Fintdef} [see Section~\ref{sec:nonlocal}]. 
We note that $\mathcal{K}$ is zero when evaluated outside of the nucleus but finite when evaluated inside of the nucleus.
In spherical coordinates, the divergence of $\bm{\Sigma}$ is given by:
\begin{equation}
    0 = \frac{\partial \Sigma_{rr}}{\partial r} - \frac{2}{r}\mathcal{G}(\rho),\label{seq:Jstresssymmetry}
\end{equation}
where we have defined the ``excess stress'':
\begin{equation}
    \mathcal{G}(\rho) = \Sigma_{tt} - \Sigma_{rr}.\label{seq:gdef}
\end{equation}
The definite integral of $\mathcal{G}(\rho)$ across the interface is precisely the mechanical surface found by Kirkwood and Buff~\cite{Kirkwood1949}:
\begin{equation}
    \upgamma_{\rm mech} = \int_0^{\infty} \left[\Sigma_{tt} - \Sigma_{rr}\right]\text{d}r.\label{seq:mechstdef}
\end{equation}
For active particles with hard-sphere interactions, this mechanical surface tension was found to be strongly \textit{negative}~\cite{Bialke2015}.

Let us consider the definite integral of Eq.~\eqref{seq:Jstresssymmetry} from $r=0$ to $\infty$:
\begin{equation}
    \mathcal{P}(\rho^{\rm in}) - \mathcal{P}(\rho^{\rm sat}) - \mathcal{K}\bigr|_{r=0}  = \int_0^{\infty}\left[\frac{2}{r}\mathcal{G}(\rho)\right]\text{d}r.\label{seq:almostlaplace}
\end{equation}
Within a region of homogeneous density, $\bm{\Sigma}$ becomes isotropic. 
Therefore $\mathcal{G}$ is only non-zero in the vicinity of the interface.
If the width of the interfacial region is small compared to the radius of the critical nucleus, then we may approximate Eq.~\eqref{seq:almostlaplace} as:
\begin{equation}
    \mathcal{P}(\rho^{\rm in}) - \mathcal{P}(\rho^{\rm sat}) - \mathcal{K}\bigr|_{r=0}  = \frac{2\upgamma_{\rm mech}}{R_C} + \mathcal{O}\left(\frac{1}{R_C^2}\right),\label{seq:laplace}
\end{equation}
which demonstrates that the Laplace pressure difference holds in active systems \textit{only without} non-local stress components and is broken by systems \textit{with} non-local stress components.
A derivation of an active Laplace pressure difference was discussed in Ref.~\cite{Solon2018GeneralizedEnsembles}, however this work did not consider the contribution of non-local stress terms.

Suppose we had taken the \textit{antiderivative} of Eq.~\eqref{seq:Jstresssymmetry} rather than the definite integral.
Doing so results in:
\begin{equation}
    \Sigma_{rr}(r) = -\mathcal{P}(\rho) + \kappa_1(\rho)\left[\frac{\partial}{\partial r} + \frac{2}{r}\right]\frac{\partial \rho}{\partial r} + \bigl(\kappa_2(\rho)+\kappa_3(\rho)\bigr)\left(\frac{\partial \rho}{\partial r}\right)^2 + \kappa_4(\rho)\frac{\partial^2\rho}{\partial r^2} + \mathcal{K} = W(r) + C,\label{seq:indefintegral} 
\end{equation}
where $C$ is a constant of integration and we have defined $W$ as:
\begin{equation}
    W(r) = \int_0^{r}\left[\frac{2\mathcal{G}}{r}\right]\text{d}r.\label{seq:Fdef}
\end{equation}
While we do not know the value of $W(r)$ for any arbitrary $r$, we have already approximated $W(\infty)$ as $2\upgamma_{\rm mech}/R_C$ when $R_C$ is much larger than the interfacial width.
In addition, L'H\^{o}pital's rule can be used to show that the integrand of $W(r)$ is zero at $r=0$, which implies that $W(0) = 0$. 
In order to determine the precise value of $C$, we recall that the density as $r\rightarrow\infty$ is $\rho^{\rm sat}$:
\begin{equation}
    \Sigma_{rr}(\infty) = W(\infty) + C = \frac{2\upgamma_{\rm mech}}{R_C} + C = -\mathcal{P}(\rho^{\rm sat})
\end{equation}
Therefore
\begin{equation}
    C = -\mathcal{P}(\rho^{\rm sat}) - \frac{2\upgamma_{\rm mech}}{R_C} 
\end{equation}
We are then left with the equation:
\begin{equation}
       \kappa_1(\rho)\left[\frac{\partial}{\partial r} + \frac{2}{r}\right]\frac{\partial \rho}{\partial r} + \bigl(\kappa_2(\rho) + \kappa_3(\rho)\bigr)\left(\frac{\partial \rho}{\partial r}\right)^2 + \kappa_4(\rho)\frac{\partial^2\rho}{\partial r^2} + \mathcal{K} = W(r)  -\frac{2\upgamma_{\rm mech}}{R_C} + \left(\mathcal{P}(\rho) - \mathcal{P}(\rho^{\rm sat})\right).\label{seq:indefintegralbc}
\end{equation}
We now introduce the generalized Maxwell construction pseudovariable~\cite{Aifantis1983TheRule,Solon2018GeneralizedMatter,Omar2023b} $\mathcal{E}$ defined by the differential equation:
\begin{equation}
    \bigl(\kappa_1(\rho) + \kappa_4(\rho)\bigr)\frac{\partial^2 \mathcal{E}}{\partial \rho^2} = \left(2\left(\kappa_2(\rho) + \kappa_3(\rho)\right) - \frac{\partial }{\partial \rho}\left(\kappa_1(\rho) + \kappa_4(\rho)\right)\right) \frac{\partial \mathcal{E}}{\partial \rho}  \label{seq:epsilondef}
\end{equation}
For passive systems, $\{\kappa_i(\rho)\}$ are dictated by the Korteweg expansion~\cite{Korteweg1904}, resulting in $\mathcal{E}\sim 1/\rho$. 
This pseudovariable was found to eliminate the need to evaluate interfacial terms when solving for the macroscopic binodals, thus providing coexistence criteria in terms of bulk equations of state.
In our case multiplying by the derivative of the pseudovariable with respect to $r$ and integrating from $r=0$ to $r=\infty$ will eliminate some, but not all, of the interfacial terms, resulting in:
\begin{align}
    \int_{0}^{\infty} \left[ \frac{2\kappa_1(\rho)}{r}\frac{\partial \mathcal{E}}{\partial \rho}\left(\frac{\partial \rho}{\partial r}\right)^2 + \mathcal{K}\frac{\partial\mathcal{E}}{\partial \rho}\frac{\partial \rho}{\partial r} \right] \text{d}r =&  \int_0^{\infty}\frac{\partial\mathcal{E}}{\partial r}W(r)\text{d}r  - \frac{2\upgamma_{\rm mech}}{R_C}\left(\mathcal{E}^{\rm sat} - \mathcal{E}^{\rm in}\right)  \nonumber \\ &  + \int_{\mathcal{E}^{\rm in}}^{\mathcal{E}^{\rm sat}} \left[\mathcal{P}(\rho) - \mathcal{P}(\rho^{\rm sat})\right]\text{d}\mathcal{E}.\label{seq:almostcoexist}
\end{align}
We now integrate the first integral on the right hand side of Eq.~\eqref{seq:almostcoexist} by parts to find:
\begin{equation}
    \int_0^{\infty}\frac{\partial\mathcal{E}}{\partial r}W(r)\text{d}r = \mathcal{E}^{\rm sat}\frac{2\upgamma_{\rm mech}}{R_C} - \int_0^{\infty}\frac{2}{r}\mathcal{E}\mathcal{G}\text{d}r.\label{seq:coexistintbyparts}
\end{equation}
We also integrate the second integral on the left hand side of Eq.~\eqref{seq:almostcoexist} by parts to find:
\begin{equation}
    \int_0^{\infty}\mathcal{K}\frac{\partial\mathcal{E}}{\partial r}\text{d}r =  - \mathcal{E}^{\rm in}\mathcal{K}\bigr|_{r=0} - \int_0^{\infty}\frac{\partial \mathcal{K}}{\partial r}\mathcal{E}\text{d}r.\label{seq:coexistintbypartsnonlocal}
\end{equation}
Substitution of Eqs.~\eqref{seq:coexistintbyparts} and ~\eqref{seq:coexistintbypartsnonlocal} into Eq.~\eqref{seq:almostcoexist} results in:
\begin{align}
    \int_{0}^{\infty}\left[\frac{2\kappa_1(\rho)}{r}\frac{\partial \mathcal{E}}{\partial \rho}\left(\frac{\partial \rho}{\partial r}\right)^2 - \frac{\partial\mathcal{K}}{\partial r}\mathcal{E}\right]\text{d}r &= \mathcal{E}^{\rm in}\mathcal{K}\bigr|_{r=0}  -\int_0^{\infty}\frac{2}{r}\mathcal{E}\mathcal{G}\text{d}r \nonumber \\ & + \frac{2\upgamma_{\rm mech}}{R_C}\mathcal{E}^{\rm in} + \int_{\mathcal{E}^{\rm in}}^{\mathcal{E}^{\rm sat}} \left[\mathcal{P}(\rho) - \mathcal{P}(\rho^{\rm sat})\right]\text{d}\mathcal{E}.\label{seq:almostcoexistsub}
\end{align}
Finally, we apply our approximation that $R_C$ is much larger than the interfacial width to find:
\begin{equation}
    \int_{\mathcal{E}^{\rm in}}^{\mathcal{E}^{\rm sat}} \left[\mathcal{P}(\rho) - \mathcal{P}(\rho^{\rm sat})\right]\text{d}\mathcal{E} +\int_0^{\infty}\left(\mathcal{E}^{\rm in}-\mathcal{E}\right)\frac{\partial\mathcal{K}}{\partial r}\text{d}r =\frac{2}{R_C}\int_{0}^{\infty}\left[\kappa_1(\rho)\frac{\partial \mathcal{E}}{\partial \rho}\left(\frac{\partial \rho}{\partial r}\right)^2 + \mathcal{G}\left(\mathcal{E}-\mathcal{E}^{\rm in}\right)\right],\label{seq:coexist2}
\end{equation}
where by making this approximation we have introduced error $\mathcal{O}\left(R_C^{-2}\right)$.
We now define the ``Ostwald'' tension~\cite{Tjhung2018} as:
\begin{equation}
    \upgamma_{\rm ost} = \frac{\rho^{\rm sat}-\rho^{\rm in}}{\rho^{\rm sat}\left(\mathcal{E}^{\rm sat} - \mathcal{E}^{\rm in}\right)}\left[\int_0^{\infty}\left(\kappa_1(\rho)\frac{\partial\mathcal{E}}{\partial \rho}\left(\frac{\partial\rho}{\partial r}\right)^2 + \mathcal{G}\left(\mathcal{E}-\mathcal{E}^{\rm in}\right)\right)\text{d}r\right],\label{seq:osttensiondef}
\end{equation}
where $\upgamma_{\rm ost}$ has the same units of mechanical surface tension. 
We show in Section~\ref{sec:fluctuatinghydroeq} that $\upgamma_{\rm ost}$ becomes identical to $\upgamma_{\rm mech}$ in a passive system where $\mathcal{E}\sim 1/\rho$.
Using the definition of the Ostwald tension, Eq.~\eqref{seq:coexist2} can be rewritten as:
\begin{equation}
    \int_{\mathcal{E}^{\rm in}}^{\mathcal{E}^{\rm sat}} \left[\mathcal{P}(\rho) - \mathcal{P}(\rho^{\rm sat})\right]\text{d}\mathcal{E}+\int_0^{\infty}\left(\mathcal{E}^{\rm in}-\mathcal{E}\right)\frac{\partial\mathcal{K}}{\partial r}\text{d}r = \frac{2\upgamma_{\rm ost}}{R_C}\frac{\rho^{\rm sat}\left(\mathcal{E}^{\rm sat} - \mathcal{E}^{\rm in}\right)}{\rho^{\rm sat} - \rho^{\rm in}} + \mathcal{O}\left(\frac{1}{R_C^2}\right).\label{seq:coexist2ost}
\end{equation}
In addition, we restate the Laplace pressure difference found above:
\begin{equation}
    \mathcal{P}(\rho^{\rm in}) - \mathcal{P}(\rho^{\rm sat}) -\mathcal{K}\bigr|_{r=0}  = \frac{2\upgamma_{\rm mech}}{R_C} + \mathcal{O}\left(\frac{1}{R_C^2}\right).\label{seq:laplacerestate}
\end{equation}
Any flux-free state will satisfy Eqs.~\eqref{seq:laplacerestate} and ~\eqref{seq:coexist2ost} simultaneously.
The solution with $R_C = \infty$ corresponds to macroscopic phase separation and reduces to the phase coexistence criteria found in Ref.~\cite{Omar2023b}.
If a system is prepared at a saturation density $\rho^{\rm sat}$ not equal to the binodal, simultaneous solution of Eqs.~\eqref{seq:laplacerestate} Eq.~\eqref{seq:coexist2} allows for the calculation of the critical nucleus density $\rho^{\rm in}$ and \textit{finite} critical radius $R_C$.

Simultaneous solution of Eqs.~\eqref{seq:laplacerestate} and ~\eqref{seq:coexist2} involves a serious complication.
The right hand sides of both equation can only be evaluated if the density profile of the critical nucleus is already known, which in turn depends on $R_C$ and $\rho^{\rm in}$. 
Therefore a \textit{self-consistent} approach must be employed, which we outline in Section~\ref{sec:coexistnumerics}.
\subsection{\label{sec:abps}Critical Nucleus of Phase-Separating Active Brownian Particles}
We consider a system of $N$ interacting ABPs in which the time variation of the position $\mathbf{r}_i$ and orientation $\mathbf{q}_i$ of the $i$th particle follow overdamped Langevin equations:
\begin{subequations}
  \label{eq:eom}
    \begin{align}
        \Dot{\mathbf{r}}_i &= U_o \mathbf{q}_i + \frac{1}{\zeta}\sum_{j\neq i}^N\mathbf{F}_{ij}\label{eq:rdot}, \\ \Dot{\mathbf{q}}_i &= \mathbf{q}_i\times \bm{\Omega}_i\label{eq:qdot},
    \end{align} 
\end{subequations}
where $\mathbf{F}_{ij}$ is the (pairwise) interparticle force, $U_o$ is the intrinsic active speed, $\zeta$ is the translational drag coefficient, and $\bm{\Omega}_i$ is a stochastic angular velocity with zero mean and variance $\langle \bm{\Omega}_i (t) \bm{\Omega}_j(t')\rangle = 2D_R\delta_{ij}\delta(t-t')\mathbf{I}$ (where $\mathbf{I}$ is the identity tensor). 
Here, $D_R$ is the rotational diffusivity, which may be athermal in origin, and is inversely related to the orientational relaxation time $\tau_R \equiv D_R^{-1}$.
One can then define an intrinsic run length, $\ell_o = U_o\tau_R$, the typical distance an ideal particle travels before reorienting.
We consider a purely repulsive form of $\mathbf{F}_{ij}$, corresponding to a sufficiently stiff interaction potential such that hard-sphere statistics are closely approximated with an effective hard-sphere diameter of $d_{\rm hs}$~\cite{Weeks1971,Omar2021}.

\begin{figure}
	\centering
	\includegraphics[width=0.75\textwidth]{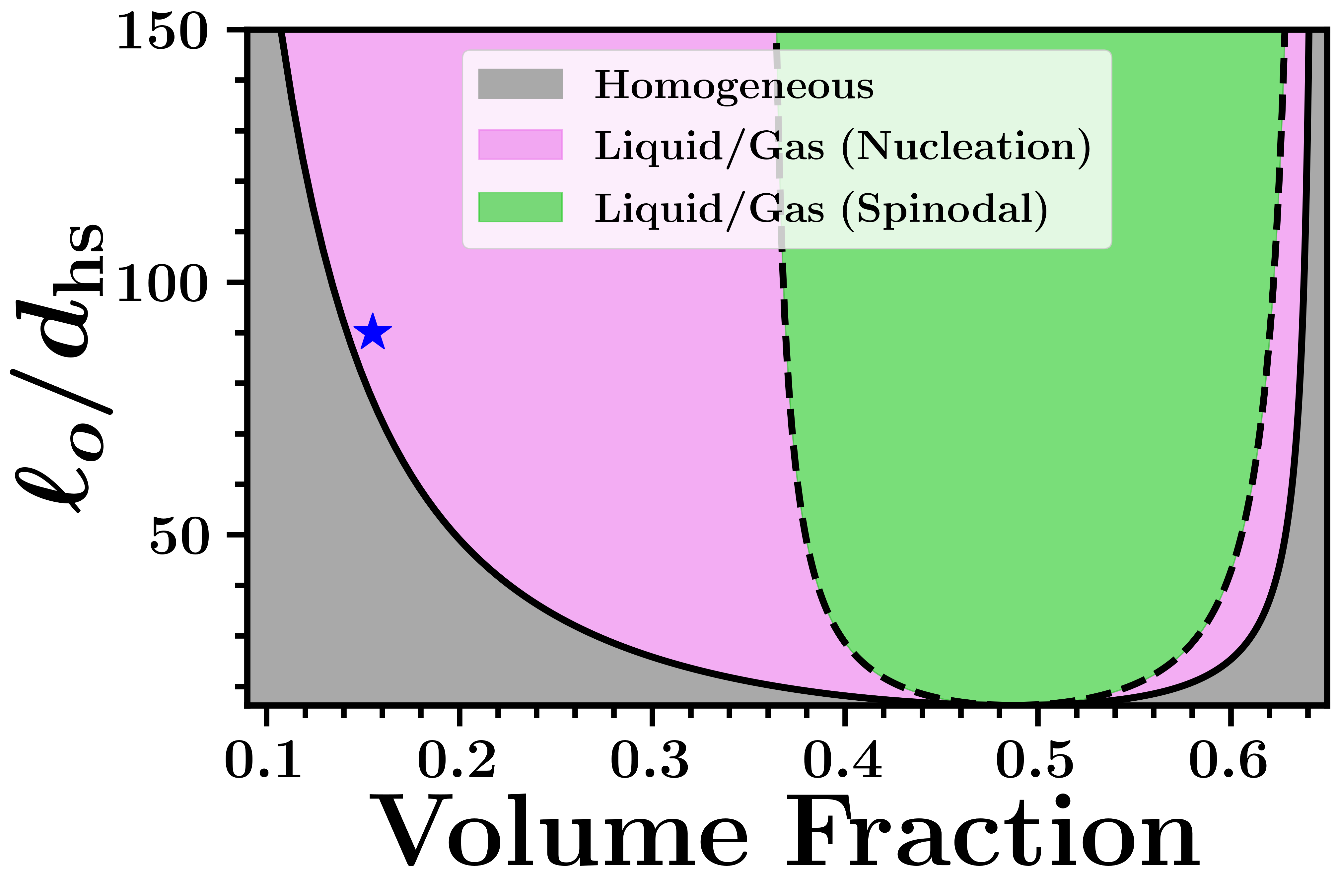}
	\caption{\protect\small{{Mechanically constructed phase diagram as calculated by Ref.~\cite{Omar2023b}. Dashed line indicates spinodals while solid line indicates binodals. Blue star included as an example point where supersaturation is small.}}}
	\label{fig:abpphasediagram}
\end{figure}

By following a similar procedure to that discussed in Ref.~\cite{Omar2023b}, one can find an expression for the flux $\mathbf{J}$ of ABPs, which is dependent on the polar order $\mathbf{m}\equiv \langle \sum_{i}\mathbf{q}_i\delta(\mathbf{r}-\mathbf{r}') \rangle$.
The flux of the polar order is then found to be dependent on the nematic order $\mathbf{Q} \equiv \langle \sum_i \mathbf{q}_i\mathbf{q}_i\delta(\mathbf{r}-\mathbf{r}_i) \rangle$. 
Thus, each orientational moment of the density field has dynamics dependent on the field one moment higher. 
Thus, finding a form for the ABP flux dependent solely on the density requires one to truncate this hierarchy of equations with some closure relation.
By setting $\mathbf{B}\equiv \langle \sum_i \mathbf{q}_i\mathbf{q}_i\mathbf{q}_i\delta(\mathbf{r}-\mathbf{r}_i) $ isotropic and demanding that $\mathbf{J}\cdot\mathbf{e}_r = \bm{0}$, we are able to solve for the form of $\mathcal{P}(\rho)$ and $\{\kappa_i(\rho)\}$ for ABPs:
\begin{subequations}
    \label{seq:abpconstitutive}
    \begin{align}
         \mathcal{P}(\rho) &= p_C(\rho) + \frac{\rho\zeta U_o\ell_o \overline{U}(\rho)}{d(d-1)}\label{seq:dynpress}, \\ \kappa_1(\rho) &= \frac{3\ell_o^2}{2d(d-1)(d+2)}\left(\overline{U}(\rho)\right)^2\frac{\partial p_C}{\partial \rho}\label{seq:adef},\\ \kappa_2(\rho)&= 0,   \\ \kappa_3(\rho) &= \frac{3\ell_o^2}{2d(d-1)(d+2)}\overline{U}(\rho)\frac{\partial}{\partial \rho}\left[\overline{U}(\rho)\frac{\partial p_C}{\partial \rho}\right]\label{seq:bdef}, \\ \kappa_4(\rho) &= 0, \\ \mathcal{K} &= 0.
    \end{align}
\end{subequations}
Here $p_C(\rho)$ is the \textit{conservative} pressure arising from particle interactions, $\overline{U}(\rho)$ is the dimensionless effective active speed, and $d(>1)$ is the system dimensionality.
In Eqs.~\eqref{seq:abpconstitutive} we have ignored the gradient contributions of the interaction potential to $\bm{\Sigma}$, which is justified when $\ell_o \gg 1$~\cite{Solon2018GeneralizedEnsembles,Omar2020,Evans2023}.
Given our expressions for $\{\kappa_i(\rho)\}$, we can solve for the generalized Maxwell construction pseudovariable as $\mathcal{E}\sim p_C(\rho)$.
The closures used to derive these forms of $\mathcal{P}(\rho)$ and $\{\kappa_i(\rho)\}$ imply that finite polar in the radial direction and nematic ordering in the tangential direction exists at interfaces in states free of radial flux.
We note that the theory derived in Section~\ref{sec:coexistcriteria} is valid for small supersaturations, i.e. the overall density of the system is close to the binodals. 
For reference, we include the phase diagram of ABPs as solved for by Ref.~\cite{Omar2023b} in Fig.~\ref{fig:abpphasediagram} and include a blue star in the figure representing a point near the binodal where our theory may hold.

\subsection{\label{sec:eos}Equations of State}
In order to numerically solve for the quantities derived in this supplemental material, we use the equations of state reported Ref.~\cite{Omar2023b} for athermal active hard spheres.
The two contributions to the bulk dynamic pressure were found to be well described by:
\begin{subequations}
    \begin{align}
        \frac{p_{\rm act}}{\zeta U_0 / ( d_{\rm{hs}}^2)} &= \phi \left(\frac{\ell_0}{d_{\rm{hs}}}\right) \frac{1}{\pi}\overline{U}\nonumber \\ &= \phi \left(\frac{\ell_0}{\mathcal{D}}\right)\frac{1}{\pi} \Biggl[1+\biggl(1 - \exp\left[-2^{7/6} \left(\frac{\ell_0}{d_{\rm{hs}}}\right)\right] \biggr)\frac{\phi}{1 - \phi / \phi_{\rm max}} \Biggr]^{-1} \  \\
        \frac{p_{\rm C}}{\zeta U_0 / ( d_{\rm{hs}}^2)} &=  6\times 2^{-7/6} \frac{1}{\pi}\frac{\phi^2}{\sqrt{1-\phi/\phi_{\rm max}}} \ , 
\end{align}
\end{subequations}
where  $p_{\rm act}$ is the active pressure, $\phi = \rho\pi d_{\rm hs}^3/6$ is the volume fraction, and $\phi_{\rm max} = 0.645$ is the maximum volume fraction of disordered hard spheres.
The active pressure above is consistent with a dimensionless active speed given by:
\begin{align}
    \overline{U} = \Biggl[1+\biggl(1 - \exp\left[-2^{7/6} \left(\frac{\ell_0}{d_{\rm hs}}\right)\right] \biggr)\frac{\phi}{1 - \phi / \phi_{\rm max}} \Biggr]^{-1}.
\end{align}
Although these equations were empirically fit to simulations conducted in the homogeneous region of the phase diagram, we assume that they remain valid in the phase separated region. 

\subsection{\label{sec:coexistnumerics}Numerical Scheme for Density Profile of Active Brownian Particles}
Before outlining the self-consistent scheme to solve Eqs.~\eqref{seq:laplacerestate} and ~\eqref{seq:coexist2} simultaneously, we describe the numerical construction of the density profile of the critical nucleus.
Solving for this density profile is a key step of the self-consistent scheme we will later outline.

While constructing a profile, we will assume that $R$, $\rho^{\rm in}$, and $\upgamma_{\rm mech}$ are already known.
Going from Eq.~\eqref{seq:indefintegralbc} to Eq.~\eqref{seq:almostcoexist} involved integrating from $r=0$ to $r=\infty$.
Suppose we had instead integrated to some arbitrary radius between $0$ and $\infty$. 
Doing so would have resulted in:
\begin{align}
    \frac{1}{2}\frac{\partial\mathcal{E}}{\partial \rho}\kappa_1(\rho)\left(\frac{\partial \rho}{\partial r}\right)^2\biggl|^{r}_0 = &\int_{\mathcal{E}^{\rm in}}^{\mathcal{E}(r)} \left[\mathcal{P}(\rho) - \mathcal{P}(\rho^{\rm in})\right]\text{d}\mathcal{E}\nonumber \\ &- \frac{2}{R_C}\left[\int_{0}^{r}\left(\kappa_1(\rho)\frac{\partial \mathcal{E}}{\partial \rho}\left(\frac{\partial \rho}{\partial r}\right)^2 + \mathcal{E}\mathcal{G}\right)\text{d}r - \mathcal{E}(r)\int_0^{r}\mathcal{G}\text{d}r\right] .\label{seq:coexist2arbr}
\end{align}
We now convert the integrals to be over $\rho$ instead of $\mathcal{E}$ or $r$:
\begin{align}
    \frac{1}{2}\frac{\partial\mathcal{E}}{\partial \rho}\kappa_1(\rho)\left(\frac{\partial \rho}{\partial r}\right)^2 = &\int_{\rho^{\rm in}}^{\rho(r)} \left[\mathcal{P}(\rho) - \mathcal{P}(\rho^{\rm in})\right]\frac{\partial\mathcal{E}}{\partial \rho}\text{d}\rho\nonumber \\ &- \frac{2}{R_C}\left[\int_{\rho^{\rm in}}^{\rho(r)}\left(\kappa_1(\rho)\frac{\partial \mathcal{E}}{\partial \rho}\left(\frac{\partial \rho}{\partial r}\right) + \mathcal{E}\mathcal{G}\left(\frac{\partial\rho}{\partial r}\right)^{-1}\right)\text{d}\rho - \mathcal{E}(\rho)\int_{\rho^{\rm in}}^{\rho(r)}\mathcal{G}\left(\frac{\partial\rho}{\partial r}\right)^{-1}\text{d}\rho\right],\label{seq:coexist2arbrho}
\end{align}
where we have used the radial symmetry condition $\partial\rho/\partial r = 0$ at $r=0$. 
We can then isolate the density gradient by rearranging and taking the square root of both sides:
\begin{align}
    \frac{\partial \rho}{\partial r} = \pm f(\rho) = &\pm\Biggl(\frac{2}{\kappa_1(\rho)}\left(\frac{\partial \mathcal{E}}{\partial \rho}\right)^{-1}\int_{\rho^{\rm in}}^{\rho(r)} \left[\mathcal{P}(\rho) - \mathcal{P}(\rho^{\rm in})\right]\frac{\partial\mathcal{E}}{\partial \rho}\text{d}\rho\nonumber \\ &- \frac{2}{R_C}\left[\int_{\rho^{\rm in}}^{\rho(r)}\left(\kappa_1(\rho)\frac{\partial \mathcal{E}}{\partial \rho}\left(\frac{\partial \rho}{\partial r}\right) + \mathcal{E}\mathcal{G}\left(\frac{\partial\rho}{\partial r}\right)^{-1}\right)\text{d}\rho - \mathcal{E}(\rho)\int_{\rho^{\rm in}}^{\rho(r)}\mathcal{G}\left(\frac{\partial\rho}{\partial r}\right)^{-1}\text{d}\rho\right]\Biggr)^{1/2}.\label{seq:drhodr}
\end{align}
Equation~\eqref{seq:drhodr} maps any value of density between $\rho^{\rm in}$ and $\rho^{\rm sat}$ to its derivative in the $r$ direction.
Using $\rho^{\rm in}$ as an initial value, this mapping between density values and $r$ derivatives allows for the construction of another mapping between density values and $r$ coordinate in the density profile:
\begin{equation}
    r(\rho) = \int_{\rho^{\rm in}}^{\rho}\frac{1}{f(\rho')}\text{d}\rho'.\label{seq:rmapping}
\end{equation}
We identify this mapping as $\upvarphi_R(r)$, the average density profile of the critical nucleus.

The above procedure which requires $R_C$, $\upgamma_{\rm mech}$, and $\rho^{\rm in}$ can be used with the following self-consistent scheme for simultaneous solution of Eqs.~\eqref{seq:laplacerestate} and ~\eqref{seq:coexist2}:
\begin{enumerate}
    \item Begin with an initial guess for the density profile, $\rho(r)$.
    \item Using $\rho(r)$, evaluate $\upgamma_{\rm mech}$ from Eq.~\eqref{seq:mechstdef}.
    \item Using a given value of $\rho^{\rm sat}$, $\rho(r)$, and $\upgamma_{\rm mech}$, find the optimal value of $R_C$ and $\rho^{\rm in}$ that solve  Eqs.~\eqref{seq:laplacerestate} and ~\eqref{seq:coexist2}.
    \item If on the first iteration of the scheme, record the Laplace pressure difference, $2\upgamma_{\rm mech}/R_C$.
    \item Solve for a new density profile, $\rho^*(r)$ using the $\upgamma_{\rm mech}$ and the newly calculated $R_C$ and $\rho^{\rm in}$.
    \item Using $\rho^*(r)$, update $\upgamma_{\rm mech}$ from Eq.~\eqref{seq:mechstdef}.
    \item Record the updated value of the Laplace pressure difference, $2\upgamma_{\rm mech}/R_C$.
    \item If the new value of the Laplace pressure is within tolerance of the old one, the procedure has converged and set $\upvarphi_R(r) = \rho^*(r)$. Return the current values of $R_c$, $\rho^{\rm in}$, and the newest density profile. Otherwise, repeat the process starting from Step 1 using $\rho^*(r)$ as the updated $\rho(r)$.
\end{enumerate}

The above numerical scheme was used to calculate the data plotted in Figs. 1 and 2 of the main text.  
The density profile corresponding to macroscopic phase separation was used as the initial guess of the above numerical scheme for the closest value of $\rho^{\rm sat}$ to the binodal, and the density profile obtained there was used as the initial guess for the next closest value of $\rho^{\rm sat}$ to the binodal, etc.

\subsection{\label{sec:noneqchempot}Connection to Nonequilibrium Chemical Potential}
As recently proposed in Ref.~\cite{Evans2023}, we define the nonequilibrium chemical potential $g$ (equivalent to that defined by Solon~\textit{et al.}~\cite{Solon2018GeneralizedEnsembles}) through a generalized Gibbs-Duhem relation:
\begin{equation}
    \text{d}g = \mathcal{E}\text{d}\mathcal{P}.
\end{equation}
This definition ensures that $g$ will be equal between two macroscopic coexisting phases, a coexistence criteria that is identical to the a generalized equal-area Maxwell construction on $\mathcal{P}$ with respect to $\mathcal{E}$.
The product rule results in:
\begin{equation}
    \text{d}g = \text{d}\left(\mathcal{P}\mathcal{E}\right)-\mathcal{P} \text{d}\mathcal{E}. \label{seq:gdiffrearrange}
\end{equation}
We may then integrate Eq.~\eqref{seq:gdiffrearrange} from the inside of the nucleus to the outside to find:
\begin{equation}
    g^{\rm sat} - g^{\rm in} = \mathcal{P}^{\rm sat}\mathcal{E}^{\rm sat} - \mathcal{P}^{\rm in}\mathcal{E}^{\rm in} - \int_{\mathcal{E}^{\rm in}}^{\mathcal{E}^{\rm sat}}\mathcal{P} \text{d}\mathcal{E}.\label{seq:gdiffrearrangeint}
\end{equation}
Substitution of Eq.~\eqref{seq:laplace} into Eq.~\eqref{seq:gdiffrearrangeint} results in:
\begin{equation}
    g^{\rm sat} - g^{\rm in} = -\frac{\mathcal{E}^{\rm in}2\upgamma_{\rm mech}}{R_C} -\mathcal{E}^{\rm in}\mathcal{K}\Bigr|_{r=0} - \int_{\mathcal{E}^{\rm in}}^{\mathcal{E}^{\rm sat}}\left[\mathcal{P}  - \mathcal{P}^{\rm sat}\right]\text{d}\mathcal{E}.\label{seq:gdiffrearrangelaplace}
\end{equation}
Substitution of Eq.~\eqref{seq:coexist2ost} into Eq.~\eqref{seq:gdiffrearrangelaplace} results in:
\begin{equation}
    g^{\rm sat} - g^{\rm in} = -\Delta g = -\frac{2}{R_C}\left[\mathcal{E}^{\rm in}\upgamma_{\rm mech} + \frac{\upgamma_{\rm ost}\rho^{\rm sat}\Delta\mathcal{E}}{\Delta\rho}\right] +\int_0^{\infty}\left(2\mathcal{E}^{\rm in}-\mathcal{E}\right)\frac{\partial\mathcal{K}}{\partial r}\text{d}r .\label{seq:gdifffinal}
\end{equation}
In general, the chemical potential within and outside the droplet will be different. 
However, as we demonstrate in Section~\ref{sec:eq}, this difference vanishes for a passive system.
We may also define a pseudopotential as $\Phi \equiv \int \mathcal{P}\text{d}\mathcal{E} $ as the antiderivative of the pressure with respect to $\mathcal{E}$, allowing indefinite integration of Eq.~\eqref{seq:gdiffrearrange} to result in:
\begin{equation}
    g = \mathcal{P}\mathcal{E} - \Phi.\label{seq:chempotform}
\end{equation}
Defining $\Phi$ allows us to express the critical radius from Eq.~\eqref{seq:coexist2ost} as:
\begin{equation}
    R_C = \frac{2\upgamma_{\rm ost}\rho^{\rm sat}\Delta\mathcal{E}}{\left(-\Delta\Phi + \Delta\mathcal{E}\mathcal{P}^{\rm sat}\right)\Delta\rho},\label{seq:critradiuspseudopotential}
\end{equation}
so long as $\bm{\Sigma}$ can be expressed fully locally. 
This form for $R_C$ will become convenient in Section~\ref{sec:potential}.
\subsection{\label{sec:nonlocal}Local and Non-local Stress Tensors}
In Section~\ref{sec:coexistcriteria} we introduced the dynamic stress tensor $\bm{\Sigma} = \bm{\sigma} + \mathbf{F}$, where $\bm{\sigma}$ is the conservative stress tensor due to interaction forces and ideal gas pressure while $\mathbf{F}$ is defined as the Coulomb integral of the body force:
\begin{equation}
    \mathbf{F} = \frac{1}{4\pi}\int \text{d}\mathbf{r}' \mathbf{b}(\mathbf{r}')\frac{\mathbf{r}-\mathbf{r}'}{|\mathbf{r}-\mathbf{r}'|^3}.\label{seq:Fdefrestate}
\end{equation}
It is straightforward to show that:
\begin{equation}
    \frac{\partial}{\partial \mathbf{r}'}\frac{1}{|\mathbf{r}-\mathbf{r}'|} = -\frac{\mathbf{r}-\mathbf{r}
    }{|\mathbf{r}-\mathbf{r}'|^3}.
\end{equation}
Then, following integration by parts, Eq.~\eqref{seq:Fdefrestate} becomes:
\begin{equation}
    \mathbf{F} = \frac{1}{4\pi}\int \text{d}\mathbf{r}'\frac{\partial}{\partial \mathbf{r}'}\mathbf{b}(\mathbf{r}') \frac{1}{|\mathbf{r}-\mathbf{r}'|}.
\end{equation}
Note that $|\mathbf{r}-\mathbf{r}'|^{-1}$ is the Green's function of the Laplace operator in three dimensions. 
Therefore solving for $\mathbf{F}$ is equivalent to finding the inverse Laplacian of $\boldsymbol{\nabla}\mathbf{b}$. 
Under conditions of spherical symmetry this can be solved for as:
\begin{align}
    \mathbf{F} = &-\int_r^{\infty} \text{d}r_2 \frac{1}{r_2^2}\int_0^{r_2}\text{d}r_1 r_1^2\frac{\partial}{\partial r_1}\left[\text{b}_r\right]\mathbf{e}_{r}\mathbf{e}_r + \int_r^{\infty}\text{d}r_2\frac{1}{r_2^2}\int_0^{r_2}\text{d}r_1 r_1 \left(\text{b}_r\right)\mathbf{e}_{\theta}\mathbf{e}_{\theta} \nonumber \\ & + \int_r^{\infty}\text{d}r_2\frac{1}{r_2^2}\int_0^{r_2}\text{d}r_1 r_1 \left(\text{b}_r\right)\mathbf{e}_{\phi}\mathbf{e}_{\phi}\label{seq:Fsolvesphere}
\end{align}
While the divergence of $\mathbf{F}$ is equal to $\mathbf{b}$, we are not guaranteed that all components of $\mathbf{F}$ are \textit{local}. 
Depending on the form of $\mathbf{b}$ one may find components of $\mathbf{F}$ that depend on integrals over the entire spatial profile, conceptually similar to the non-local chemical potential found in Ref.~\cite{Tjhung2018}. 
In order to obey the required symmetries of the flux, an expansion of $\mathbf{b}$ in gradients of the density may only contain odd gradients in the density. 
Any first order term will have the form $b_o(\rho)\boldsymbol{\nabla}\rho$, which is trivially the divergence of $\left[\int b_o(\rho)\text{d}\rho\right]\mathbf{I}$. 
Thus first order terms in $\mathbf{b}$ will always result in bulk isotropic terms of $\mathbf{F}$, which can be expressed locally. 
However, the third order terms in $\mathbf{b}$ will result in second order terms of $\mathbf{F}$ that may or may not have non-local contributions.
The exact form of these stress components in spherical coordinates may be computed by substituting the third order terms of an expansion of $\mathbf{b}$ into Eq.~\eqref{seq:Fsolvesphere}.
For a spatially isotropic system, the third order gradient expansion of $\mathbf{b}$ is:
\begin{equation}
    \mathbf{b} = \beta_0(\rho)\boldsymbol{\nabla}\rho + \beta_1(\rho)\boldsymbol{\nabla}\rho\boldsymbol{\nabla}^2\rho + \beta_2(\rho)\boldsymbol{\nabla}|\boldsymbol{\nabla}\rho|^2 + \beta_3(\rho)\boldsymbol{\nabla}\boldsymbol{\nabla}^2\rho + \beta_4(\rho)|\boldsymbol{\nabla}\rho|^2\boldsymbol{\nabla}\rho\label{seq:bodyforcethirdorder}
\end{equation}
When the density field and the forces are spherically symmetric, i.e. $\boldsymbol{\nabla}\rho$ is parallel to $\mathbf{e}_r$, $\mathbf{F}$ can be expressed as:
\begin{equation}
    \mathbf{F} = \left[\int\beta_0(\rho)\text{d}\rho + \beta_3(\rho)\boldsymbol{\nabla}^2\rho + \left(\frac{1}{2}\frac{\partial\beta_3}{\partial\rho} + \beta_2(\rho) - \frac{1}{2}\beta_1(\rho)\right)|\boldsymbol{\nabla}\rho|^2 + \mathcal{K}(r)\right]\mathbf{I} +\left(\beta_1(\rho) - \frac{\partial\beta_3}{\partial\rho}\right)\boldsymbol{\nabla}\rho\boldsymbol{\nabla}\rho,\label{seq:Ffrombodyforce}
\end{equation}
where $\mathcal{K}$ is a non-local term defined as:
\begin{equation}
    \mathcal{K} = \int_r^{\infty}\left[-\beta_4(\rho) + \frac{\partial\beta_2}{\partial\rho} + \frac{1}{2}\frac{\partial\beta_1}{\partial\rho} - \frac{1}{2}\frac{\partial^2\beta_3}{\partial\rho^2}\right]\left(\frac{\partial\rho}{\partial r'}\right)^3\text{d}r'.
\end{equation}
By taking the divergence of $\mathbf{F}$ one can verify that $\mathbf{b}$ is recovered.
We note that $\mathbf{F}$ is not unique; it may be altered by divergence-free terms.

In certain cases, such as ABPs, the body force can naturally be found as the divergence of a stress tensor, and in these cases no non-local terms appear in $\mathbf{F}$~\cite{Omar2020, Omar2023b}.
The condition for $\mathbf{F}$ to be fully local is:
\begin{equation}
    \beta_4(\rho) = \frac{\partial\beta_2}{\partial\rho} + \frac{1}{2}\frac{\partial\beta_1}{\partial\rho} - \frac{1}{2}\frac{\partial^2\beta_3}{\partial\rho^2}.
\end{equation}

\subsection{\label{sec:ambconnect}Connection to Active Model B+}
With the development of our mechanical perspective, we can now revisit the work of Tjhung \textit{et al.}~\cite{Tjhung2018} which derive the criteria for the critical nucleus for model AMB+ (see Eqs.~16-22 in Ref.~\cite{Tjhung2018}) and determine if there is a mechanical interpretation for their criteria. 
As frequently noted by the authors of Ref.~\cite{Tjhung2018}, while AMB+ is a general phenomenological description of active systems, it remains unclear if the quantities that emerge from this perspective, such as the pseudotension and pseudopressure, are connected to the actual mechanical tension and pressure, respectively. 
Here, we will formally map our criteria to that of Ref.~\cite{Tjhung2018}.
In doing so, we demonstrate that the mapping is not unique, and therefore the interpretation of pseudotension and pseudopressure are indeed arbitrary.

We make a few modifications to the version of model AMB+ presented in Ref.~\cite{Tjhung2018} to facilitate comparison to our work. 
These include the consideration of a nonconstant mobility (the mobility is often presumed constant to reduce the stochastic contribution of the flux from multiplicative to additive) and we will not specify a particular form (e.g., the chemical potential stemming from the canonical quartic free energy)  of the ``bulk'' AMB+ chemical potential, $\mu_o$ below.  
We also replace the linearly scaled (by the critical density) density ($\phi$ in Ref.~\cite{Tjhung2018}) with the physical number density, $\rho$.
Finally, as we consider nucleation far from the critical point, we did not limit our perspective to small density departures from the critical density. 
As a result, our flux expression is $\mathcal{O}(\boldsymbol{\nabla}^3)$ while that of AMB+ (as presented in Ref.~\cite{Tjhung2018}) is $\mathcal{O}(\boldsymbol{\nabla}^3\rho^2)$.
The restriction to quadratic density in this form of AMB+ leads to the absence of some terms allowable by symmetry but higher order in density, such as those considered in Ref.~\cite{Nardini2017EntropyMatter}.
We nevertheless consider the version of AMB+ without the additional terms in order to recover the criteria obtained in Ref.~\cite{Tjhung2018}.

We begin with a summary of the derivation of finite-size coexistence criteria presented in Ref.~\cite{Tjhung2018}.
We adopt all notation used in Ref.~\cite{Tjhung2018}, except we replace the AMB+ parameter $\zeta$ with $\mathcal{Z}$ in order to avoid confusion with the microscopic drag and replace the density derivative of the AMB+ effective free energy $f'(\rho)$ with $\mu_o$ for additional clarity.

The deterministic component of the flux in AMB+ is given by:
\begin{multline}
    \mathbf{J}/M[\rho] = -\frac{\partial \mu_o}{\partial \rho}\boldsymbol{\nabla}\rho + K(\rho)\boldsymbol{\nabla}\boldsymbol{\nabla}^2\rho + \frac{\partial K}{\partial\rho}\boldsymbol{\nabla}\rho\boldsymbol{\nabla}^2\rho + \frac{1}{2}\frac{\partial K}{\partial \rho}\boldsymbol{\nabla}|\boldsymbol{\nabla}\rho|^2\\ + \frac{1}{2}\frac{\partial^2 K(\rho)}{\partial\rho^2}\boldsymbol{\nabla}\rho|\boldsymbol{\nabla}\rho|^2 - \lambda(\rho)\boldsymbol{\nabla}|\boldsymbol{\nabla}\rho|^2 - \frac{\partial\lambda}{\partial\rho}\boldsymbol{\nabla}\rho|\boldsymbol{\nabla}\rho|^2 + \mathcal{Z}(\rho)\boldsymbol{\nabla}\rho\boldsymbol{\nabla}^2\rho.
\end{multline}
For a spherically symmetric system, the radial flux can be rewritten as:
\begin{equation}
    \frac{\mathbf{J}\cdot\mathbf{e}_r}{M[\rho]} = -\frac{\partial \mu}{\partial r},
\end{equation}
where the effective chemical potential is given by:
\begin{multline}
    \mu = \mu_o - K(\rho)\frac{\partial^2\rho}{\partial r^2} - \frac{2}{r}K(\rho)\frac{\partial\rho}{\partial r} + \left(\lambda - \frac{1}{2}\frac{\partial K}{\partial \rho} - \frac{1}{2}\mathcal{Z}(\rho)\right)\left(\frac{\partial\rho}{\partial r}\right)^2 \\- \frac{1}{2}\int_r^{\infty}\frac{\partial\mathcal{Z}}{\partial\rho}\left(\frac{\partial\rho}{\partial r'}\right)^3\text{d}r' + \int_r^{\infty}\frac{2\mathcal{Z}}{r'}\left(\frac{\partial \rho}{\partial r'}\right)^2\text{d}r',\label{seq:ambchempot}
\end{multline}
Again, a key difference between this expression and that found in Ref.~\cite{Tjhung2018} is that we are allowing for density dependence on all coefficients and retaining terms in the flux at the order of $\boldsymbol{\nabla}^3$.
The finite size coexistence criteria can be found by setting the radial flux to zero. 
This implies that $\mu$ is a constant, which means one coexistence criteria can be immediately read as:
\begin{equation}
    \mu_0(\rho^{\rm sat}) = \mu_0(\rho^{\rm in}) - \frac{1}{2}\int_r^{\infty}\frac{\partial\mathcal{Z}}{\partial\rho}\left(\frac{\partial\rho}{\partial r'}\right)^3\text{d}r' + \frac{2}{R_C}\int_r^{\infty}\mathcal{Z}(\rho)\left(\frac{\partial\rho}{\partial r'}\right)^2\text{d}r'\label{seq:ambcoexist1}.
\end{equation}
Therefore the chemical potential is constant everywhere, with a jump in the bulk component of the chemical potential occurring at the interface.
The second coexistence criteria can be found by introducing the pseudovariable which satisfies the following differential equation:
\begin{equation}
    \frac{\partial^2\psi}{\partial \rho^2} = \frac{\mathcal{Z} - 2\lambda }{K}\frac{\partial\psi}{\partial \rho}.\label{seq:ambpseudodef}
\end{equation}
We now multiply Eq.~\eqref{seq:ambchempot} by $\partial_r\psi$ and integrate across all space to find:
\begin{multline}
    \mu_o(\rho^{\rm sat})\left(\psi^{\rm sat} - \psi^{\rm in}\right) - \int_0^{\infty} \mu_o(\rho)\frac{\partial \psi}{\partial r}\text{d}r = -\int_0^{\infty} \frac{2}{r}K(\rho)\frac{\partial\psi}{\partial \rho}\left(\frac{\partial \rho}{\partial r}\right)^2 \\ - \frac{1}{2}\int_0^{\infty}\frac{\partial\psi}{\partial r}\int_r^{\infty}\frac{\partial\mathcal{Z}}{\partial\rho}\left(\frac{\partial\rho}{\partial r'}\right)^3\text{d}r\text{d}r'+ 2\int_0^{\infty}\frac{\partial \psi}{\partial r}\int_r^{\infty}\frac{\mathcal{Z}(\rho)}{r'}\left(\frac{\partial\rho}{\partial r'}\right)^2 \text{d}r'\text{d}r.\label{seq:coexistambintermediate}
\end{multline}
Integration by parts and algebraic manipulation of Eq.~\eqref{seq:coexistambintermediate} results in:
\begin{equation}
    \int_{\psi^{\rm in}}^{\psi^{\rm sat}}\left[\mu_0 - \mu_0^{\rm sat}\right]\text{d}\psi +  \frac{1}{2}\int_0^{\infty}(\psi^{\rm in}-\psi)\frac{\partial\mathcal{Z}}{\partial\rho}\left(\frac{\partial\rho}{\partial r'}\right)^3\text{d}r = \frac{2\sigma}{R_C} \label{seq:ambcoexist2},
\end{equation}
where the pseudotension $\sigma$ is defined as:
\begin{equation}
    \sigma = \int_0^{\infty} K(\rho)\frac{\partial \psi}{\partial \rho}\left( \frac{\partial \rho}{\partial r}\right)^2\text{d}r - \int_0^{\infty} \mathcal{Z}(\rho)\psi(\rho)\left(\frac{\partial \rho}{\partial r}\right)^2\text{d}r + \psi^{\rm in}\int_0^{\infty}\mathcal{Z}(\rho)\left(\frac{\partial \rho}{\partial r}\right)^2\text{d}r.\label{seq:pseudotensiondef}
\end{equation}
Together, Eqs.~\eqref{seq:ambcoexist1} and~\eqref{seq:ambcoexist2} comprise the finite-size coexistence criteria for AMB+.
The \textit{mechanical} finite-size coexistence criteria derived in this work were found by setting the radial component of the divergence of the dynamic stress tensor to zero.
No constitutive model for flux was specifically invoked.
However, we can map AMB+ to an equivalent mechanical model in which we assume that the flux is proportional to a body force, i.e. $\mathbf{J} \propto \mathbf{b}$. 
We can expand this body force $\mathbf{b}$ using Eq.~\eqref{seq:bodyforcethirdorder}, which then allows us to map the expansion coefficients $\{\beta_i\}$ to AMB+ coefficients as:
\begin{subequations}
    \label{seq:ambtobodyforce}
    \begin{equation}
        \beta_0(\rho) = -M[\rho]\frac{\partial\mu_o}{\partial\rho},
    \end{equation}
    \begin{equation}
        \beta_1(\rho) = M[\rho]\mathcal{Z}(\rho) + M[\rho]\frac{\partial K}{\partial\rho},
    \end{equation}
    \begin{equation}
        \beta_2(\rho) = -M[\rho]\lambda(\rho) + \frac{1}{2}M[\rho]\frac{\partial K}{\partial\rho},
    \end{equation}
    \begin{equation}
        \beta_3(\rho) = M[\rho]K(\rho),
    \end{equation}
    \begin{equation}
        \beta_4(\rho) = -M[\rho]\frac{\partial\lambda}{\partial\rho} + \frac{1}{2}M[\rho]\frac{\partial^2K}{\partial\rho^2}.
    \end{equation}
\end{subequations}
One can then use Eq.~\eqref{seq:Ffrombodyforce} to identify the stress expansion coefficients $\{\kappa_i\}$:
\begin{subequations}
    \label{seq:ambtostresscoeff}
    \begin{equation}
        \mathcal{P} = \int M[\rho]\frac{\partial\mu_o}{\partial\rho}\text{d}\rho,
    \end{equation}
    \begin{equation}
        \kappa_1(\rho) = M[\rho]K(\rho),
    \end{equation}
    \begin{equation}
        \kappa_2(\rho) = -M[\rho]\lambda(\rho) + \frac{1}{2}M[\rho]\frac{\partial K}{\partial\rho} + \frac{1}{2}\frac{\partial M}{\partial\rho}K(\rho)-\frac{1}{2}M[\rho]\mathcal{Z}(\rho)        ,
    \end{equation}
    \begin{equation}
        \kappa_3(\rho) = M[\rho]\mathcal{Z}(\rho)- K(\rho)\frac{\partial M}{\partial\rho},
    \end{equation}
    \begin{equation}
        \kappa_4(\rho) = 0,
    \end{equation}
    \begin{equation}
        \mathcal{K} = \frac{1}{2}\int_r^{\infty}\left[-\frac{\partial^2 M}{\partial\rho^2}-\frac{2\lambda(\rho) - \mathcal{Z}(\rho) + \frac{\partial K}{\partial\rho}}{K(\rho)}\frac{\partial M}{\partial\rho} +\frac{\frac{\partial\mathcal{Z}}{\partial\rho}}{K(\rho)}M[\rho] \right]K(\rho)\left(\frac{\partial\rho}{\partial r}\right)^3\text{d}r.
    \end{equation}
\end{subequations}
We can now appreciate that $M[\rho]$ is a degree of freedom in this mapping, and depending on the choice of $M[\rho]$ the physical interpretation of quantities such as $\mu_o$, $\psi$, and $\sigma$ may change.
If one insists that $\bm{\Sigma}$ be written fully locally, one can solve for a mobility $M[\rho]$ that satisfies this demand by solving the differential equation:
\begin{equation}
    -\frac{\partial^2 M}{\partial\rho^2}-\frac{2\lambda(\rho) - \mathcal{Z}(\rho) + \frac{\partial K}{\partial\rho}}{K(\rho)}\frac{\partial M}{\partial\rho} +\frac{\frac{\partial\mathcal{Z}}{\partial\rho}}{K(\rho)}M[\rho] = 0. 
\end{equation}
Solving this equation is difficult without further information regarding $K$, $\lambda$, and $\zeta$. 
In the remainder of this Section, we will discuss the physical interpretations of AMB+ quantities (e.g. chemical potential, pseudotension) and coexistence criteria for particular choices of $M[\rho]$.

One choice for $M[\rho]$ is to set as proportional to some constant, i.e. $M[\rho]\sim 1$. 
For this choice, one can use Eqs.~\eqref{seq:ambtostresscoeff} to find:
\begin{subequations}
    \label{seq:ambtostresscoeffmob1}
    \begin{equation}
        \mathcal{P} \sim \mu_0,
    \end{equation}
    \begin{equation}
        \kappa_1(\rho) \sim K(\rho),
    \end{equation}
    \begin{equation}
        \kappa_2(\rho) \sim -\lambda(\rho) + \frac{1}{2}\frac{\partial K}{\partial\rho} -\frac{1}{2}\mathcal{Z}(\rho)        ,
    \end{equation}
    \begin{equation}
        \kappa_3(\rho) \sim \mathcal{Z}(\rho),
    \end{equation}
    \begin{equation}
        \kappa_4(\rho) = 0,
    \end{equation}
    \begin{equation}
        \mathcal{K} \sim \frac{1}{2}\int_r^{\infty}\frac{\partial\mathcal{Z}}{\partial\rho}\left(\frac{\partial\rho}{\partial r}\right)^3\text{d}r.
    \end{equation}
\end{subequations}
We can now see that when $M[\rho]\sim 1$, the dynamic pressure is analogous to the AMB+ chemical potential. 
We now ask how to relate $\mathcal{E}$ with $\psi$. 
For convenience, we restate the definition of $\mathcal{E}$ given by Eq.~\eqref{seq:epsilondef}:
\begin{equation}
    \frac{\partial^2\mathcal{E}}{\partial \rho^2} = \frac{2(\kappa_2(\rho) + \kappa_3(\rho)) - \frac{\partial}{\partial \rho}\left(\kappa_1(\rho) + \kappa_4(\rho)\right)}{\kappa_1(\rho) + \kappa_4(\rho)}\frac{\partial\mathcal{E}}{\partial\rho}.
\end{equation}
Substitution of Eqs.~\eqref{seq:ambtostresscoeffmob1} into Eq.~\eqref{seq:epsilondef} results in:
\begin{equation}
    \frac{\partial^2\mathcal{E}}{\partial\rho^2} = -\frac{2\lambda - \mathcal{Z}}{K}\frac{\partial\mathcal{E}}{\partial \rho}.
\end{equation}
This result, combined with Eq.~\eqref{seq:ambpseudodef}, implies that $\psi\sim \mathcal{E}$ when $M[\rho]\sim 1$. 
We can now compare the mechanically derived coexistence criteria to that of AMB+.
The Laplace pressure difference found in this work is:
\begin{equation}
    \mathcal{P}^{\rm in}-\mathcal{P}^{\rm sat} = \mathcal{K}\Bigr|_{r=0} + \frac{2\upgamma_{\rm mech}}{R_C}.
\end{equation}
The mechanical surface tension under a $M[\rho]\sim 1$ mapping is equal to:
\begin{equation}
    \upgamma_{\rm mech} = \int_0^{\infty}\left[-\kappa_3(\rho)\left(\frac{\partial\rho}{\partial r}\right)^2 - \kappa_4(\rho)\frac{\partial^2\rho}{\partial r^2}\right]\text{d}r = -\int_0^{\infty}\mathcal{Z}(\rho)\left(\frac{\partial\rho}{\partial r}\right)^2\text{d}r.\label{seq:mechtensionambmob1}
\end{equation}
Therefore our mechanically derived Laplace pressure difference is:
\begin{equation}
    \mathcal{P}^{\rm in}-\mathcal{P}^{\rm sat} = \frac{1}{2}\int_0^{r}\frac{\partial\mathcal{Z}}{\partial\rho}\left(\frac{\partial\rho}{\partial r}\right)^3\text{d}r - \frac{2}{R_C}\int_0^{\infty}\mathcal{Z}(\rho)\left(\frac{\partial\rho}{\partial r}\right)^2\text{d}r,
\end{equation}
which is identical to Eq.~\eqref{seq:ambcoexist1} because $\mathcal{P}\sim\mu_0$.
The second coexistence criteria found by this work is given by:
\begin{equation}
    \int_{\mathcal{E}^{\rm in}}^{\mathcal{E}^{\rm sat}}\left[\mathcal{P} - \mathcal{P}^{\rm sat}\right]\text{d}\mathcal{E} + \int_0^{\infty} \left(\mathcal{E}^{\rm in}-\mathcal{E}^{\rm out}\right)\left(\frac{\partial\mathcal{K}}{\partial r}\right)\text{d}r = \frac{2}{R_C}\int_0^{\infty}\left(\kappa_1(\rho)\frac{\partial\mathcal{E}}{\partial\rho}\left(\frac{\partial\rho}{\partial r}\right)^2 + \mathcal{G}\left(\mathcal{E}-\mathcal{E}^{\rm in}\right)\right)\text{d}r.
\end{equation}
Substitution of the relations to AMB+ coefficients in our constant mobility mapping results in:
\begin{multline}
    \int_{\mathcal{E}^{\rm in}}^{\mathcal{E}^{\rm sat}}\left[\mathcal{P} - \mathcal{P}^{\rm sat}\right]\text{d}\mathcal{E}-\mathcal{E}^{\rm in}\frac{1}{2}\int_0^{\infty}\frac{\partial\mathcal{Z}}{\partial\rho}\left(\frac{\partial\rho}{\partial r}\right)^3\text{d}r + \frac{1}{2}\int_0^{\infty}\mathcal{E}\frac{\partial\mathcal{Z}}{\partial\rho}\left(\frac{\partial\rho}{\partial r}\right)^3\text{d}r \\ = \frac{2}{R_C}\int_0^{\infty}\left(K(\rho)\frac{\partial\mathcal{E}}{\partial\rho}\left(\frac{\partial\rho}{\partial r}\right)^2 + \mathcal{E}^{\rm in}\mathcal{Z}(\rho)\left(\frac{\partial\rho}{\partial r}\right)^2 - \mathcal{E}\mathcal{Z}(\rho)\left(\frac{\partial\rho}{\partial r}\right)^2\right)\text{d}r.
\end{multline}
Noting that under this mapping $\mathcal{P}\sim\mu_0$ and $\psi \sim \mathcal{E}$, we can see that the right hand side of the above equation reduces to $2\sigma/R_C$ and the second AMB+ coexistence criteria Eq.~\eqref{seq:ambcoexist2} is recovered.

Another usefeul choice for $M[\rho]$ is to set it to be proportional to the inverse of $\mathcal{E}$, i.e. $M[\rho]\sim 1/\mathcal{E}$. For this choice, one can use Eqs.~\eqref{seq:ambtostresscoeff} to find:
\begin{subequations}
    \label{seq:ambtostresscoeffmobE}
    \begin{equation}
        \mathcal{P} = \int \frac{1}{\mathcal{E}}\frac{\partial\mu_o}{\partial\rho}\text{d}\rho,
    \end{equation}
    \begin{equation}
        \kappa_1(\rho) = \frac{1}{\mathcal{E}}K(\rho),
    \end{equation}
    \begin{equation}
        \kappa_2(\rho) = -\frac{1}{\mathcal{E}}\lambda(\rho) + \frac{1}{2}\frac{1}{\mathcal{E}}\frac{\partial K}{\partial\rho} - \frac{1}{2}\frac{1}{\mathcal{E}^2}\frac{\partial\mathcal{E}}{\partial\rho}K(\rho)-\frac{1}{2}\frac{1}{\mathcal{E}}\mathcal{Z}(\rho)        ,
    \end{equation}
    \begin{equation}
        \kappa_3(\rho) = \frac{1}{\mathcal{E}}\mathcal{Z}(\rho)+\frac{1}{\mathcal{E}^2}\frac{\partial\mathcal{E}}{\partial\rho} K(\rho),
    \end{equation}
    \begin{equation}
        \kappa_4(\rho) = 0,
    \end{equation}
    \begin{equation}
        \mathcal{K} \sim \frac{1}{2}\int_r^{\infty}\frac{1}{\mathcal{E}}\frac{\partial\mathcal{Z}}{\partial\rho}\left(\frac{\partial\rho}{\partial r}\right)^3\text{d}r.
    \end{equation}
\end{subequations}
From Eqs.~\eqref{seq:ambtostresscoeffmobE}, one can immediately see that $\mathcal{E}\text{d}\mathcal{P} = \text{d}\mu_0$, which implies that the effective chemical potential $g$ identified by this work is proportional to the bulk AMB+ chemical potential $\mu_0$ when $M[\rho]\sim 1/\mathcal{E}$.
Further, one can substitute the mapping Eqs.~\eqref{seq:ambtostresscoeffmobE} into the pseudovariable definitions Eq.~\eqref{seq:epsilondef} and Eq.~\eqref{seq:ambpseudodef} to find that $\mathcal{E}\sim 1/\psi$.
We now consider the coexistence criteria under this $M[\rho]\sim 1/\mathcal{E}$ mapping.
We start with the mechanical Maxwell construction:
\begin{multline}
    \int_{\mathcal{E}^{\rm in}}^{\mathcal{E}^{\rm sat}}\left[\mathcal{P}-\mathcal{P}^{\rm sat}\right]\text{d}\mathcal{E}\\   = -\frac{1}{2}\int_0^{\infty}\left(\mathcal{E}^{\rm in} - \mathcal{E}\right)\frac{1}{\mathcal{E}}\frac{\partial\mathcal{Z}}{\partial\rho}\left(\frac{\partial\rho}{\partial r}\right)^3\text{d}r+ \frac{2}{R_C}\int_0^{\infty}\left(\kappa_1(\rho)\frac{\partial\mathcal{E}}{\partial\rho}\left(\frac{\partial\rho}{\partial r}\right)^2 + \mathcal{G}\left(\mathcal{E}-\mathcal{E}^{\rm in}\right)\right)\text{d}r.
\end{multline}
Substitution of Eq.~\eqref{seq:gdiffrearrangelaplace} results in:
\begin{multline}
    g^{\rm in}-g^{\rm out} + \frac{2}{R_C}\mathcal{E}^{\rm in}\int_0^{\infty}\left[\frac{1}{\mathcal{E}}\mathcal{Z}(\rho)\left(\frac{\partial\rho}{\partial r}\right)^2 + K(\rho)\frac{1}{\mathcal{E}^2}\frac{\partial\mathcal{E}}{\partial\rho}\left(\frac{\partial\rho}{\partial r}\right)^2\right] \text{d}r -\frac{1}{2}\mathcal{E}^{\rm in}\int_0^{\infty}\frac{1}{\mathcal{E}}\frac{\partial\mathcal{Z}}{\partial\rho}\left(\frac{\partial\rho}{\partial r}\right)^3\text{d}r\\  = -\frac{1}{2}\int_0^{\infty}\left(\mathcal{E}^{\rm in} - \mathcal{E}\right)\frac{1}{\mathcal{E}}\frac{\partial\mathcal{Z}}{\partial\rho}\left(\frac{\partial\rho}{\partial r}\right)^3\text{d}r+ \frac{2}{R_C}\int_0^{\infty}\left(\kappa_1(\rho)\frac{\partial\mathcal{E}}{\partial\rho}\left(\frac{\partial\rho}{\partial r}\right)^2 + \mathcal{G}\left(\mathcal{E}-\mathcal{E}^{\rm in}\right)\right)\text{d}r.
\end{multline}
Substitution of Eqs.~\eqref{seq:ambtostresscoeffmobE} results in:
\begin{equation}
    g^{\rm in}-g^{\rm out}  = \frac{1}{2}\int_0^{\infty}\frac{\partial\mathcal{Z}}{\partial\rho}\left(\frac{\partial\rho}{\partial r}\right)^3\text{d}r- \frac{2}{R_C}\int_0^{\infty}\mathcal{Z}(\rho)\left(\frac{\partial\rho}{\partial r}\right)^2\text{d}r
\end{equation}
Therefore, the AMB+ coexistence criteria Eq.~\eqref{seq:ambcoexist1} is recovered (up to a $(\partial_r\rho)^3$ contribution).
We now consider the mechanically derived Laplace pressure difference, with Eqs.~\eqref{seq:ambtostresscoeffmobE} substituted:
\begin{equation}
    \mathcal{P}^{\rm in}-\mathcal{P}^{\rm sat} =\frac{1}{2} \int_0^{\infty}\frac{1}{\mathcal{E}}\frac{\partial\mathcal{Z}}{\partial\rho}\left(\frac{\partial\rho}{\partial r}\right)^3\text{d}r - \frac{2}{R_C}\int_0^{\infty}\left[\frac{1}{\mathcal{E}}\mathcal{Z}(\rho)\left(\frac{\partial\rho}{\partial r}\right)^2 + \frac{1}{\mathcal{E}^2}\frac{\partial\mathcal{E}}{\partial\rho}K(\rho)\left(\frac{\partial\rho}{\partial r}\right)^2\right]\text{d}r.
\end{equation}
Recall the generalized Gibbs-Duhem relation $\text{d}g = \mathcal{E}\text{d}\mathcal{P}$. 
Under the choice $M[\rho]\sim 1/\mathcal{E}$, we know that $\psi \sim 1/\mathcal{E}$ and $g\sim \mu_0$, which implies that $\psi \text{d}\mu_o = \text{d}\mathcal{P} = \text{d}\left(\mu_o\psi\right) - \mu_o\text{d}\psi$.
Therefore:
\begin{equation}
    \mathcal{P}^{\rm in} - \mathcal{P}^{\rm sat} = \mu_0^{\rm in}\psi^{\rm in} - \mu_0^{\rm sat}\psi^{\rm sat} + \int_{\psi^{\rm in}}^{\psi^{\rm sat}}\mu_0\text{d}\psi.
\end{equation}
From Eq.~\eqref{seq:ambcoexist1} this can be rewritten as:
\begin{equation}
    \mathcal{P}^{\rm in}-\mathcal{P}^{\rm sat} =   \frac{1}{2}\psi^{\rm in}\int_0^{\infty}\frac{\partial\mathcal{Z}}{\partial\rho}\left(\frac{\partial\rho}{\partial r}\right)^3 \text{d}r - \psi^{\rm in} \frac{2}{R_C}\int_0^{\infty}  \mathcal{Z}(\rho)\left(\frac{\partial\rho}{\partial r}\right)^2 \text{d}r  + \int_{\psi^{\rm in}}^{\psi^{\rm sat}}\left[\mu_0 - \mu_o^{\rm sat}\right]\text{d}\psi.\label{seq:ambrecoveralmostmobE}
\end{equation}
We can then substitute Eq.~\eqref{seq:ambrecoveralmostmobE} into the Laplace pressure difference to find that:
\begin{equation}
     \int_{\psi^{\rm in}}^{\psi^{\rm sat}}\left[\mu_0 - \mu_o^{\rm sat}\right]\text{d}\psi +  \frac{1}{2}\int_0^{\infty}(\psi^{\rm in}-\psi)\frac{\partial\mathcal{Z}}{\partial\rho}\left(\frac{\partial\rho}{\partial r}\right)^3 \text{d}r = \frac{2\sigma}{R_C},
\end{equation}
which recovers the second coexistence criteria Eq.~\eqref{seq:ambcoexist2}.

We emphasize that choosing $M[\rho]$ to be constant or proportional to $1/\mathcal{E}$ facilitates a clear physical interpretation for AMB+, but will in general lead to nonlocal terms.
Choosing $M[\rho]$ such that the mapping is fully local may lead to difficulty in physically interpreting AMB+ quantities. 
For example, consider the values of $\{\kappa_i\}$ found for ABPs in Eqs.~\eqref{seq:abpconstitutive}. 
Substitution of these expressions into Eqs.~\eqref{seq:ambtostresscoeff} results in constraining $\mathcal{K}=0$, which then leads to a solution for $M[\rho]$ as:
\begin{equation}
    M[\rho]\sim \text{exp}\left[-\int \frac{\left[\frac{\partial}{\partial\rho}\left(\overline{U}\frac{\partial}{\partial\rho}\left[\overline{U}\frac{\partial p_C}{\partial\rho}\right]\right) \right]}{\overline{U}^2\frac{\partial^2 p_C}{\partial \rho^2}}\text{d}\rho\right].
\end{equation}
This solution for $M[\rho]$ leads to us identifying $\mu_0$ as:
\begin{equation}
    \mu_0 \sim \int \text{exp}\left[\int \frac{\left[\frac{\partial}{\partial\rho}\left(\overline{U}\frac{\partial}{\partial\rho}\left[\overline{U}\frac{\partial p_C}{\partial\rho}\right]\right) \right]}{\overline{U}^2\frac{\partial^2 p_C}{\partial \rho^2}}\text{d}\rho\right]\text{d}\mathcal{P},
\end{equation}
which has a less clear physical interpretation than when $M[\rho]$ is constant or proportional to $1/\mathcal{E}$. 
This implies that, despite the generic mapping between mechanics and AMB+ presented here, directly comparing quantities such as pressure, chemical potential, or surface tension within our model of ABP transport to AMB+ is non-trivial. \newpage
\section{\label{sec:dynamic}Dynamics of an Active Nucleus}
In this Section, we start from the fluctuating hydrodynamics derived in our recent work~\cite{Langford2023} and derive the Langevin dynamics of a nucleus using the route outlined by Cates and Nardini~\cite{Cates2023ClassicalSeparation}. 
We then present a detailed demonstration of how the noise statistics associated with the Langevin dynamics were calculated.
Third, we discuss and numerically justify the existence of an effective potential barrier governing the nucleation dynamics.
Finally, we discuss the implication of this effective potential for predicting the rate of nucleation.

\subsection{\label{sec:flucthydro}Langevin Dynamics of a Spherical Nucleus}
We aim to derive the Langevin equation for the radius of a nucleus of active Brownian particles in the spherically symmetric geometry described by Section~\ref{sec:coexistcriteria}. 
We assume the dynamics of the nucleus to proceed quasistatically: any given nucleus with radius $R$ (not necessarily equal to $R_C$) will be in a local mechanical balance with its surroundings.
This implies that the density immediately outside of the nucleus $\rho^{\rm out}\neq \rho^{\rm sat}$, with diffusive transport in the majority phase driven by this inequality.
Then, as diffusion will cause $\rho^{\rm out}$ to change in time, the local mechanical balance will demand $R$ to change in time as well.
The starting point to derive the dynamics described above is the fluctuating hydrodynamics which we recently derived~\cite{Langford2023}:
\begin{subequations}
    \label{seq:flucthydro}
    \begin{align}
        \frac{\partial \rho}{\partial t} &= -\boldsymbol{\nabla}\cdot\mathbf{J}\label{seq:continuity},\\ \mathbf{J} &= \frac{1}{\zeta}\boldsymbol{\nabla}\cdot\bm{\Sigma} + \bm{\eta}^{\rm act},\label{seq:hydroflux}
    \end{align}
where $\bm{\Sigma}$ is defined identically as the dynamic stress tensor in Eqs.~\eqref{seq:dynstressform} and~\eqref{seq:abpconstitutive}, while $\bm{\eta^{\rm act}}$ is an entirely athermal stochastic contribution to the flux with zero mean and a variance of:
\begin{align}
     \langle \bm{\eta}^{\rm act}(\mathbf{r},t)\bm{\eta}^{\rm act}(\mathbf{r}',t') \rangle = 2\frac{k_BT^{\rm act}}{\zeta}\left(\rho\mathbf{I} - \frac{d}{d-1}\mathbf{Q}'\right)\delta(t-t')\delta(\mathbf{r}-\mathbf{r}') \ \label{seq:etavar},
\end{align}
\end{subequations}
Here ${k_B T^{\rm act} \equiv \zeta U_o \ell_o / d(d-1)}$ is the energy scale of the athermal active fluctuations and $\mathbf{Q}'$ is the traceless nematic order tensor.
Substitution of Eqs.~\eqref{seq:dynstressform} and Eq.~\eqref{seq:hydroflux} into Eq.~\eqref{seq:continuity} results in:
\begin{equation}
    \zeta\frac{\partial \rho}{\partial t} = -\nabla^2\Sigma_{rr} + \frac{2}{r}\frac{\partial\mathcal{G}}{\partial r} + \frac{2}{r^2}\mathcal{G} - \zeta\boldsymbol{\nabla}\cdot\bm{\eta}^{\rm act}.\label{seq:densedynamicssphere}
\end{equation}
From here, we proceed in a similar fashion as Cates and Nardini~\cite{Cates2023ClassicalSeparation}.
We recently established a surface-area minimizing principle for liquid-gas interfaces of ABPs resulting from a positive capillary-wave tension, $\upgamma_{\rm cw}$, which resists fluctuations to the area-minimizing interfacial interfacial shape.
The strength of the fluctuations are set by $k_BT^{\rm act}$~\cite{Langford2023}, while the ratio of the capillary tension to $k_BT^{\rm act}$ -- the active interfacial stiffness -- controls the amplitude of long-wavelength interfacial fluctuations.
A fluctuating nucleus can be well-approximated as a sphere when the average radius $R$ satisfies  $(R^2\upgamma_{\rm cw}/k_BT^{\rm act})^{1/2} \gg 1$.
Therefore we introduce the radialy symmetric ansatz connecting the density field and the radius of a nucleus $R$ at time $t$, which was proposed by Bray~\cite{Bray1994TheoryKinetics}:
\begin{equation}
    \rho = \upvarphi_{R}\left[r - R(t)\right],\label{seq:ansatz}
\end{equation}
where $\upvarphi_R$ is the density profile calculated in Section~\ref{sec:coexistnumerics}.
Substitution of Eq.~\eqref{seq:ansatz} into Eq.~\eqref{seq:densedynamicssphere} results in:
\begin{equation}
    -\zeta\upvarphi_R'\frac{\partial R}{\partial t} = -\nabla^2\Sigma_{rr} + \frac{2}{r}\frac{\partial\mathcal{G}}{\partial r} + \frac{2}{r^2}\mathcal{G} - \zeta\boldsymbol{\nabla}\cdot\bm{\eta}^{\rm act}\label{seq:beforeinvlaplace},
\end{equation}
where $\upvarphi_R'$ is the $r$ derivative of $\upvarphi_R$.
We now define the inverse Laplace operator as:
\begin{equation}
    \nabla^{-2}f(r) = -\int_{r}^{\infty}\int_{0}^{r_2}\text{d}r_2\text{d}r_1 \left(\frac{r_1}{r_2}\right)^{2}f(r_1),\label{seq:invlaplacedef}
\end{equation}
where $f(r)$ is a function that vanishes in the limit of $r\to\infty$.
We apply the inverse Laplace operator to both sides of Eq.~\eqref{seq:beforeinvlaplace}.
The effect of the inverse Laplace operator on the second term on the left hand side is:
\begin{equation}
    \nabla^{-2}\left[\frac{2}{r}\frac{\partial \mathcal{G}}{\partial r}\right] = - \int_r^{\infty}\text{d}r_2\frac{1}{r_2^2}\int_{0}^{r_2}\text{d}r_12r_1\frac{\partial \mathcal{G}}{\partial r_1}.\label{seq:invlaplaceexample}
\end{equation}
We now integrate Eq.~\eqref{seq:invlaplaceexample} by parts while choosing:
\begin{align*}
    u = \int_{0}^{r_2}\text{d}r_1 2r_1\frac{\partial\mathcal{G}}{\partial r_1}&,\quad du = 2r_2\frac{\partial\mathcal{G}}{\partial r_2}\text{d}r_2,\\ dv = -\frac{1}{r_2^2}\text{d}r_2&,\quad v = \frac{1}{r_2}.
\end{align*}
Then Eq.~\eqref{seq:invlaplaceexample} can equivalently be expressed as
\begin{equation}
    \nabla^{-2}\left[\frac{2}{r}\frac{\partial \mathcal{G}}{\partial r}\right] = - \frac{2}{r}\int_0^r\text{d}r_1 r_1\frac{\partial\mathcal{G}}{\partial r_1} + 2 \mathcal{G}(r).
\end{equation}
A second integration by parts results in:
\begin{align}
    \nabla^{-2}\left[\frac{2}{r}\frac{\partial \mathcal{G}}{\partial r}\right] &= - \frac{2}{r}\left(r\mathcal{G}(r) - \int_0^{r}\mathcal{G}\text{d}r_1\right)+ 2 \mathcal{G}(r),\nonumber \\ &= \frac{2}{r}\int_0^{r}\mathcal{G}\text{d}r_1.\label{seq:invlaplaceexampleresult}
\end{align}
Applying a similar procedure to the other terms in Eq.~\eqref{seq:beforeinvlaplace} results in:
\begin{equation}
    \nabla^{-2}\left[\frac{2}{r^2}\mathcal{G}\right] = -\frac{2}{r}\int_0^{r}\mathcal{G}\text{d}r_1 - 2\int_{r}^{\infty} \frac{\mathcal{G}}{r_1}\text{d}r_1,\label{seq:othergterm}
\end{equation}
\begin{equation}
    \nabla^{-2}\left[-\zeta\upvarphi'_R\frac{\partial R}{\partial t}\right] = \zeta R\frac{\partial R}{\partial t}\left(\rho^{\rm sat}-\rho^{\rm in}\right) + \mathcal{O}\left(\frac{1}{R}\right),\label{seq:invlaplacetimeterm}
\end{equation}
\begin{equation}
    \nabla^{-2}\left[-\nabla^2\Sigma_{rr}\right] = -\Sigma_{rr} - \mathcal{P}(\rho^{\rm sat}),\label{seq:invlaplacetrivial}
\end{equation}
where the $\mathcal{P}(\rho^{\rm sat})$ in Eq.~\eqref{seq:invlaplacetrivial} comes from the boundary term of $\Sigma_{rr}$ as $r\to\infty$. 
From the above equations one can see that applying the inverse Laplace operator to both sides of Eq.~\eqref{seq:beforeinvlaplace} results in:
\begin{equation}
    \zeta R\frac{\partial R}{\partial t}\Delta\rho + \mathcal{O}\left(\frac{1}{R}\right) = \Sigma_{rr} + \mathcal{P}(\rho^{\rm sat}) + \int_{r}^{\infty}\text{d}r_1\frac{2}{r_1}\mathcal{G} - \nabla^{-2}\left[\zeta \boldsymbol{\nabla}\cdot\bm{\eta}^{\rm act}\right],\label{seq:afterinvlaplace}
\end{equation}
where we have defined the difference between an arbitrary property $a$ at the interior and exterior of the bubble/droplet as $\Delta a \equiv a^{\rm in} - a^{\rm sat}$.
We now multiply both sides of the equation by $\partial\mathcal{E}/\partial r$ and integrate from $r=0$ to $r=\infty$:
\begin{align}
     -\zeta R\frac{\partial R}{\partial t}\Delta\rho\Delta\mathcal{E}+ \mathcal{O}\left(\frac{1}{R}\right) =& \int_0^{\infty}\Sigma_{rr}\frac{\partial\mathcal{E}}{\partial r}\text{d}r + \mathcal{P}(\rho^{\rm sat})\left(\mathcal{E}^{\rm sat} - \mathcal{E}^{\rm in}\right) \nonumber \\ &+\int_{0}^{\infty}\text{d}r\frac{\partial\mathcal{E}}{\partial r}\int_{r}^{\infty}\text{d}r_1\frac{2\mathcal{G}}{r_1} - \int_{0}^{\infty}\text{d}r\frac{\partial\mathcal{E}}{\partial r}\nabla^{-2}\left[\zeta\boldsymbol{\nabla}\cdot\bm{\eta}^{\rm act}\right].\label{seq:afterpseudo}
\end{align}
The double integral can be integrating by parts with:
\begin{align*}
    u = \int_{r}^{\infty}\frac{2\mathcal{G}}{r_2}\text{d}r_1&,\quad du = -\frac{2\mathcal{G}}{r}\text{d}r,\\ dv = \frac{\partial\mathcal{E}}{\partial r}\text{d}r&,\quad v = \mathcal{E}
\end{align*}
\begin{align}
     -\zeta R\frac{\partial R}{\partial t}\Delta\rho\Delta\mathcal{E} + \mathcal{O}\left(\frac{1}{R}\right) =& \int_0^{\infty}\Sigma_{rr}\frac{\partial\mathcal{E}}{\partial r}\text{d}r + \mathcal{P}(\rho^{\rm sat})\left(\mathcal{E}^{\rm sat} - \mathcal{E}^{\rm in}\right)-\frac{2\upgamma_{\rm mech}}{R}\mathcal{E}^{\rm in} \nonumber \\ & + \frac{2}{R}\int_0^{\infty}\mathcal{E}\mathcal{G}\text{d}r - \int_{0}^{\infty}\text{d}r\frac{\partial\mathcal{E}}{\partial r}\nabla^{-2}\left[\zeta\boldsymbol{\nabla}\cdot\bm{\eta}^{\rm act}\right].\label{seq:afterpseudobyparts}
\end{align}
Now consider the first integral on the right hand side.
Substituting in the constitutive expression for $\Sigma_{rr}$ from Eq.~\eqref{seq:stationary} and integrating by parts results in
\begin{equation}
    \int_0^{\infty}\Sigma_{rr}\frac{\partial\mathcal{E}}{\partial r}\text{d}r = -\int_{\mathcal{E}^{\rm in}}^{\mathcal{E}^{\rm sat}}\text{d}\mathcal{E}\mathcal{P}(\rho) + \int_{0}^{\infty}\frac{2}{r}a(\rho)\frac{\partial \mathcal{E}}{\partial \rho}\left(\upvarphi'_R\right)^2.\label{seq:sigmarrbyparts}
\end{equation}
Substitution of Eq.~\eqref{seq:sigmarrbyparts} into Eq.~\eqref{seq:afterpseudobyparts} gives:
\begin{align}
     -\zeta R\frac{\partial R}{\partial t}\Delta\rho\Delta\mathcal{E} + \mathcal{O}\left(\frac{1}{R}\right) = &-\int_{\mathcal{E}^{\rm in}}^{\mathcal{E}^{\rm sat}}\text{d}\mathcal{E}\mathcal{P}(\rho) + \mathcal{P}(\rho^{\rm sat})\left(\mathcal{E}^{\rm sat} - \mathcal{E}^{\rm in}\right)-\frac{2\upgamma_{\rm mech}}{R}\mathcal{E}^{\rm in} \nonumber \\ & + \frac{2}{R}\int_0^{\infty}\left(\mathcal{E}\mathcal{G} + a(\rho)\frac{\partial\mathcal{E}}{\partial \rho}\left(\frac{\partial\upvarphi_R}{\partial r}\right)^2\right)\text{d}r\nonumber \\ & - \int_{0}^{\infty}\text{d}r\frac{\partial\mathcal{E}}{\partial r}\nabla^{-2}\left[\zeta\boldsymbol{\nabla}\cdot\bm{\eta}^{\rm act}\right].\label{seq:afterpseudosigmarrbyparts}
\end{align}
From Eq.~\eqref{seq:osttensiondef} we can rewrite Eq.~\eqref{seq:afterpseudosigmarrbyparts} as:
\begin{align}
     -\zeta R\frac{\partial R}{\partial t}\Delta\rho\Delta\mathcal{E} + \mathcal{O}\left(\frac{1}{R}\right) = &-\int_{\mathcal{E}^{\rm in}}^{\mathcal{E}^{\rm sat}}\text{d}\mathcal{E}\left[\mathcal{P}(\rho) - \mathcal{P}(\rho^{\rm sat}) \right]+\frac{2\upgamma_{\rm ost}}{R}\frac{\rho^{\rm sat}\Delta\mathcal{E}}{\Delta\rho} \nonumber \\ & - \int_{0}^{\infty}\text{d}r\frac{\partial\mathcal{E}}{\partial r}\nabla^{-2}\left[\zeta\boldsymbol{\nabla}\cdot\bm{\eta}^{\rm act}\right].\label{seq:afterpseudosigmarrbypartsost}
\end{align}
Recognizing that the first integral on the right-hand-side of Eq.~\eqref{seq:afterpseudosigmarrbypartsost} is a constant, we can rearrange:
\begin{align}
     \zeta R\frac{\partial R}{\partial t}\Delta\rho\Delta\mathcal{E} + \mathcal{O}\left(\frac{1}{R}\right) = \frac{2\upgamma_{\rm ost}\rho^{\rm sat}\Delta\mathcal{E}}{\Delta\rho}\left[\frac{1}{R_C} - \frac{1}{R}\right] + \int_{0}^{\infty}\text{d}r\frac{\partial\mathcal{E}}{\partial r}\nabla^{-2}\left[\zeta\boldsymbol{\nabla}\cdot\bm{\eta}^{\rm act}\right],\label{seq:Revolveinit}
\end{align}
where the critical radius is defined as:
\begin{equation}
    R_C = \frac{2\upgamma_{\rm ost}\rho^{\rm sat}\Delta\mathcal{E}}{\Delta\rho\int_{\mathcal{E}^{\rm in}}^{\mathcal{E}^{\rm sat}}\left[\mathcal{P}(\rho) - \mathcal{P}(\rho^{\rm sat})\right]\text{d}\mathcal{E}},\label{seq:Rcdef}
\end{equation}
which is identical to the critical radius found in Section~\ref{sec:coexistcriteria} [see Eq.~\eqref{seq:coexist2ost}], highlighting the consistency of the two approaches.

The final evolution equation for the radial growth can be compactly written as:
\begin{equation}
    \zeta_{\rm eff}\frac{\partial R}{\partial t} = F^D(R) + \tilde{\chi}(t),\label{seq:drdt}
\end{equation}
where we have defined an effective drag force:
\begin{equation}
    \zeta_{\rm eff} = \frac{4\pi\left(\rho^{\rm sat}-\rho^{\rm in}\right)^2R^3\zeta}{\rho^{\rm sat}},\label{seq:defmobility}
\end{equation}
as well as a deterministic force:
\begin{equation}
    F^D(R) = 8\pi R^2\upgamma_{\rm ost}\left[\frac{1}{R_C} - \frac{1}{R}\right].\label{seq:Fdetermindef}
\end{equation}
$\tilde{\chi}(t)$ represents the stochastic force with:
\begin{equation}
    \tilde{\chi}(t) = \frac{\zeta_{\rm eff}}{R\Delta\rho\Delta\mathcal{E}}\nabla^{-2}\left[\boldsymbol{\nabla}\cdot\bm{\eta}^{\rm act}\right].\label{seq:chidef}
\end{equation}
In Section~\ref{sec:noise}, we derive the statistics of $\tilde{\chi}(t)$ finding that it has zero mean and a variance given by:
\begin{equation}
    \langle \tilde{\chi}(t)\tilde{\chi}(t') \rangle = 2k_BT^{\rm act} \zeta_{\rm eff}\delta(t-t').\label{seq:chivar}
\end{equation}
From Eqs.~\eqref{seq:drdt} and ~\eqref{seq:chivar} we can see that the dynamics of the nucleus obey the fluctuation dissipation theorem. 
In addition, the deterministic term can be integrated to define an effective potential governing nucleation as further discussed in Section~\ref{sec:potential}.

\subsection{\label{sec:noise}Noise Statistics}
We now seek to derive the statistics of $\tilde{\chi}(t)$ [Eq.~\eqref{seq:chivar}].
We start by rewriting the statistics of $\bm{\eta}^{\rm act}$ [Eq.~\eqref{seq:etavar}] to be compatible with our spherical symmetry conditions:
\begin{equation}
    \langle \bm{\eta}^{\rm act}(\mathbf{r},t)\bm{\eta}^{\rm act}(\mathbf{r}',t') \rangle = 2\frac{k_BT^{\rm act}}{4\pi\zeta}\left(\rho\mathbf{I} - \frac{d}{d-1}\mathbf{Q}'\right)\delta(t-t')\left(\frac{\delta(r-r')}{r^2}\right).\label{seq:etavarsphere}
\end{equation}
As found in Ref.~\cite{Langford2024TheoryPhases}, $\mathbf{Q}'$ is given by:
\begin{align}
    \mathbf{Q}' = -\frac{d-1}{\ell_o^2 D_R \zeta \overline{U}(\rho)}\left[\kappa_1(\rho)\nabla^2\rho \mathbf{I} + \kappa_3(\rho)\boldsymbol{\nabla}\rho\boldsymbol{\nabla}\rho\right].
\end{align}
Thus, $\bm{\eta}^{\rm act}$ may have anisotropic contributions to its correlator, particularly at interfaces where gradients in the density are large.
This anisotropy is due to $\bm{\eta}^{\rm act}$'s microscopic roots in the aggregate of individual $\bm{\Omega}_i\times\mathbf{q}_i$.
In regions of space with nematic alignment, the result of the cross product of an isotropic noise with each orientation $\mathbf{q}_i$ will be biased towards directions orthogonal to that alignment.
Before finding the statistics of $\tilde{\chi}$, we first find the statistics of $\xi_R(\mathbf{r},t)$, which we define as
\begin{equation}
    \xi_{R}(\mathbf{r},t) = -\boldsymbol{\nabla}\cdot\bm{\eta}^{\rm act}.\label{seq:xidef}
\end{equation}
We split the correlations of $\xi_R$ into three contributions,
\begin{equation}
    \langle \xi_{R}(\mathbf{r},t)\xi_{R}(\mathbf{r}',t') \rangle = \langle \xi_{R}(\mathbf{r},t)\xi_{R}(\mathbf{r}',t') \rangle^{\rm iso} + \langle \xi_{R}(\mathbf{r},t)\xi_{R}(\mathbf{r}',t') \rangle^{\rm aniso}_1 + \langle \xi_{R}(\mathbf{r},t)\xi_{R}(\mathbf{r}',t') \rangle^{\rm aniso}_2,\label{seq:xisplit}
\end{equation}
where the contribution labeled $\rm{iso}$ originates from the component of the $\bm{\eta}^{\rm act}$ correlator proportional to $\rho\mathbf{I}$ while the contributions labeled $\rm{aniso}$ originates from the component of the $\bm{\eta}^{\rm act}$ correlator proportional to $\mathbf{Q}'$. Each contribution is given by:
\begin{equation}
    \langle \xi_{R}(\mathbf{r},t)\xi_{R}(\mathbf{r}',t') \rangle^{\rm iso} = -\frac{2k_BT^{\rm act}}{4\pi\zeta}\delta(t-t')\left(\frac{\partial}{\partial r} + \frac{2}{r}\right)\left[\rho\frac{\partial}{\partial r}\left(\frac{\delta(r-r')}{r^2}\right)\right],\label{seq:xiisodef}
\end{equation}
\begin{align}
    \langle \xi_{R}(\mathbf{r},t)\xi_{R}(\mathbf{r}',t') \rangle^{\rm aniso}_1 = &-\frac{2k_BT^{\rm act}}{4\pi\zeta}\delta(t-t')\left(\frac{\partial }{\partial r} + \frac{2}{r}\right)\nonumber \\ &\left[\left(\frac{1}{\ell_o^2 D_R\zeta \overline{U}}\left[ \kappa_1(\rho)\left(\frac{\partial }{\partial r} + \frac{2}{r}\right)\frac{\partial \rho}{\partial r} - \kappa_3(\rho)\left(\frac{\partial \rho}{\partial r}\right)^2\right]\right)\frac{\partial}{\partial r}\left(\frac{\delta(r-r')}{r^2}\right)\right],\label{seq:xianiso1def}
\end{align}
and
\begin{equation}
    \langle \xi_{R}(\mathbf{r},t)\xi_{R}(\mathbf{r}',t') \rangle^{\rm aniso}_2 = \frac{2k_BT^{\rm act}}{4\pi\zeta}\delta(t-t')\left(\frac{\partial}{\partial r} + \frac{2}{r}\right)\left[\left( \frac{1}{r'}\frac{1}{\ell_o^2 D_R\zeta \overline{U}}\left[ \kappa_3(\rho)\left(\frac{\partial\rho}{\partial r}\right)^2\right]\right)\frac{\delta(r-r')}{r^2}\right].\label{seq:xianiso2def}
\end{equation}
We now define $\chi(t)$ as:
\begin{equation}
    \chi(t) = -\int_0^{\infty}\text{d}r \frac{\partial\mathcal{E}}{\partial r}\int_r^{\infty}\text{d}r_2\frac{1}{r_2^2}\int_0^{r_2}r_1^2\xi_R(\mathbf{r}_1,t).
\end{equation}
The correlations of $\chi(t)$ are then given by
\begin{equation}
    \langle \chi(t)\chi(t')\rangle = \left\langle\int_0^{\infty}\text{d}r \frac{\partial\mathcal{E}}{\partial r}\int_r^{\infty}\text{d}r_2\frac{1}{r_2^2}\int_0^{r_2}r_1^2\xi_R(\mathbf{r}_1,t) \int_0^{\infty}\text{d}r' \frac{\partial\mathcal{E}}{\partial r'}\int_r^{\infty}\text{d}r_4\frac{1}{r_4^2}\int_0^{r_4}r_3^2\xi_R(\mathbf{r}_3,t')\right \rangle,
\end{equation}
which can be rewritten as:
\begin{equation}
    \langle \chi(t)\chi(t')\rangle = \int_0^{\infty}\int_0^{\infty}\text{d}r\text{d}r'\frac{\partial\mathcal{E}}{\partial r}\frac{\partial\mathcal{E}}{\partial r'}\int_{r}^{\infty}\int_{r'}^{\infty}\int_0^{r_2}\int_0^{r_4}\text{d}r_1\text{d}r_2\text{d}r_3\text{d}r_4\left(\frac{r_1r_3}{r_2r_4}\right)^2\langle \xi_R(\mathbf{r}_1,t)\xi_R(\mathbf{r}_3,t')\rangle.
\end{equation}
Now we similarly divide $\chi(t)$ into contributions:
\begin{equation}
    \langle \chi(t)\chi(t')\rangle = \langle \chi(t)\chi(t')\rangle^{\rm iso} + \langle \chi(t)\chi(t')\rangle^{\rm aniso}_1+ \langle \chi(t)\chi(t')\rangle^{\rm aniso}_2,\label{seq:chisplit}
\end{equation}
where
\begin{align}
    \langle \chi(t)\chi(t')\rangle^{\rm iso} = &\int_0^{\infty}\int_0^{\infty}\text{d}r\text{d}r'\frac{\partial\mathcal{E}}{\partial r}\frac{\partial\mathcal{E}}{\partial r'}\nonumber \\ &\times\int_{r}^{\infty}\int_{r'}^{\infty}\int_0^{r_2}\int_0^{r_4}\text{d}r_1\text{d}r_2\text{d}r_3\text{d}r_4\left(\frac{r_1r_3}{r_2r_4}\right)^2\langle \xi_R(\mathbf{r}_1,t)\xi_R(\mathbf{r}_3,t')\rangle^{\rm iso},\label{seq:chiiso}
\end{align}
\begin{align}
    \langle \chi(t)\chi(t')\rangle^{\rm aniso}_1 = &\int_0^{\infty}\int_0^{\infty}dr\text{d}r'\frac{\partial\mathcal{E}}{\partial r}\frac{\partial\mathcal{E}}{\partial r'}\nonumber \\ &\times\int_{r}^{\infty}\int_{r'}^{\infty}\int_0^{r_2}\int_0^{r_4}\text{d}r_1\text{d}r_2\text{d}r_3\text{d}r_4\left(\frac{r_1r_3}{r_2r_4}\right)^2\langle \xi_R(\mathbf{r}_1,t)\xi_R(\mathbf{r}_3,t')\rangle^{\rm aniso}_1,\label{seq:chianiso1}
\end{align}
and 
\begin{align}
    \langle \chi(t)\chi(t')\rangle^{\rm aniso}_2 = &\int_0^{\infty}\int_0^{\infty}\text{d}r\text{d}r'\frac{\partial\mathcal{E}}{\partial r}\frac{\partial\mathcal{E}}{\partial r'}\nonumber \\ &\times\int_{r}^{\infty}\int_{r'}^{\infty}\int_0^{r_2}\int_0^{r_4}\text{d}r_1\text{d}r_2\text{d}r_3\text{d}r_4\left(\frac{r_1r_3}{r_2r_4}\right)^2\langle \xi_R(\mathbf{r}_1,t)\xi_R(\mathbf{r}_3,t')\rangle^{\rm aniso}_2.\label{seq:chianiso2}
\end{align}
We will first solve for $\langle \chi(t)\chi(t')\rangle^{\rm iso}$.
Substitution of Eq.~\eqref{seq:xiisodef} into Eq.~\eqref{seq:chiiso} results in:
\begin{align}
    \langle \chi(t)\chi(t')\rangle^{\rm iso} = &\frac{2k_BT^{\rm act}}{4\pi\zeta}\delta(t-t')\int_0^{\infty}\int_0^{\infty}\text{d}r\text{d}r'\frac{\partial\mathcal{E}}{\partial r}\frac{\partial\mathcal{E}}{\partial r'}\nonumber \\ &\times\int_{r}^{\infty}\int_{r'}^{\infty}\int_0^{r_2}\int_0^{r_4}\text{d}r_1\text{d}r_2\text{d}r_3\text{d}r_4\left(\frac{r_1r_3}{r_2r_4}\right)^2\left(\frac{\partial}{\partial r_1} + \frac{2}{r_1}\right)\left[\rho(r_1)\frac{\partial}{\partial r_1}\left(\frac{\delta(r_1-r_3)}{r_1^2}\right)\right].\label{seq:chiisosub}
\end{align}
We now distribute the divergence in $r_1$ to find:
\begin{align}
    \langle \chi(t)\chi(t')\rangle^{\rm iso} = &\frac{2k_BT^{\rm act}}{4\pi\zeta}\delta(t-t')\int_0^{\infty}\int_0^{\infty}\text{d}r\text{d}r'\frac{\partial\mathcal{E}}{\partial r}\frac{\partial\mathcal{E}}{\partial r'}\nonumber \\ &\times\int_{r}^{\infty}\int_{r'}^{\infty}\int_0^{r_2}\int_0^{r_4}\text{d}r_1\text{d}r_2\text{d}r_3\text{d}r_4\left(\frac{r_1r_3}{r_2r_4}\right)^2\frac{\partial}{\partial r_1}\left[\rho(r_1)\frac{\partial}{\partial r_1}\left(\frac{\delta(r_1-r_3)}{r_1^2}\right)\right] \nonumber \\ &+\frac{2k_BT^{\rm act}}{4\pi\zeta}\delta(t-t')\int_0^{\infty}\int_0^{\infty}dr\text{d}r'\frac{\partial\mathcal{E}}{\partial r}\frac{\partial\mathcal{E}}{\partial r'}\nonumber \\ &\times\int_{r}^{\infty}\int_{r'}^{\infty}\int_0^{r_2}\int_0^{r_4}\text{d}r_1\text{d}r_2\text{d}r_3\text{d}r_4\left(\frac{r_3}{r_2r_4}\right)^2 2r_1\left[\rho(r_1)\frac{\partial}{\partial r_1}\left(\frac{\delta(r_1-r_3)}{r_1^2}\right)\right].\label{seq:chiisosubsplit}
\end{align}
We now integrate the top integral by parts, which results in a boundary term as well as an integral which cancels out with the bottom integral.
We are then left with:
\begin{align}
    \langle \chi(t)\chi(t')\rangle^{\rm iso} = &-\frac{2k_BT^{\rm act}}{4\pi\zeta}\delta(t-t')\int_0^{\infty}\int_0^{\infty}\text{d}r\text{d}r'\frac{\partial\mathcal{E}}{\partial r}\frac{\partial\mathcal{E}}{\partial r'}\nonumber \\ &\times \int_{r}^{\infty}\int_{r'}^{\infty}\int_0^{r_4}\text{d}r_2\text{d}r_3\text{d}r_4\left(\frac{r_1r_3}{r_2r_4}\right)^2\left(\rho(r_1)\frac{\partial}{\partial r_1}\left(\frac{\delta(r_1-r_3)}{r_1^2}\right)\right)\Biggr|^{r_1 = r_2}_{r_1 = 0}
\end{align}
Evaluation of the limits must be done with care, particularly for the $r_1=0$ term.
Before evaluating these limits, we rearrange such that there are no derivatives in $r_1$:
\begin{align}
    \langle \chi(t)\chi(t')\rangle^{\rm iso} = &\frac{2k_BT^{\rm act}}{4\pi\zeta}\delta(t-t')\int_0^{\infty}\int_0^{\infty}\text{d}r\text{d}r'\frac{\partial\mathcal{E}}{\partial r}\frac{\partial\mathcal{E}}{\partial r'}\nonumber \\ &\times \int_{r}^{\infty}\int_{r'}^{\infty}\int_0^{r_4}\text{d}r_2\text{d}r_3\text{d}r_4\left(\frac{r_3}{r_2r_4}\right)^2\rho(r_1)\frac{2}{r_1}\delta(r_1-r_3)\Biggr|^{r_1 = r_2}_{r_1 = 0}\nonumber \\ &+\frac{2k_BT^{\rm act}}{4\pi\zeta}\delta(t-t')\int_0^{\infty}\int_0^{\infty}\text{d}r\text{d}r'\frac{\partial\mathcal{E}}{\partial r}\frac{\partial\mathcal{E}}{\partial r'}\nonumber \\ &\times \int_{r}^{\infty}\int_{r'}^{\infty}\int_0^{r_4}\text{d}r_2\text{d}r_3\text{d}r_4\left(\frac{r_3}{r_2r_4}\right)^2\left(\rho(r_1)\frac{\partial}{\partial r_3}\delta(r_1-r_3)\right)\Biggr|^{r_1 = r_2}_{r_1 = 0}.
\end{align}
We now evaluate the limits, noting that the $r_1\to0$ limits will vanish:
\begin{align}
    \langle \chi(t)\chi(t')\rangle^{\rm iso} = &\frac{2k_BT^{\rm act}}{4\pi\zeta}\delta(t-t')\int_0^{\infty}\int_0^{\infty}\text{d}r\text{d}r'\frac{\partial\mathcal{E}}{\partial r}\frac{\partial\mathcal{E}}{\partial r'}\nonumber \\ &\times \int_{r}^{\infty}\int_{r'}^{\infty}\int_0^{r_4}\text{d}r_2\text{d}r_3\text{d}r_4\left(\frac{r_3}{r_2r_4}\right)^2\rho(r_2)\frac{2}{r_2}\delta(r_3-r_2)\nonumber \\ &+\frac{2k_BT^{\rm act}}{4\pi\zeta}\delta(t-t')\int_0^{\infty}\int_0^{\infty}\text{d}r\text{d}r'\frac{\partial\mathcal{E}}{\partial r}\frac{\partial\mathcal{E}}{\partial r'}\nonumber \\ &\times \int_{r}^{\infty}\int_{r'}^{\infty}\int_0^{r_4}\text{d}r_2\text{d}r_3\text{d}r_4\left(\frac{r_3}{r_2r_4}\right)^2\left(\rho(r_2)\frac{\partial}{\partial r_3}\delta(r_2-r_3)\right).
\end{align}
We now integrate the first integral with respect to $r_3$.
Because the integral with respect to $r_3$ goes from $r_3=0$ to $r_3=r_4$, the integral will be zero if $r_4<r_2$.
However, if $r_4>r_2$, we can apply the sifting property of the delta function.
We can handle both cases by applying the sifting function while applying up a factor of $H(r_4-r_2)$,
\begin{align}
    \langle \chi(t)\chi(t')\rangle^{\rm iso} = &\frac{2k_BT^{\rm act}}{4\pi\zeta}\delta(t-t')\int_0^{\infty}\int_0^{\infty}\text{d}r\text{d}r'\frac{\partial\mathcal{E}}{\partial r}\frac{\partial\mathcal{E}}{\partial r'} \int_{r}^{\infty}\int_{r'}^{\infty}\text{d}r_2\text{d}r_4\left(\frac{1}{r_4}\right)^2\rho(r_2)\frac{2}{r_2}H(r_4-r_2)\nonumber \\ &+\frac{2k_BT^{\rm act}}{4\pi\zeta}\delta(t-t')\int_0^{\infty}\int_0^{\infty}\text{d}r\text{d}r'\frac{\partial\mathcal{E}}{\partial r}\frac{\partial\mathcal{E}}{\partial r'}\nonumber \\ &\times \int_{r}^{\infty}\int_{r'}^{\infty}\int_0^{r_4}\text{d}r_2\text{d}r_3\text{d}r_4\left(\frac{r_3}{r_2r_4}\right)^2\left(\rho(r_2)\frac{\partial}{\partial r_3}\delta(r_2-r_3)\right),
\end{align}
where $H$ is the Heaviside step function.
We can now integrate the bottom integral by parts which results in:
\begin{align}
    \langle \chi(t)\chi(t')\rangle^{\rm iso} = &\frac{2k_BT^{\rm act}}{4\pi\zeta}\delta(t-t')\int_0^{\infty}\int_0^{\infty}\text{d}r\text{d}r'\frac{\partial\mathcal{E}}{\partial r}\frac{\partial\mathcal{E}}{\partial r'} \int_{r}^{\infty}\int_{r'}^{\infty}\text{d}r_2\text{d}r_4\left(\frac{1}{r_4}\right)^2\rho(r_2)\frac{2}{r_2}H(r_4-r_2)\nonumber \\ &+\frac{2k_BT^{\rm act}}{4\pi\zeta}\delta(t-t')\int_0^{\infty}\int_0^{\infty}\text{d}r\text{d}r'\frac{\partial\mathcal{E}}{\partial r}\frac{\partial\mathcal{E}}{\partial r'}\nonumber \\ &\times \int_{r}^{\infty}\int_{r'}^{\infty}\text{d}r_2\text{d}r_4\left(\frac{r_3}{r_2r_4}\right)^2\left(\rho(r_2)\delta(r_2-r_3)\right)\Biggr|^{r_3=r_4}_{r_3=0}\nonumber \\ &-\frac{2k_BT^{\rm act}}{4\pi\zeta}\delta(t-t')\int_0^{\infty}\int_0^{\infty}\text{d}r\text{d}r'\frac{\partial\mathcal{E}}{\partial r}\frac{\partial\mathcal{E}}{\partial r'}\nonumber \\ &\times \int_{r}^{\infty}\int_{r'}^{\infty}\int_0^{r_4}\text{d}r_2\text{d}r_3\text{d}r_4\left(\frac{1}{r_2r_4}\right)^2\left(2r_3\rho(r_2)\delta(r_2-r_3)\right).
\end{align}
Evaluate the integral in $r_3$ of the bottom integral leads to a term which cancels with the top integral, resulting in:
\begin{align}
    \langle \chi(t)\chi(t')\rangle^{\rm iso} = &\frac{2k_BT^{\rm act}}{4\pi\zeta}\delta(t-t')\int_0^{\infty}\int_0^{\infty}\text{d}r\text{d}r'\frac{\partial\mathcal{E}}{\partial r}\frac{\partial\mathcal{E}}{\partial r'}\nonumber \\ &\times \int_{r}^{\infty}\int_{r'}^{\infty}\text{d}r_2\text{d}r_4\left(\frac{r_3}{r_2r_4}\right)^2\left(\rho(r_2)\delta(r_2-r_3)\right)\Biggr|^{r_3=r_4}_{r_3=0}.
\end{align}
We can now evaluate the limits,
\begin{align}
    \langle \chi(t)\chi(t')\rangle^{\rm iso} = &\frac{2k_BT^{\rm act}}{4\pi\zeta}\delta(t-t')\int_0^{\infty}\int_0^{\infty}\text{d}r\text{d}r'\frac{\partial\mathcal{E}}{\partial r}\frac{\partial\mathcal{E}}{\partial r'}\int_{r}^{\infty}\text{d}r_2\left(\frac{1}{r_2}\right)^2\rho(r_2)\int_{r'}^{\infty}\text{d}r_4\delta(r_2-r_4).
\end{align}
Evaluating the integral in $r_4$ now depends on $r'$.
If $r' < r_2$ we may apply the sifting property of the delta function.
If $r'>r_2$ the integral vanishes.
Therefore we apply the sifting property and gain a factor of the Heaviside step function:
\begin{align}
    \langle \chi(t)\chi(t')\rangle^{\rm iso} = &\frac{2k_BT^{\rm act}}{4\pi\zeta}\delta(t-t')\int_0^{\infty}\int_0^{\infty}\text{d}r\text{d}r'\frac{\partial\mathcal{E}}{\partial r}\frac{\partial\mathcal{E}}{\partial r'}\int_{r}^{\infty}\text{d}r_2\left(\frac{1}{r_2}\right)^2\rho(r_2)H(r_2-r').
\end{align}
If $r<r'$ then the integral will be zero until $r_2$ reaches $r'$.
Then we may absorb the step function into the limits of integration:
\begin{align}
    \langle \chi(t)\chi(t')\rangle^{\rm iso} = &\frac{2k_BT^{\rm act}}{4\pi\zeta}\delta(t-t')\int_0^{\infty}\int_0^{\infty}\text{d}r\text{d}r'\frac{\partial\mathcal{E}}{\partial r}\frac{\partial\mathcal{E}}{\partial r'}\int_{\text{max}(r,r')}^{\infty}\text{d}r_2\left(\frac{1}{r_2}\right)^2\rho(r_2).
\end{align}
Integration by parts results in:
\begin{align}
    \langle \chi(t)\chi(t')\rangle^{\rm iso} = &\frac{2k_BT^{\rm act}}{4\pi\zeta}\delta(t-t')\int_0^{\infty}\int_0^{\infty}\text{d}r\text{d}r'\frac{\partial\mathcal{E}}{\partial r}\frac{\partial\mathcal{E}}{\partial r'}\frac{\rho(\text{max}(r,r'))}{\text{max}(r,r')} \nonumber \\ &+\frac{2k_BT^{\rm act}}{4\pi\zeta}\delta(t-t')\int_0^{\infty}\int_0^{\infty}\text{d}r\text{d}r'\frac{\partial\mathcal{E}}{\partial r}\frac{\partial\mathcal{E}}{\partial r'}\int_{\text{max}(r,r')}^{\infty}\text{d}r_2\left(\frac{1}{r_2}\right)\frac{\partial \rho}{\partial r_2}.
\end{align}
The $\text{max}(r,r')$ functions are cumbersome to deal with as is.
We therefore divide the integrals over $r'$ into sections where $r'<r$ or $r'>r$:
\begin{align}
    \langle \chi(t)\chi(t')\rangle^{\rm iso} = &\frac{2k_BT^{\rm act}}{4\pi\zeta}\delta(t-t')\left(\int_0^{\infty}\int_0^{r}\text{d}r\text{d}r'\frac{\partial\mathcal{E}}{\partial r}\frac{\partial\mathcal{E}}{\partial r'}\frac{\rho(r)}{r} +\int_0^{\infty}\int_{r}^{\infty}\text{d}r\text{d}r'\frac{\partial\mathcal{E}}{\partial r}\frac{\partial\mathcal{E}}{\partial r'}\frac{\rho(r')}{r'}\right) \nonumber \\ &+\frac{2k_BT^{\rm act}}{4\pi\zeta}\delta(t-t')\int_0^{\infty}\int_0^{r}\text{d}r\text{d}r'\frac{\partial\mathcal{E}}{\partial r}\frac{\partial\mathcal{E}}{\partial r'}\int_{r}^{\infty}\text{d}r_2\left(\frac{1}{r_2}\right)\frac{\partial \rho}{\partial r_2}\nonumber \\ &+\frac{2k_BT^{\rm act}}{4\pi\zeta}\delta(t-t')\int_0^{\infty}\int_{r}^{\infty}\text{d}r\text{d}r'\frac{\partial\mathcal{E}}{\partial r}\frac{\partial\mathcal{E}}{\partial r'}\int_{r'}^{\infty}\text{d}r_2\left(\frac{1}{r_2}\right)\frac{\partial \rho}{\partial r_2}.
\end{align}
We now integrate the last line by parts in $r'$, directly evaluate the integral over $r'$ in the second line, and combine like terms.  
This results in:
\begin{align}
    \langle \chi(t)\chi(t')\rangle^{\rm iso} = &\frac{2k_BT^{\rm act}}{4\pi\zeta}\delta(t-t')\left(\int_0^{\infty}\int_0^{r}\text{d}r\text{d}r'\frac{\partial\mathcal{E}}{\partial r}\frac{\partial\mathcal{E}}{\partial r'}\frac{\rho(r)}{r} +\int_0^{\infty}\int_{r}^{\infty}\text{d}r\text{d}r'\frac{\partial\mathcal{E}}{\partial r}\frac{\partial\mathcal{E}}{\partial r'}\frac{\rho(r')}{r'}\right) \nonumber \\ &-\frac{2k_BT^{\rm act}}{4\pi\zeta}\delta(t-t')\mathcal{E}^{\rm in}\int_0^{\infty}\text{d}r\frac{\partial\mathcal{E}}{\partial r} \int_{r}^{\infty}\text{d}r_2\left(\frac{1}{r_2}\right)\frac{\partial \rho}{\partial r_2}\nonumber \\ &+\frac{2k_BT^{\rm act}}{4\pi\zeta}\delta(t-t')\int_0^{\infty}\text{d}r\frac{\partial\mathcal{E}}{\partial r}\int_{r}^{\infty}\text{d}r'\mathcal{E}\left(\frac{1}{r'}\right)\frac{\partial \rho}{\partial r'}.
\end{align}
We now integrate the second integral of the first line by parts and directly evaluate the $r'$ integral in the first term of the first line, which results in:
\begin{align}
    \langle \chi(t)\chi(t')\rangle^{\rm iso} = &\frac{4k_BT^{\rm act}}{4\pi\zeta}\delta(t-t')\int_0^{\infty}\text{d}r\frac{\partial\mathcal{E}}{\partial r}\mathcal{E}\frac{\rho(r)}{r}- \frac{4k_BT^{\rm act}}{4\pi\zeta}\delta(t-t')\mathcal{E}^{\rm in}\int_0^{\infty}\text{d}r\frac{\partial\mathcal{E}}{\partial r}\frac{\rho(r)}{r}  \nonumber \\ &-\frac{2k_BT^{\rm act}}{4\pi\zeta}\delta(t-t')\mathcal{E}^{\rm in}\int_0^{\infty}\text{d}r\frac{\partial\mathcal{E}}{\partial r} \int_{r}^{\infty}\text{d}r_2\left(\frac{1}{r_2}\right)\frac{\partial \rho}{\partial r_2}\nonumber \\ &+\frac{2k_BT^{\rm act}}{4\pi\zeta}\delta(t-t')\int_0^{\infty}\text{d}r\frac{\partial\mathcal{E}}{\partial r}\int_{r}^{\infty}\text{d}r'\mathcal{E}\left(\frac{1}{r'}\right)\frac{\partial \rho}{\partial r'}.
\end{align}
We now integrate both the second and third line by parts, then combine like terms, which results in:
\begin{align}
    \langle \chi(t)\chi(t')\rangle^{\rm iso} = &\frac{4k_BT^{\rm act}}{4\pi\zeta}\delta(t-t')\int_0^{\infty}\text{d}r\frac{\partial\mathcal{E}}{\partial r}\mathcal{E}\frac{\rho(r)}{r}- \frac{4k_BT^{\rm act}}{4\pi\zeta}\delta(t-t')\mathcal{E}^{\rm in}\int_0^{\infty}\text{d}r\frac{\partial\mathcal{E}}{\partial r}\frac{\rho(r)}{r}  \nonumber \\ &+\frac{2k_BT^{\rm act}}{4\pi\zeta}\delta(t-t')\left(\mathcal{E}^{\rm in}\right)^2\int_0^{\infty}\text{d}r\left(\frac{1}{r}\right)\frac{\partial \rho}{\partial r} - \frac{4k_BT^{\rm act}}{4\pi\zeta}\delta(t-t')\mathcal{E}^{\rm in}\int_0^{\infty}\text{d}r\mathcal{E} \left(\frac{1}{r}\right)\frac{\partial \rho}{\partial r}\nonumber \\ &+\frac{2k_BT^{\rm act}}{4\pi\zeta}\delta(t-t')\int_0^{\infty}\text{d}r\left(\mathcal{E}(r)\right)^2\left(\frac{1}{r}\right)\frac{\partial \rho}{\partial r}.
\end{align}
Now, as was assumed before, we note that each of the above integrals is non-zero only near the interface.
Then if $R$ is much larger than the width of the interface, we may rewrite each integral as:
\begin{align}
    \langle \chi(t)\chi(t')\rangle^{\rm iso} = &\frac{4k_BT^{\rm act}}{4R\pi\zeta}\delta(t-t')\int_0^{\infty}\text{d}r\frac{\partial\mathcal{E}}{\partial r}\mathcal{E}\rho(r)- \frac{4k_BT^{\rm act}}{4R\pi\zeta}\delta(t-t')\mathcal{E}^{\rm in}\int_0^{\infty}\text{d}r\frac{\partial\mathcal{E}}{\partial r}\rho(r) \nonumber \\ &+\frac{2k_BT^{\rm act}}{4R\pi\zeta}\delta(t-t')\left(\mathcal{E}^{\rm in}\right)^2\int_0^{\infty}\text{d}r\frac{\partial \rho}{\partial r} - \frac{4k_BT^{\rm act}}{4R\pi\zeta}\delta(t-t')\mathcal{E}^{\rm in}\int_0^{\infty}\text{d}r\mathcal{E} \frac{\partial \rho}{\partial r}\nonumber \\ &+\frac{2k_BT^{\rm act}}{4R\pi\zeta}\delta(t-t')\int_0^{\infty}\text{d}r\left(\mathcal{E}(r)\right)^2\frac{\partial \rho}{\partial r} + \mathcal{O}(1/R^2).
\end{align}
The first integral may be rewritten as:
\begin{align}
    \langle \chi(t)\chi(t')\rangle^{\rm iso} = &\frac{2k_BT^{\rm act}}{4R\pi\zeta}\delta(t-t')\int_0^{\infty}\text{d}r\frac{\partial}{\partial r}\left[\left(\mathcal{E}\right)^2\right]\rho(r)- \frac{4k_BT^{\rm act}}{4R\pi\zeta}\delta(t-t')\mathcal{E}^{\rm in}\int_0^{\infty}\text{d}r\frac{\partial\mathcal{E}}{\partial r}\rho(r) \nonumber \\ &+\frac{2k_BT^{\rm act}}{4R\pi\zeta}\delta(t-t')\left(\mathcal{E}^{\rm in}\right)^2\int_0^{\infty}\text{d}r\frac{\partial \rho}{\partial r} - \frac{4k_BT^{\rm act}}{4R\pi\zeta}\delta(t-t')\mathcal{E}^{\rm in}\int_0^{\infty}\text{d}r\mathcal{E} \frac{\partial \rho}{\partial r}\nonumber \\ &+\frac{2k_BT^{\rm act}}{4R\pi\zeta}\delta(t-t')\int_0^{\infty}\text{d}r\left(\mathcal{E}(r)\right)^2\frac{\partial \rho}{\partial r} + \mathcal{O}(1/R^2).
\end{align}
Then the above integrals may be compactly rewritten as:
\begin{align}
    \langle \chi(t)\chi(t')\rangle^{\rm iso} = &\frac{2k_BT^{\rm act}}{4R\pi\zeta}\delta(t-t')\int_0^{\infty}\text{d}r\frac{\partial}{\partial r}\left[\left(\mathcal{E}\right)^2\rho(r)\right]- \frac{4k_BT^{\rm act}}{4R\pi\zeta}\delta(t-t')\mathcal{E}^{\rm in}\int_0^{\infty}\text{d}r\frac{\partial }{\partial r}\left[\mathcal{E}\rho\right] \nonumber \\ &+\frac{2k_BT^{\rm act}}{4R\pi\zeta}\delta(t-t')\left(\mathcal{E}^{\rm in}\right)^2\int_0^{\infty}\text{d}r\frac{\partial \rho}{\partial r}  + \mathcal{O}(1/R^2).
\end{align}
All integrals may now be evaluated directly and simplify to:
\begin{align}
    \langle \chi(t)\chi(t')\rangle^{\rm iso} = &\frac{2k_BT^{\rm act}}{4R\pi\zeta}\delta(t-t')\rho^{\rm sat}\left(\mathcal{E}^{\rm sat} - \mathcal{E}^{\rm in}\right)^2  + \mathcal{O}(1/R^2).
\end{align}
$\langle \chi(t)\chi(t')\rangle^{\rm aniso}_1$ and $\langle \chi(t)\chi(t')\rangle^{\rm aniso}_2$ can be solved for in a similar manner as $\langle \chi(t)\chi(t')\rangle^{\rm iso}$, and we find that $\langle \chi(t)\chi(t')\rangle^{\rm aniso}_1$ is \textit{zero} while $\langle \chi(t)\chi(t')\rangle^{\rm aniso}_2$ is:
\begin{equation*}
    \langle \chi(t)\chi(t')\rangle^{\rm aniso}_2 = \frac{4k_BT^{\rm act}}{4\pi D_R\zeta\ell_o^2R^2}\Delta\mathcal{E}\delta(t-t')\left[\int_0^{\infty}dr\frac{\kappa_3(\rho)}{\overline{U}(\rho)}\left(\frac{\partial \rho}{\partial r}\right)^2\left(\mathcal{E}^{\rm in}-\mathcal{E}\right) \right] + O\left(\frac{1}{R^3}\right),
\end{equation*}
which implies that the most significant contributions to the anisotropic noise correlations are $\mathcal{O}(1/R^2)$ or smaller.
Therefore,
\begin{align}
    \langle \chi(t)\chi(t')\rangle = &\frac{2k_BT^{\rm act}}{4R\pi\zeta}\delta(t-t')\rho^{\rm sat}\left(\mathcal{E}^{\rm sat} - \mathcal{E}^{\rm in}\right)^2  + \mathcal{O}(1/R^2).
\end{align}
We now define $\tilde{\chi}$ as:
\begin{equation}
    \tilde{\chi}(t) = -\frac{\zeta_{\rm eff}}{R\left(\rho^{\rm sat}-\rho^{\rm in}\right)\left(\mathcal{E}^{\rm sat} - \mathcal{E}^{\rm in}\right)}\chi(t),
\end{equation}
which will have zero mean and correlations:
\begin{equation}
    \langle\tilde{\chi}(t)\tilde{\chi}(t') \rangle = 2k_BT^{\rm act}\zeta_{\rm eff}\delta(t-t').
\end{equation}
\subsection{\label{sec:potential}Effective Potential}
\begin{figure}
	\centering
	\includegraphics[width=0.75\textwidth]{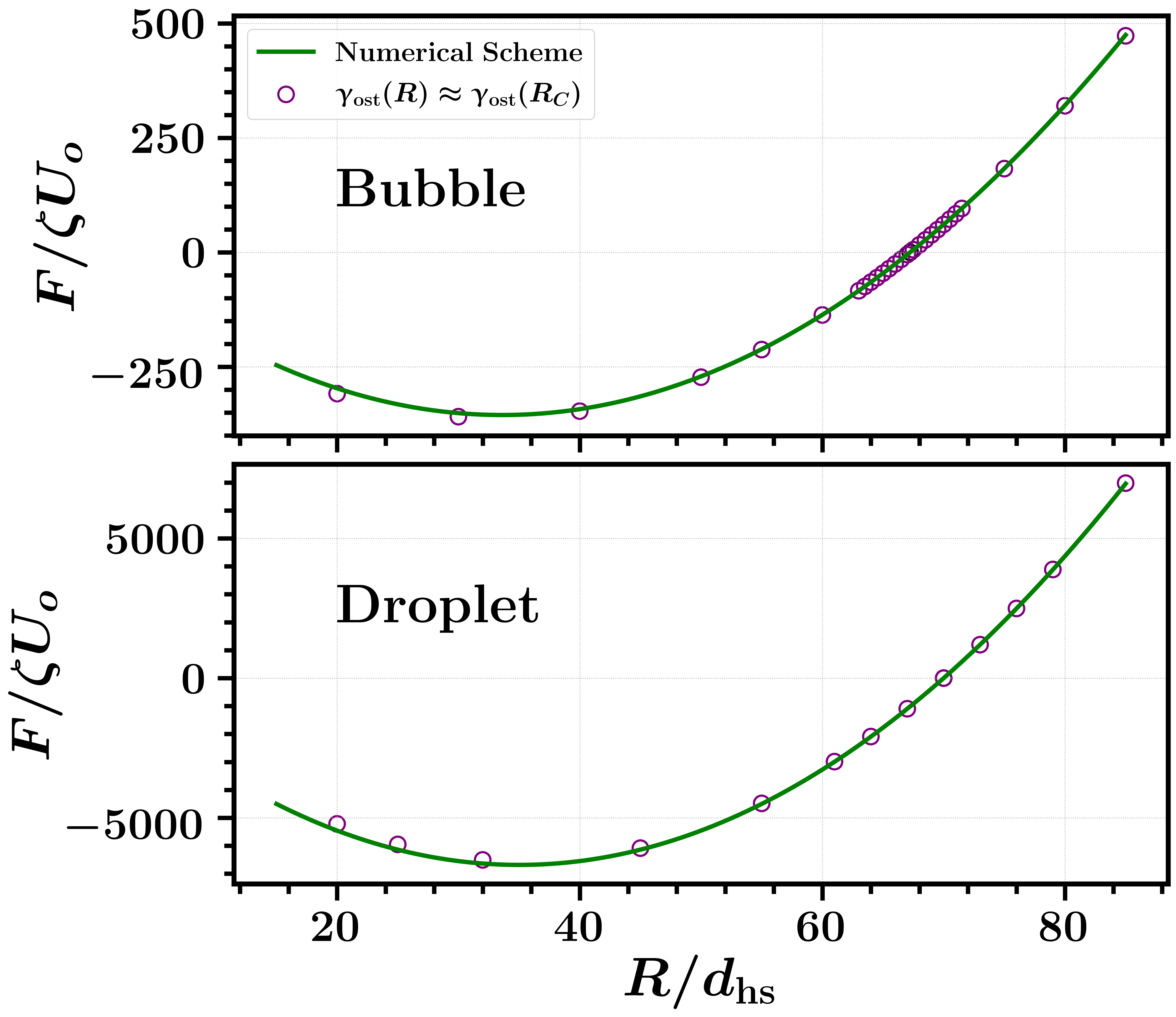}
	\caption{\protect\small{{Deterministic force as defined by the negative derivative in $R$ of Eq.~\eqref{seq:Udef} as evaluated via the numerical scheme described above or treating coefficients as constant in $R$. The bubble forces were evaluated for a system at $\ell_o/d_{\rm hs} = 70.7$ and $R_C/d_{\rm hs} = 67$. The droplet forces were evaluated for a system at $\ell_o/d_{\rm hs} = 70.7$ and $R_C/d_{\rm hs} = 70$. }}}
	\label{fig:approxcompare}
\end{figure}

Cates and Nardini~\cite{Cates2023ClassicalSeparation} have argued that the nucleation of active matter must have a defineable potential barrier because Eqs.~\eqref{seq:drdt} and ~\eqref{seq:chivar} satisfy an effective fluctuation dissipation theorem.
This nucleation potential is defined such that $F^D(R) = -\partial U(R)/\partial R$:
\begin{equation}
    U(R) = \int \text{d}R 8\pi R^2\upgamma_{\rm ost}\left[\frac{1}{R}-\frac{1}{R_C}\right],\label{seq:Udef}
\end{equation}
where $F^D(R)$ is defined by Eq.~\eqref{seq:Fdetermindef}.
In general, evaluating this integral analytically is difficult because $\upgamma_{\rm ost}$ is implicitly a function of $R$. 
Cates and Nardini proposed \textit{neglecting} the $R$ dependence of $\upgamma_{\rm ost}$ and treating it as a constant:
\begin{equation}
    U(R) =  8\pi \upgamma_{\rm ost}\left[\frac{R^2}{2} -\frac{R^3}{3R_C} \right].\label{seq:Uapprox}
\end{equation}
We verify the validity of this approximation by evaluating the \textit{integrand} of Eq.~\eqref{seq:Udef} both numerically and via the approximation proposed by Cates and Nardini. 
We follow the same numerical scheme as discussed in Section~\ref{sec:coexistnumerics}, except now $R$ is taken as an input and the two unknown values solved for are $\rho^{\rm in}$ and $\rho^{\rm out}$. 
The result of calculating $-\partial U(R)/\partial R$ both from our numerical scheme and the approximation of Cates and Nardini is plotted in Fig.~\ref{fig:approxcompare}.

The difference between ignoring and accounting for the $R$ dependence of $\upgamma_{\rm ost}$ is marginal for both bubbles and droplets, as shown by Fig.~\ref{fig:approxcompare}.
Therefore evaluating coefficients at $R=R_C$ and treating them as constant is a robust approximation, justifying the use of Eq.~\eqref{seq:Uapprox}.
We can then define a potential barrier as $U(R_C)$, which is given by:
\begin{equation}
    U(R_C) = \frac{4}{3}\pi \upgamma_{\rm ost}R_C^2.\label{seq:potentialbarrier}
\end{equation}
We use Eq.~\eqref{seq:potentialbarrier} to calculate the nucleation barriers presented in the main text.

We may also substitute $R_C$ as defined by Eq.~\eqref{seq:critradiuspseudopotential} into Eq.~\eqref{seq:Uapprox} to find a clean separation between terms proportional to surface area and volume:
\begin{equation}
    U(R) = 4\pi\upgamma_{\rm ost}R^2 - \frac{4}{3}\pi R^3 \mathcal{V}\label{seq:nucbarrierintuit},
\end{equation}
where $\mathcal{V}$ has units of energy per volume and has form:
\begin{equation}
    \mathcal{V} = \frac{\mathcal{P}(\rho^{\rm sat})\Delta\rho}{\rho^{\rm sat}} - \frac{\Delta\Phi\Delta\rho}{\rho^{\rm sat}\Delta\mathcal{E}}.
\end{equation}
\subsection{Becker-D\"{o}ring Nucleation Rate Calculation}
\begin{figure}
	\centering
	\includegraphics[width=0.75\textwidth]{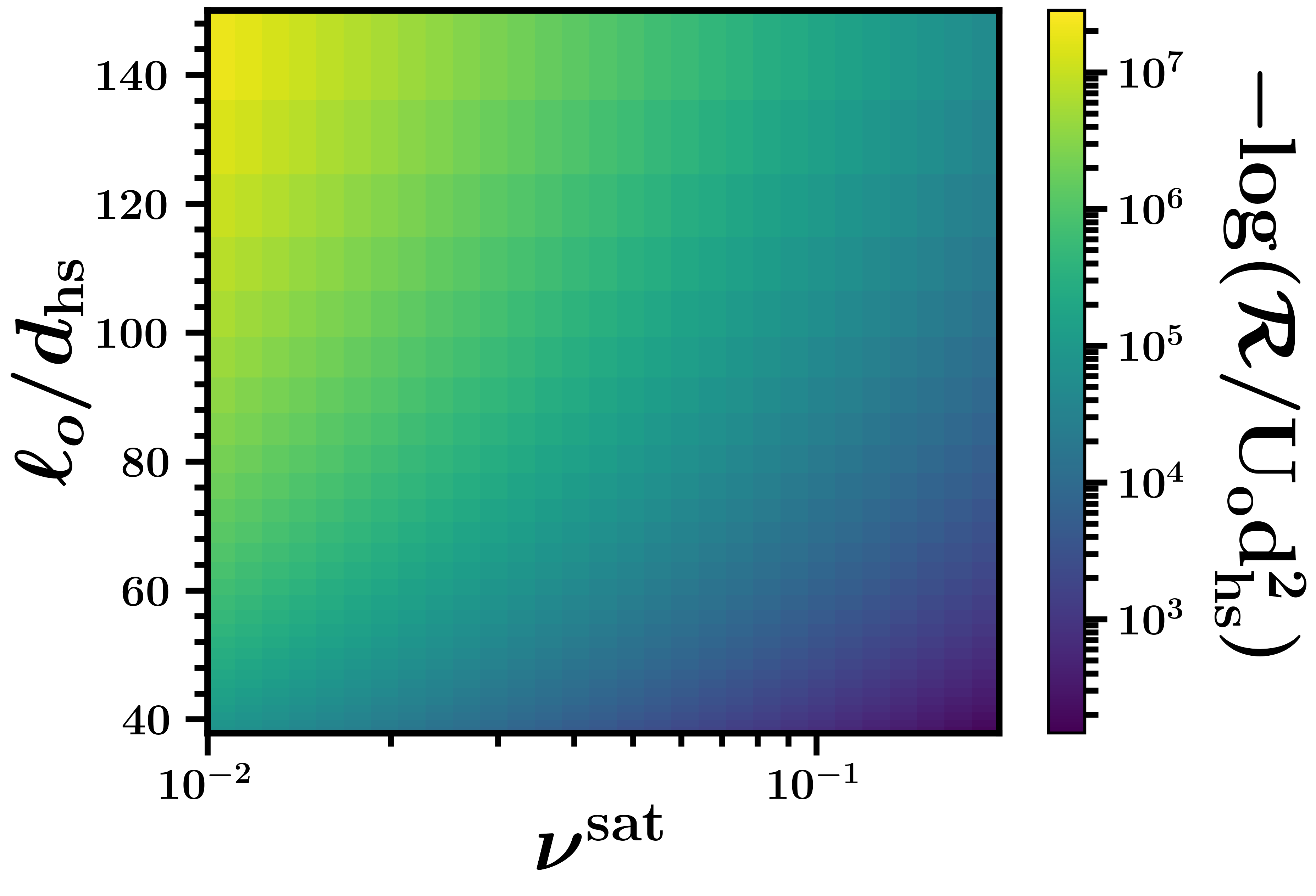}
	\caption{\protect\small{{Nucleation rate of droplets as a function of activity and supersaturation, as predicted by Eq.~\eqref{seq:nucleationrate} }}}
	\label{fig:nucleationratemap}
\end{figure}

Using the barriers calculated in the previous Section we can calculate the nucleation rate outlined in Appendix A of the main text, which we reproduce here as:
\begin{equation}
     \mathcal{R} = \left(\rho^{\rm sat}\right)^2 \pi U_o \sqrt{\frac{\upgamma_{\rm ost}}{2\pi(\rho^{\rm in})^2 k_BT^{\rm act}}} \text{exp}\left[-\frac{U(\ell^*)}{k_BT^{\rm act}}\right].\label{seq:nucleationrate}
\end{equation}
$\mathcal{R}$ is calculated for ABP droplets and the results are plotted in Fig.~\ref{fig:nucleationratemap}.
At constant $\nu^{\rm sat}$, while $\upgamma_{\rm ost}/k_BT^{\rm act}$ scales roughly linearly with run length, the absolute value of $\rho^{\rm sat}$ also decreases.
At constant $\ell_o$, $\rho^{\rm sat}$ increases linearly with $\nu^{\rm sat}$.
These dependencies on activity and saturation within the prefactor are far weaker than the exponential dependence of a barrier which depends strongly on both activity and saturation. 
Thus the qualitative behavior of Fig.~\ref{fig:nucleationratemap} is dominated by the strength of the nucleation barrier.
 \newpage
\section{\label{sec:eq}Equilibrium Nucleus Coexistence}

In this Section we first derive the finite-sized coexistence of an equilibrium system through a constrained free energy minimization. We then show that a mechanical theory constructed in the same manner as Section~\ref{sec:static} is equivalent to equilibrium nucleation theory when the underlying dynamics are passive. For brevity we will skip many steps of the derivation as this procedure will be entirely analogous to those outlined in the previous two Sections.

\subsection{\label{sec:thermo}Thermodynamics of Classical Nucleation}
We consider a nucleus of radius $R$ and density $\rho^{\rm in}$ coexisting with a parent phase of density $\rho^{\rm out}$.
The system has an overall density of $\rho^{\rm sat}$ and volume $V$.
The Helmholtz free energy of this system is then given by:
\begin{equation}
    \mathcal{F} = \frac{4}{3}\pi R^3f(\rho^{\rm in}) + \left(V - \frac{4}{3}\pi R^3\right)f(\rho^{\rm out}) + 4\pi R^2\upgamma\label{seq:free-energy},
\end{equation}
where $\upgamma$ is the surface tension and $f(\rho)$ is a bulk Helmholtz free energy density of a system with uniform density $\rho$.
The total number of particles in the system is conserved with:
\begin{equation}
\label{seq:Nconservation}
    \rho^{\rm in}\frac{4}{3}\pi R^3 + \rho^{\rm out}\left(V - \frac{4}{3}\pi R^3\right) = \rho^{\rm sat}V.
\end{equation}
Our free energy has dependencies on three adjustable internal variables, $\rho^{\rm in}$, $\rho^{\rm out}$, and $R$. 
The properties associated with a critical nucleus can be found by performing a constrained extremization with respect to $\rho^{\rm in}$, $\rho^{\rm out}$, and $R$ and looking for solutions where the radius is finite and not macroscopic. 
We construct the following Lagrangian:
\begin{equation}
    \mathcal{W} = \mathcal{F}(\rho^{\rm in}, \rho^{\rm out}, R) - \lambda\left(\rho^{\rm in}\frac{4}{3}\pi R^3 + \rho^{\rm out}\left(V - \frac{4}{3}\pi R^3\right) - \rho^{\rm sat}V\right),
\end{equation}
where $\lambda$ is a Lagrange multiplier. 
We extremize $\mathcal{W}$ by setting its partial derivatives with respect to $\rho^{\rm in}$, $\rho^{\rm out}$, $\lambda$, and $R$ (and evaluating these variables at their unknown critical values) to zero:
\begin{subequations}
    \label{seq:Wextremize}
    \begin{align}
        0 &= \frac{\partial \mathcal{W}}{\partial \rho^{\rm in}}\Bigr|_{\rho^{\rm in} = \rho^{\rm in*}} = \frac{\partial f}{\partial \rho}\Bigr|_{\rho = \rho^{\rm in^*}} - \lambda^*\label{seq:muinWpartialderiv}, \\  
        0 &= \frac{\partial \mathcal{W}}{\partial \rho^{\rm out}}\Bigr|_{\rho^{\rm out} = \rho^{\rm out*}} = \frac{\partial f}{\partial \rho}\Bigr|_{\rho = \rho^{\rm out^*}} - \lambda^*\label{seq:muoutWpartialderiv}, \\ 
        0 &= \frac{\partial \mathcal{W}}{\partial R}\Bigr|_{R=R_C} = f(\rho^{\rm in}) - f(\rho^{\rm out}) - \lambda^*\left(\rho^{\rm in} - \rho^{\rm out}\right) + \frac{2\upgamma}{R_C}\label{seq:pressWpartialderiv}, \\
        0 &= \frac{\partial\mathcal{W}}{\partial \lambda}\Bigr|_{\lambda = \lambda^*} = (\rho^{\rm in} - \rho^{\rm out})\frac{4}{3}\pi R^3 + (\rho^{\rm out} - \rho^{\rm sat})V\label{seq:massconsWpartial}. 
    \end{align}
\end{subequations}
Here $\rho^{\rm in*}$, $\rho^{\rm out*}$, $\lambda^*$, and $R_C$ denote the extremized values of $\rho^{\rm in}$, $\rho^{\rm out}$, $\lambda$, and $R$, respectively.
Defining the chemical potential as $\mu(\rho)\equiv \partial f/\partial \rho$ allows us to appreciate that the extremized value of the Lagrange multiplier, $\lambda^*$, is simply the coexistence chemical potential with $\lambda^* \equiv \mu^{\rm coexist} = \mu(\rho^{\rm in*}) = \mu(\rho^{\rm out*})$ by combining Eqs.~\eqref{seq:muinWpartialderiv}~and~\eqref{seq:muoutWpartialderiv}.
By defining the thermodynamic pressure, $p(\rho) \equiv -f(\rho) + \rho\mu(\rho)$, we can express the coexistence criteria with the following three equations:
\begin{subequations}
\label{seq:lagrange}
    \begin{align}
        \mu(\rho^{\rm in*}) &= \mu(\rho^{\rm out*}), \\
        p(\rho^{\rm in*}) &= p(\rho^{\rm out*}) + \frac{2\upgamma}{R_C}, \\
        \rho^{\rm sat}V &= \rho^{\rm in*}\frac{4}{3}\pi R_C^3 + \rho^{\rm out*}\left(V - \frac{4}{3}\pi R_C^3\right).
    \end{align}
\end{subequations}
Equations~\eqref{seq:lagrange} allow for the determination of the unknown densities and critical radius. 

As we are interested in solutions in which the radius takes on large but non-macroscopic values, we consider the limit of $R^3 \ll V$.
In this limit, the particle conservation equation [Eq.~\eqref{seq:Nconservation}] reduces to $\rho^{\rm out} \approx \rho^{\rm sat}$ and the nucleus in contact with a infinite particle reservoir with chemical potential $\mu(\rho^{\rm sat}) \equiv \mu^{\rm sat}$. 
The Lagrangian then takes the following form:
\begin{equation}
    \mathcal{W} \approx \frac{4}{3}\pi R^3\left(w(\rho^{\rm in}) - w(\rho^{\rm sat})\right) + 4\pi R^2\upgamma,\label{seq:nucbarriereq}
\end{equation}
where $w(\rho) \equiv f(\rho) - \mu^{\rm sat}\rho$ is the grand potential energy density (also equivalent to the negative pressure, $-p$) of a system with density $\rho$ in contact with a particle reservoir with chemical potential $\mu^{\rm sat}$.
We can plainly identify that the simplified Lagrangian contains two familiar terms in classical nucleation theory: a \textit{negative volumetric term} proportional to the grand potential energy density difference between the nucleus and the parent phase and \textit{positive areal term} proportional to the surface tension.
The volumetric driving term defined in the main text [i.e.,~${\mathcal{W} = -\frac{4}{3}\pi R^3\mathcal{V} + 4\pi R^2\upgamma}$] has several equivalent forms ${\mathcal{V} \equiv -\Delta w \equiv \Delta p \equiv -\Delta f + \mu^{\rm sat}\Delta\rho}$ where we have again used the notation of ${\Delta a = a(\rho^{\rm in}) - a(\rho^{\rm sat})}$.
Extremization of $\mathcal{W}$ with respect to $R$ and $\rho^{\rm in}$ results in the following coexistence criteria:
\begin{subequations}
\label{seq:thermocoexist}
\begin{align}
    \mu(\rho^{\rm in*}) &= \mu^{\rm sat}, 
    \\ p(\rho^{\rm in*}) &= p(\rho^{\rm sat}) + \frac{2\upgamma}{R_C}.
\end{align}
Note that these criteria could also have been obtained by directly taking the limit of $R_C \ll V$ in Eqs.~\eqref{seq:lagrange}.
The value of the Lagrangian evaluated at the critical radius and interior density is the \textit{nucleation barrier}, $F^{\dag} \equiv \mathcal{W}(\rho^{\rm in}=\rho^{\rm in*}, R=R_C)$
Finally, Eqs.~\eqref{seq:thermocoexist} can be trivially rearranged to isolate $R_C$ with:
\begin{align}
    R_C &= \frac{2\upgamma}{ w(\rho^{\rm sat}) - w(\rho^{\rm in*})}.
\end{align}
\end{subequations}
We note that in the main text and for most of this SI we denote $\rho^{\rm in*}$ simply as $\rho^{\rm in}$ for notational convenience.
The stability of the solutions to Eqs.~\eqref{seq:thermocoexist} can be determined by computing the Hessian matrix of $\mathcal{W}$ and evaluating it at the solutions.
A positive surface energy and volumetric driving force for growth will generally result in the critical nucleus representing a unstable fixed point. 

\subsection{\label{sec:fluctuatinghydroeq}Equilibrium Density Field} 
We consider particles which follow an overdamped Langevin equation with an additional translational Brownian stochastic noise that satisfies the fluctuation-dissipation theorem.
The overdamped Langevin equation for the $i$th particle of this system is:
\begin{align}
    \Dot{\mathbf{r}}_i = \frac{1}{\zeta}\sum_{j\neq i}^{N}\mathbf{F}_{ij} + \bm{\eta}_i(t),\label{seq:eqparticleequations}
\end{align}
where $\bm{\eta}_i$ is the translational Brownian noise with zero average and variance $\langle \bm{\eta}_i(t)\bm{\eta}_j(t') = 2(k_BT/\zeta)\delta(t-t')\delta_{ij}\mathbf{I}\rangle$.
For this system to be passive, we require that the interparticle forces $\mathbf{F}_{ij}$ are reciprocal, i.e. $\mathbf{F}_{ij} = -\mathbf{F}_{ji}$.
We assume that the interparticle forces, overall system density, and temperature are such that the system is phase separated. 
Following the Irving-Kirkwood-Noll~\cite{Irving1950,Lehoucq2010} procedure for a system described by Eq.~\eqref{seq:eqparticleequations} results in the following density field dynamics at the mean-field level:
\begin{subequations}
\label{seq:eqhydro}
\begin{align}
    \frac{\partial\rho}{\partial t} &= -\boldsymbol{\nabla} \cdot \mathbf{J},\  \label{seq:continuityeq} \\ \mathbf{J} &= \frac{1}{\zeta}\boldsymbol{\nabla}\cdot \left(\bm{\sigma}^C + \bm{\sigma}^B\right),\ \\ 
    \bm{\sigma}^C &= \left[-p_C(\rho) + \rho\kappa(\rho)\nabla^2\rho + \frac{\left(\kappa(\rho)+ \rho\frac{\partial\kappa}{\partial \rho}\right)}{2}|\boldsymbol{\nabla}\rho|^2\right]\mathbf{I}  - \kappa(\rho)\boldsymbol{\nabla}\rho\boldsymbol{\nabla}\rho,\  \label{eq:cstress} \\
    \bm{\sigma}^B &= -k_BT\rho\mathbf{I},\  \label{seq:idealgasstress}
\end{align}
\end{subequations}
where $\bm{\sigma}^B$ is the ``ideal gas'' stress generated by the stochastic translational Brownian noise (absent in our treatment of athermal active matter) and we define $p(\rho) = p_C(\rho) + k_BT\rho$. 
We also note that in this model of equilibrium dynamics the stress tensor is fully local.

Within these passive dynamics we may identify ${\kappa_1(\rho)=\rho\kappa(\rho)}$, ${\kappa_2(\rho)=1/2\left(\rho\partial\kappa/\partial \rho + \kappa(\rho)\right)}$, ${\kappa_3(\rho)=-\kappa(\rho)}$, and ${\kappa_4(\rho)=0}$. 
This result was first identified by Korteweg~\cite{Korteweg1904}, and these relations have been demonstrated as equivalent to assuming variational dynamics of some Ginzburg-Landau free energy functional~\cite{Shang2011FluctuatingWaves}.
In this case Eq.~\eqref{seq:epsilondef} reduces to:
\begin{equation}
    \frac{\partial^2\mathcal{E}}{\partial \rho^2} = -\frac{1}{\rho}\frac{\partial\mathcal{E}}{\partial\rho}
\end{equation}

\noindent In this case, (within inconsequential constants of integration,) the identity of the pseudovariable becomes $\mathcal{E}\sim 1/\rho = v$, the molecular volume.
By substituting Eq.~\eqref{seq:eqhydro} into Eq.~\eqref{seq:mechstdef}, we may simplify the mechanical surface tension to:
\begin{equation}
    \upgamma_{\rm mech} = \int_{0}^{\infty}dr \mathcal{G}(\rho) = \int_{0}^{\infty}dr \kappa(\rho)\left(\frac{\partial \rho}{\partial r}\right)^2.
\end{equation}
By substituting $\mathcal{E} = 1/\rho$ into Eq.~\eqref{seq:osttensiondef}, we may simplify the Ostwald tension to:
\begin{align}
    \upgamma_{\rm ost} &= \frac{v^{\rm sat}\left(\rho^{\rm sat}-\rho^{\rm in}\right)}{\left(v^{\rm sat} - v^{\rm in}\right)}\left[\int_0^{\infty}\left(-\rho\kappa(\rho)\frac{1}{\rho^2}\left(\frac{\partial\rho}{\partial r}\right)^2 + \kappa(\rho)\left(\frac{\partial \rho}{\partial r}\right)^2\left(\frac{1}{\rho}-v^{\rm in}\right)\right)\text{d}r\right], \nonumber \\ &= -\frac{v^{\rm sat}v^{\rm in}\left(\rho^{\rm sat}-\rho^{\rm in}\right)}{\left(v^{\rm sat} - v^{\rm in}\right)}\left[\int_0^{\infty}\kappa(\rho)\left(\frac{\partial\rho}{\partial r}\right)^2\text{d}r\right], \nonumber \\ &= -\frac{\left(v^{\rm in}\rho^{\rm sat}-1\right)}{\left(1 - \rho^{\rm sat}v^{\rm in}\right)}\left[\int_0^{\infty}\kappa(\rho)\left(\frac{\partial\rho}{\partial r}\right)^2\text{d}r\right],\nonumber \\ &= \int_0^{\infty}\kappa(\rho)\left(\frac{\partial\rho}{\partial r}\right)^2\text{d}r = \upgamma_{\rm mech},
\end{align}
where we have used $v = 1/\rho$.
Therefore the equilibrium Ostwald tension is identical to the equilibrium mechanical surface tension.

We now discuss the equilibrium relationship between $\mathcal{E}$ and the AMB+ pseudovaribale $\psi$ in order to avoid any potential confusion. 
As discussed in Section 1.7, AMB+ can be mapped to the divergence of some dynamic stress tensor $\bm{\Sigma}$, with the mapping dependent on the choice of mobility $M[\rho]$. 
Choosing $M[\rho] = \rho$ and setting the nonequilibrium AMB+ constants $(\lambda,\mathcal{Z})$ will result in a $\bm{\Sigma}$ that exactly recovers the Korteweg stress tensor from Eq.~\eqref{seq:eqhydro}. 
This mapping also results in $\psi \sim \rho$. 
One may also deduce that $\psi \sim \rho$ from Section 1.7 after noting that the choice of $M[\rho] = \rho$ is an instance of the $M[\rho] \sim 1/\mathcal{E}$ case already discussed in detail.

\subsection{\label{sec:phasecoexisteq}Equivalence of Thermodynamic and Mechanical Perspectives}
We now demonstrate that the mechanically-derived generalized critical nucleus criteria [Eqs.~\eqref{seq:laplacerestate} and~\eqref{seq:coexist2ost}] applied to passive systems recover the criteria obtained through the variational method described in Section~\ref{sec:thermo}. 
The pseudovariable variable for passive systems~\cite{Aifantis1983TheRule, Omar2023b} is simply $\mathcal{E} \sim v$ where $v = 1/\rho$ is the specific volume and the equivalence of the Ostwald and mechanical tensions $\upgamma_{\rm ost} = \upgamma_{\rm mech} \equiv \upgamma$ results in the generalized criteria now taking the following form for passive systems:
\begin{subequations}
\label{seq:eqcoexist}
\begin{equation}
    p(\rho^{\rm in}) - p(\rho^{\rm sat}) = \frac{2\upgamma}{R_C},\label{seq:laplaceeq}
    \end{equation}
    \begin{equation}
    \int_{v^{\rm in}}^{v^{\rm sat}} \left[p(v) - p(v^{\rm sat})\right]\text{d}v +\frac{2\upgamma v^{\rm in}}{R_C} = 0,\label{seq:coexist2osteq}
    \end{equation}
\end{subequations}
where we have also recognized that the dynamic pressure is simply the static thermodynamic pressure. 
Eq.~\eqref{seq:laplaceeq} is the familiar Laplace pressure difference criterion that was recovered in Section~\ref{sec:thermo} using thermodynamics.
All that is required to establish the equivalence of the criteria obtained from our mechanical approach [Eqs.~\eqref{seq:eqcoexist}] and the thermodynamically obtained criteria [Eqs.~\eqref{seq:thermocoexist}] is to demonstrate the equivalence of Eq.~\eqref{seq:coexist2osteq} with equality of chemical potential. 
We first eliminate the surface tension from Eq.~\eqref{seq:coexist2osteq} by using the Laplace pressure difference, resulting in:
\begin{equation}
    \int_{v^{\rm in}}^{v^{\rm sat}} p(v) dv - \left( p(v^{\rm sat})v^{\rm sat} - p(v^{\rm in})v^{\rm in} \right) = 0.
\end{equation}
Integration by parts demonstrates that the left-hand-side of the above equation is simply $\int_{p^{\rm in}}^{p^{\rm sat}} v dp$, which is equivalent to  $\int_{\mu^{\rm in}}^{\mu^{\rm sat}}d\mu$ via the Gibbs-Duhem relation, $vdp = d\mu$. 
We thus arrive at the expected:
\begin{equation}
\mu(\rho^{\rm in}) = \mu^{\rm sat},
\end{equation}
demonstrating the equivalence of our mechanically derived criteria with those derived from thermodynamic considerations. 

In Section~\ref{sec:potential} we identified a general expression for the \textit{volumetric} contribution to the effective nucleation barrier:
\begin{equation}
    \mathcal{V} = \frac{\mathcal{P}(\rho^{\rm sat})\Delta\rho}{\rho^{\rm sat}} - \frac{\Delta\Phi\Delta\rho}{\rho^{\rm sat}\Delta\mathcal{E}}.
\end{equation}
With  $\Phi = \int p d\mathcal{E}$ and again noting $\upgamma_{\rm ost} = \upgamma_{\rm mech} \equiv \upgamma$ and $\mathcal{E} \sim v$, one can straightforwardly show:
\begin{equation}
    \mathcal{V} = \mu^{\rm sat}\Delta\rho - \Delta f = -\Delta w.
\end{equation}
Therefore the mechanical theory recovers the expected thermodynamic driving force for nucleation found in Eq.~\eqref{seq:nucbarriereq}.

\subsection{\label{sec:vdw}Application to van der Waals Fluid}
Let us consider an equilibrium system where $\kappa(\rho)$ is a constant and $p(\rho)$ is given by the van der Waals equation of state:
\begin{equation}
    \left(p + \rho^2 a\right)\left(\frac{1}{\rho} - b\right) = k_BT,\label{seq:vdweos}
\end{equation}
where $a$ is a measure of the average attraction between molecules and $b$ is the volume excluded by a single particle.
The critical point of this system is given by:
\begin{equation}
    p^C = \frac{a}{27b^2},\quad k_BT^C = \frac{8a}{27b},\quad \rho^C = \frac{1}{3b}.\label{seq:vdwcrit}
\end{equation}
We now non-dimensionalize variables by their value at the critical point:
\begin{equation}
    p' = \frac{p}{p^C},\quad T' =\frac{T}{T^C},\quad \rho' = \frac{\rho}{\rho^C}.\label{seq:vdwprime}
\end{equation}
Then the dimensionless van der Waals equation of state is given by:
\begin{equation}
    p' = \frac{8T'}{\frac{3}{\rho'}-1} - 3(\rho')^2.\label{seq:vdwND}
\end{equation}
Equation~\eqref{seq:vdwprime} results in the following dimensionless form of $\kappa$:
\begin{equation}
    \kappa' = \kappa \frac{\left(\rho^C\right)^{8/3}}{p^C} 
\end{equation}
We note that this non-dimensionalized form of the van der Waals equation of state leads to two different natural energy scales: $k_BT^C$ and $p^C/\rho^C$. 
When calculating the mechanical surface tension of a phase separated Van der Waals fluid via Eq.~\eqref{seq:mechstdef}, our above non-dimensionalization results in the latter energy scale. 
In order to convert it to have units of $k_BT^C$ one can simply multiply the final result by a factor of $3/8$.

\begin{figure}
	\centering
	\includegraphics[width=0.75\textwidth]{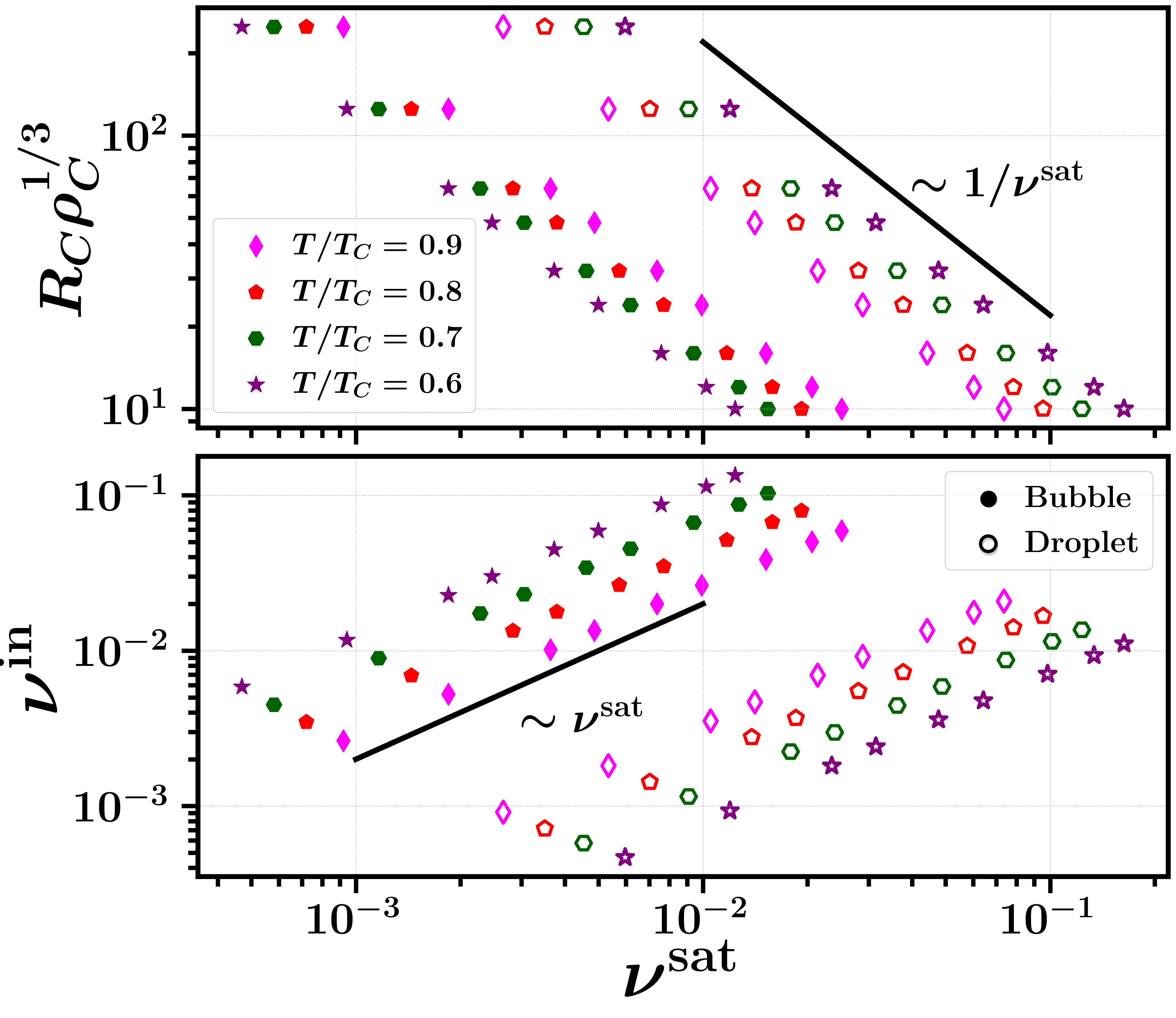}
	\caption{\protect\small{{Critical radius and relative change in nucleus density ($\nu^{\rm in} \equiv |\rho^{\rm in} - \rho^{\rm gas, \rm liq}| / \rho^{\rm gas, \rm liq}$) as a function of supersaturation ($\nu^{\rm sat} \equiv |\rho^{\rm sat} - \rho^{\rm gas, \rm liq}| / \rho^{\rm gas, \rm liq}$).}}}
	\label{fig:fig2equilibrium}
\end{figure}

\begin{figure}
	\centering
	\includegraphics[width=0.75\textwidth]{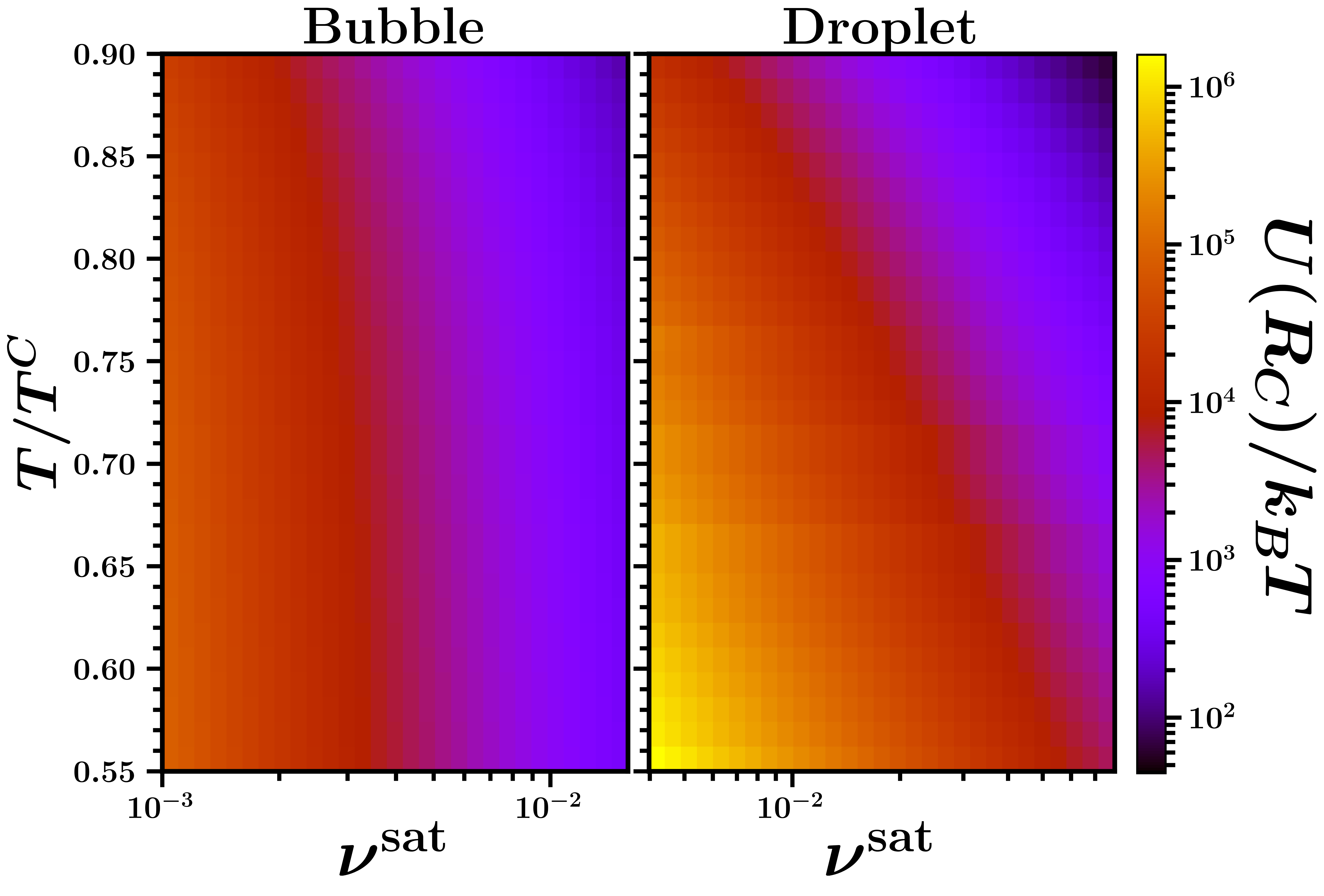}
	\caption{\protect\small{{Barrier $U(R_C)$ relative to thermal energy as a function of temperature and supersaturation.}}}
	\label{fig:fig3equilibrium}
\end{figure}

We now set $\kappa'=1$ and calculate the coexisting density and radius of the critical nuclei for van der Waals fluids.
This calculation was carried out similarly to those done for the active nuclei, only now with the van der Waals equation of state.
The results of this calculation are plotted in Fig.~\ref{fig:fig2equilibrium}.
In addition, we may calculate the nucleation barrier as a function of saturation and temperature using Eq.~\eqref{seq:potentialbarrier}, noting that $\upgamma_{\rm ost} = \upgamma_{\rm mech}$ for this passive system. 
The results of this calculation are plotted in Fig.~\ref{fig:fig3equilibrium}.
 \newpage
\section{\label{sec:coarsen}Active Coarsening Numerics}
\begin{figure}
	\centering
	\includegraphics[width=0.75\textwidth]{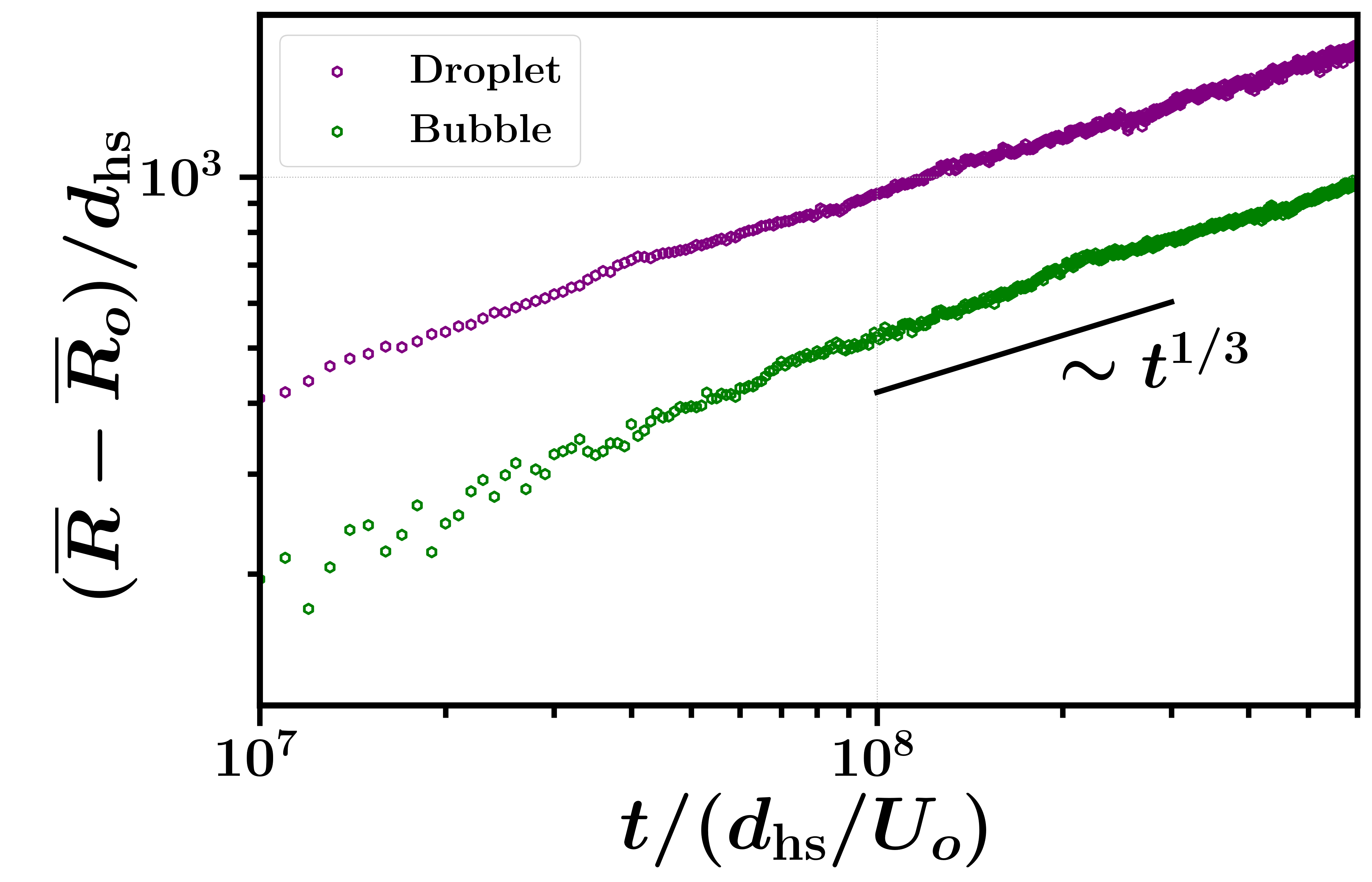}
	\caption{\protect\small{{Average radius growth as a function of time for nuclei intialized with a Gaussian distribution with mean $750 d_{\rm hs}$.}}}
	\label{fig:avgradiusgauss}
\end{figure}

\begin{figure}
	\centering
	\includegraphics[width=0.75\textwidth]{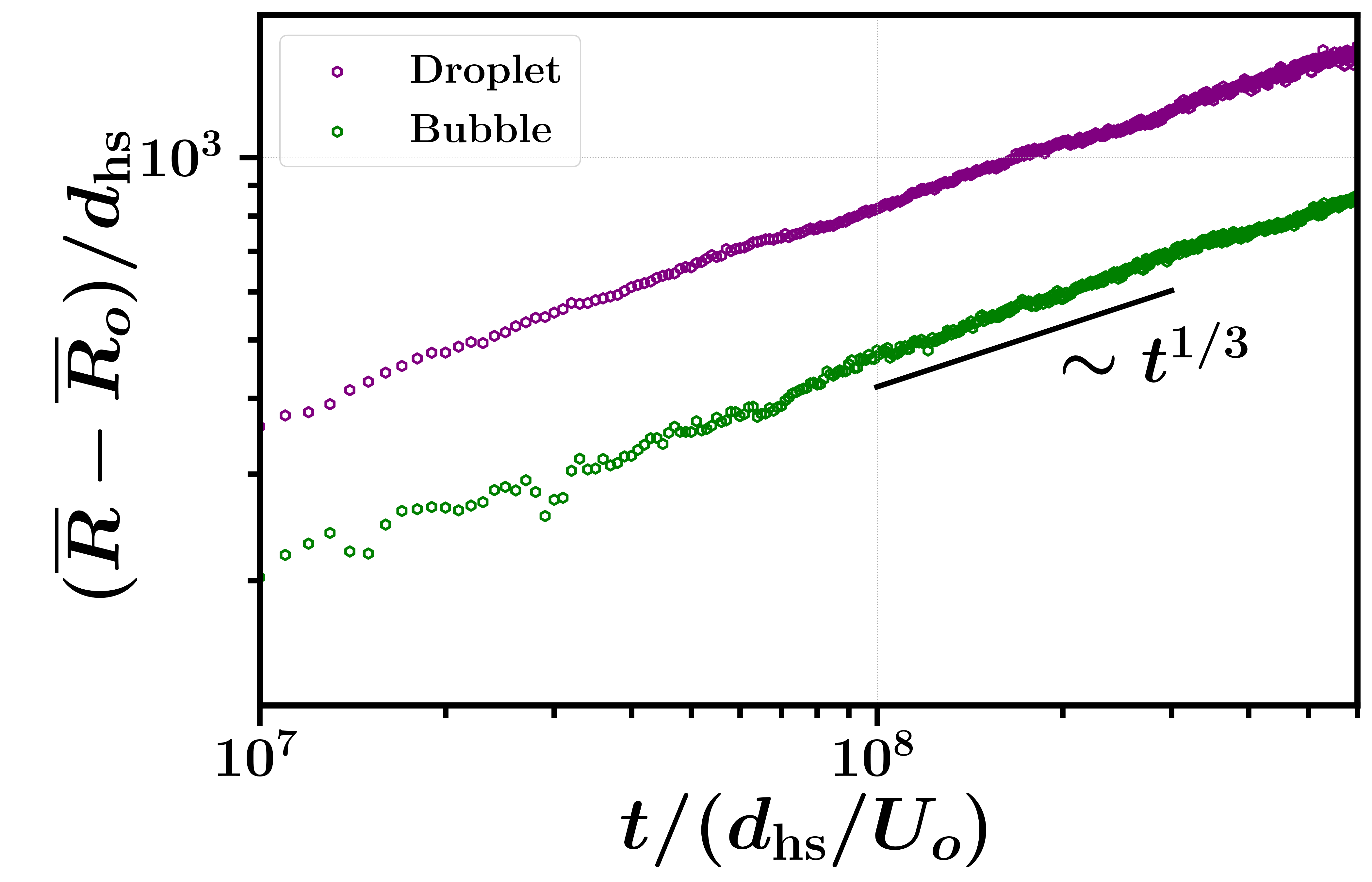}
	\caption{\protect\small{{Average radius growth as a function of time for nuclei initialized with an LSW distribution with mean $750 d_{\rm hs}$.}}}
	\label{fig:avgradiuslsw}
\end{figure}

\begin{figure}
	\centering
	\includegraphics[width=0.75\textwidth]{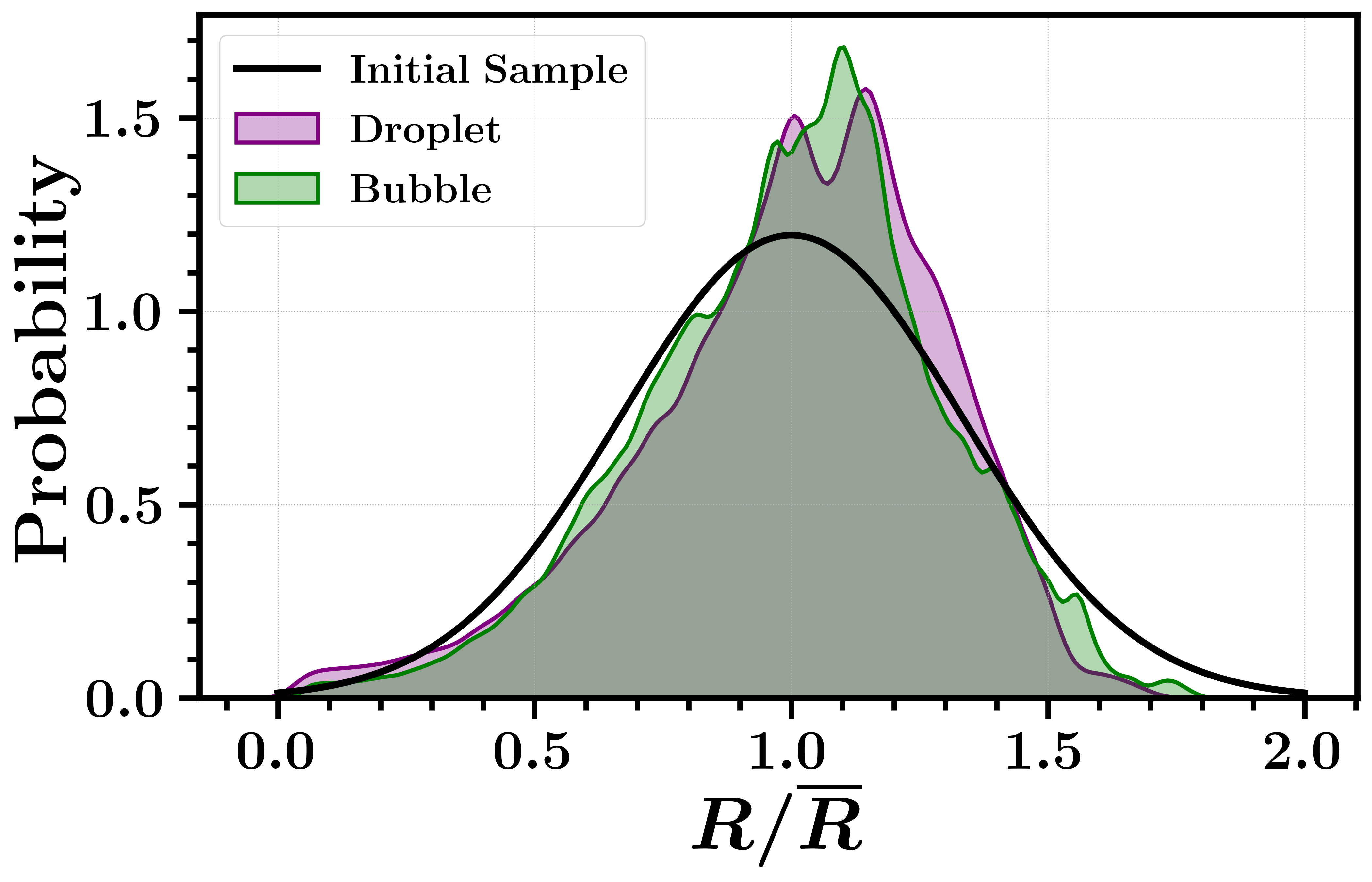}
	\caption{\protect\small{{Nucleus size distribution sampled from last third of simulation steps for particles initialized with Gaussian distribution.}}}
	\label{fig:gaussdistribution}
\end{figure}

\begin{figure}
	\centering
	\includegraphics[width=0.75\textwidth]{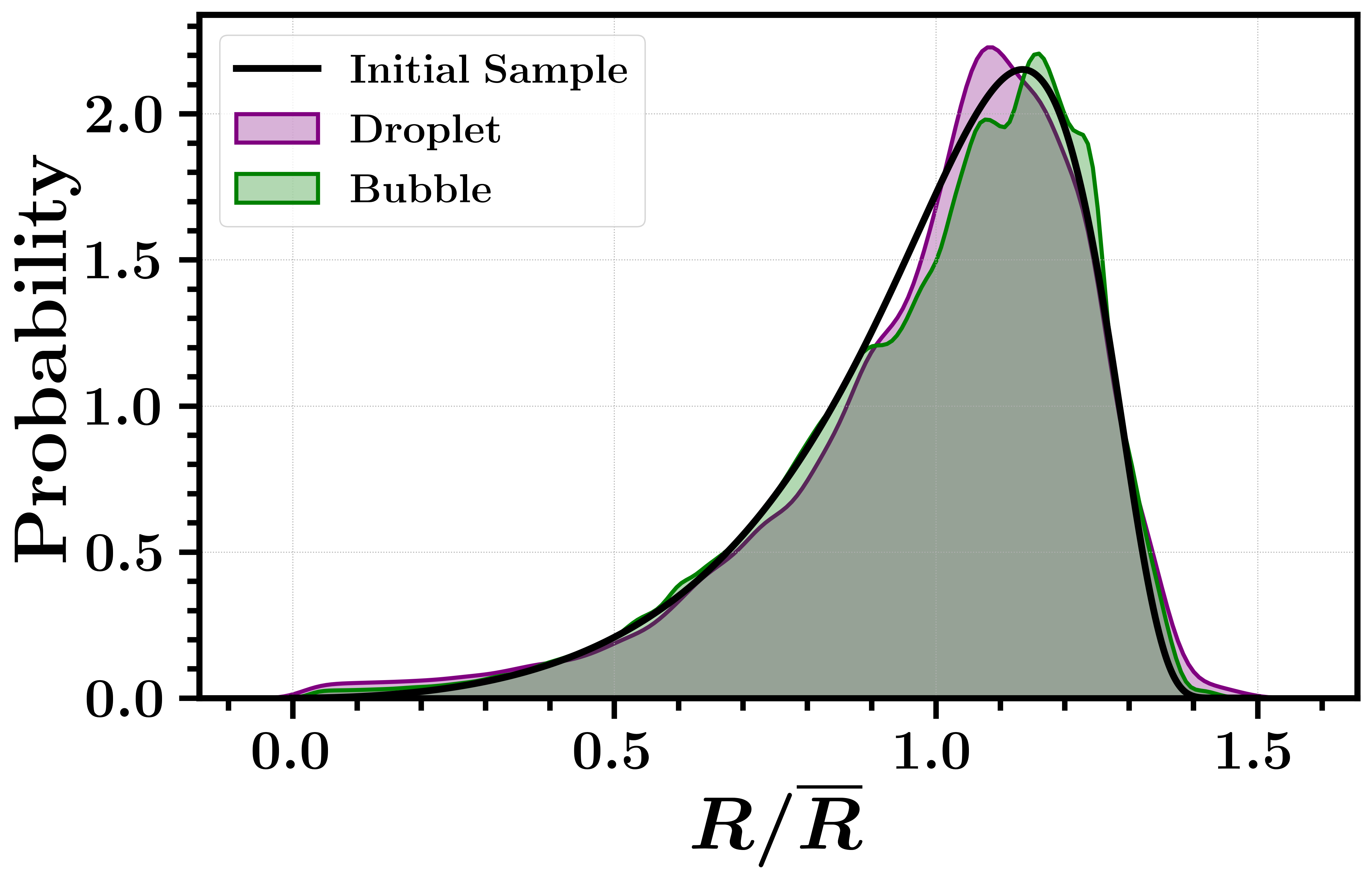}
	\caption{\protect\small{{Nucleus size distribution sampled from last third of simulation steps for particles initialized with LSW distribution.}}}
	\label{fig:lswdistribution}
\end{figure}

In this Section we report numerical results for the coarsening of a highly dilute ABP system, using the numerical method outlined in Appendix C, demonstrating the recovery of LSW theory~\cite{Lifshitz1961TheSolutions}. 
\noindent Given the periodic cell dimensions, nucleus locations, and nucleus radii, it is straightforward to set up the necessary matrix equations to find the coarsening rates and numerically integrate. 
We conduct numerical simulation of periodic cells of 250 nuclei radii randomly drawn from a distribution. 
The PACKMOL software package~\cite{Martinez2009PACKMOL:Simulations} was used to give random locations within the periodic cells such that all nuclei had a separation of at least two times the largest radius.
Particle sizes were sampled such that they had an average radius of $750d_{\rm hs}$ and a cubic cell of width $a_o = 2.1\times10^5 d_{\rm hs}$, corresponding to an overall volume fraction of $2\times10^{-7}$. 
For the overall phase fractions simulated, we expect the results of LSW theory~\cite{Lifshitz1961TheSolutions} to be recovered. 
The main predictions of LSW theory are the average nucleus size growing with the cube root of time and a steady-state distribution of nucleus sizes $P(R) \propto \mathcal{N}(x)$ at long times, where $x$ is the ratio of the nucleus radius to the average nucleus radius and $\mathcal{N}$ is proportional to:
\[ \begin{cases} 
      x^2(3+x)^{-7/3}\left(\frac{3}{2}-x\right)^{-11/3} \text{exp}\left(-\frac{3}{3-2x}\right)& x< 3/2 \\
      0 & x\geq 3/2 .
   \end{cases}
\]
We therefore initialized all simulations with radii sampled from a Gaussian distribution as well as initial radii sampled from the distribution predicted by LSW. 
Simulations were integrated in time for a duration of $6\times10^{8}d_{hs}/U_o$. 
Simulation results were averaged over 10 independent randomly generated initial positions and radii for both of the intial distributions.
We plot the average nucleus radius cubed as a function of time in Figs.~\ref{fig:avgradiusgauss} and~\ref{fig:avgradiuslsw}.
Growth rates demonstrate excellent agreement with the scaling predicted by LSW theory. 
Droplets coarsen significantly faster than bubbles due to the higher $\upgamma_{\rm ost}$ and therefore larger driving force for coarsening.
The distribution of nucleus sizes was also measured in the final third of the simulation time (i.e. times from $4-6\times10^{8}$) and are given by Figs.~\ref{fig:gaussdistribution} and~\ref{fig:lswdistribution}.
The nuclei initialized from Gaussian distribution veer from their initial sampling significantly by the end of the simulation time while nuclei initialized from the LSW distribution stay close to the LSW distribution, although with some error due to the finite amount of sampling.  \newpage
\clearpage

\addcontentsline{toc}{section}{References}
\bibliographystyle{bibStyle}